\documentclass[aps,showpacs,nofootinbib,superscriptaddress]{revtex4-1}
\usepackage{epsf}
\usepackage{graphicx}
\usepackage{amsmath}
\usepackage{mathrsfs}
\usepackage{epstopdf}
\usepackage{slashed}
\usepackage{wrapfig}
\usepackage{hyperref}
\hypersetup{
    colorlinks=true,
    linkcolor=blue,
    filecolor=magenta,      
    urlcolor=cyan,
    citecolor=blue,
}
\usepackage{cleveref}
\begin{document}

\title{In medium dispersion relation effects in nuclear inclusive reactions at intermediate and low energies}
\date{\today} 

\author{Juan Nieves}
\affiliation{Instituto de F\'\i sica Corpuscular (IFIC), Centro Mixto
CSIC-Universidad de Valencia, Institutos de Investigaci\'on de
Paterna, Apartado 22085, E-46071 Valencia, Spain} 
\author{Joanna E. Sobczyk}
\affiliation{Faculty of Physics and Astronomy, Wroclaw University, Wroclaw, Poland}

\begin{abstract}
In a well-established many-body framework, successful in modeling a great variety of nuclear processes, we analyze the role of the spectral functions (SFs) accounting for the modifications of  the dispersion relation of nucleons embedded in a nuclear medium. We concentrate in processes mostly governed by one-body mechanisms, and study possible 
approximations to evaluate the particle-hole propagator using SFs. We also investigate how to include together SFs and long-range RPA-correlation corrections in the evaluation of  nuclear response functions, discussing  
the existing interplay between both type of nuclear effects. At low energy transfers ($\le 50$ MeV), we compare our predictions for inclusive muon and radiative pion captures in nuclei, and charge-current (CC) neutrino-nucleus cross sections with experimental results. We also present an analysis of intermediate energy quasi-elastic neutrino scattering for various targets and both neutrino and antineutrino CC driven processes. In all cases, we pay special attention to estimate the uncertainties affecting  the theoretical predictions.  In particular, we show that errors on the $\sigma_\mu/\sigma_e$ ratio are much smaller than 5\%, and also  much smaller than the size of the SF+RPA nuclear corrections,  which produce significant effects, not only in the individual cross sections, but also in their ratio for neutrino energies below 
400 MeV. These latter nuclear corrections, beyond Pauli blocking, turn out to be thus essential  to achieve a correct theoretical understanding of this ratio of cross sections of interest for appearance neutrino oscillation experiments. We  also briefly compare our  SF and RPA results  to predictions obtained within other representative approaches.
\end{abstract}
\maketitle

\section*{Introduction}

The description of inclusive lepton-nucleus processes has attracted a lot of attention in the last years. The topic has become especially important in the context of neutrino physics~\cite{Gallagher:2011zza, Morfin:2012kn,Formaggio:2013kya, Alvarez-Ruso:2014bla, Mosel:2016cwa, Katori:2016yel}, where highly accurate theoretical  predictions are essential to conduct the analysis of neutrino properties aiming at making new discoveries possible, like the CP violation in the leptonic sector. For nuclear physics,  neutrino cross sections
incorporate richer information than electron-scattering ones, providing an excellent testing ground for nuclear structure, many-body
mechanisms and reaction models. In addition, neutrino cross-section measurements allow  to investigate the axial structure of the
nucleon and baryon resonances, enlarging the  views of hadron structure
beyond what is presently known from experiments with hadronic and
electromagnetic probes. Thus and besides the large activity in the last 15 years (see for instance the reviews cited above), a new wave of neutrino-nucleus theoretical works and detailed  analysis have recently become available \cite{Ankowski:2014yfa, Meucci:2014bva, Meucci:2015bea, Rocco:2015cil, Gallmeister:2016dnq, Martini:2016eec, Megias:2016fjk, Nakamura:2016cnn, Ankowski:2016jdd, Vagnoni:2017hll}.

 Neutrino and antineutrino scattering on nuclei without pions
exiting the nucleus is a fundamental detection channel for long-baseline neutrino experiments, such as T2K, MINOS, NOvA and the
future DUNE. At intermediate  energies, a  microscopical description of the interaction of the neutrinos with the nuclei, that form part of the detectors, 
should at least account for three distinctive nuclear corrections, in addition to the well-established  Pauli-blocking effects.  These are long-range collective RPA\footnote{RPA stands for the 
random phase approximation to compute the effects of long-range nucleon-nucleon correlations.} and in medium nucleon dispersion relation effects, and multinucleon absorption modes. In this work, we will focus in the first two ones, since  we will study processes mostly governed by one-body mechanisms.  There exists an abundant literature addressing multinucleon contributions to the pion-less quasi-elastic (QE) cross section in the context of the so-called MiniBooNE axial mass puzzle and the problem of the neutrino energy reconstruction~\cite{Morfin:2012kn,Alvarez-Ruso:2014bla,Katori:2016yel, Martini:2009uj,Martini:2010ex, Nieves:2011pp, Nieves:2011yp, Martini:2011wp,  Nieves:2012yz, Nieves:2013fr, Martini:2013sha,Gran:2013kda, Amaro:2011aa, Simo:2014esa, Megias:2014qva, RuizSimo:2016ikw}, and we refer the reader to these works for details. We would only like to mention that this topic has become quite  relevant in neutrino reactions since the neutrino 
beams are not monochromatic but wide-band~\cite{Benhar:2011ef, Nieves:2014lpa}.

Spectral functions (SFs) account for the modifications of  the dispersion relation of nucleons embedded in a nuclear medium, while medium polarization or collective RPA correlations 
do for the change of the electroweak coupling strengths, from their free nucleon values, due to the presence of strongly interacting nucleons. The latter take into account the absorption of the gauge boson, mediator of the interaction, by the nucleus as a whole instead of by an individual nucleon, and their importance decreases as the gauge boson wave-length becomes much shorter than the nuclear size. In medium dispersion relation effects associated to the hit nucleon are always evaluated  for bound nucleons, and thus their impact should be rather independent of the neutrino kinematics. However, one should expect that SF effects become less important in the case of the ejected nucleon, when the energy and momentum transfers are much larger than those accessible close to the Fermi sea level. Both SF~\cite{Benhar:2005dj, Benhar:2006nr, Benhar:2009wi, Benhar:2010nx, Vagnoni:2017hll, Leitner:2008ue, Nieves:2004wx, Nieves:2005rq, Ivanov:2013saa, Ivanov:2015wpa} and RPA, \cite{Singh:1992dc, Singh:1993rg, 
Singh:1998md,Volpe:2000zn,  Kolbe:2003ys,  
Nieves:2004wx, Nieves:2005rq, Valverde:2006zn, 
Martini:2009uj, Martini:2010ex, Martini:2011wp,  Nieves:2011pp, Nieves:2011yp, Jachowicz:2002rr, Pandey:2013cca, Pandey:2014tza, Martini:2016eec, Pandey:2016jju} corrections have been implemented in the calculation of neutrino QE cross sections at low and intermediate energies, following approaches previously tested in electro-nuclear reactions~\cite{Benhar:1991af, Benhar:1994hw, Benhar:2006wy, Rocco:2015cil, Alberico:1981sz, Leitner:2008ue, Gil:1997bm, Gil:1997jg, Antonov:2011bi, Pandey:2014tza}, and their relevance has been clearly shown. 

The theoretical concept of superscaling (a very weak dependence of the reduced cross section on the momentum transfer $q$ at excitation
energies below the QE peak for large enough $q$ and no dependence on the mass number) was introduced in Refs.~\cite{Alberico:1988bv, Barbaro:1998gu, Donnelly:1999sw} analyzing $(e,e')$ inclusive data. 
Though RPA effects cannot be taken into account within the superscaling approach (SuSA), at high momentum transfers it certainly incorporates SF corrections  based on the analysis of electron-nucleus scattering data. Thus, SuSA has been also used for analyses of neutrino-nucleus processes in numerous studies~\cite{Amaro:2004bs, Amaro:2006pr, Amaro:2010sd, Amaro:2011qb, GonzalezJimenez:2012bz, Gonzalez-Jimenez:2014eqa} providing a set of interesting predictions.

From a microscopical perspective, the combined effect of both SF and RPA corrections in neutrino reactions has been studied only within the 
model employed in Refs.~\cite{Nieves:2004wx, Nieves:2005rq}, and there SFs were implemented within certain approximations, which amount to neglect the width of the hole states.  Moreover, the low energy results (inclusive muon capture rates and   $^{12}{\rm C}(\nu_\mu,\mu^-)X$
 and $^{12}{\rm C}(\nu_e,e^-)X$ cross sections near threshold) presented in \cite{Nieves:2004wx} did not include SF effects. In this work,  we perform a careful analysis of RPA and SF nuclear effects, paying special attention to the existing interplay between them. Moreover, at low energies, we use full SF response functions and we also study  the inclusive radiative pion capture reaction, which  at the nucleon level is a much simpler process
 and thus, it better illustrates the role played by RPA and SF corrections and the possible deficiencies of the model of Ref.~\cite{Nieves:2004wx}, when tested at  very low excitation energies almost beyond its scope of applicability. The  many-body model used in this work has been  successfully applied in the past to describe photon,
electron, pion, kaon, $\Lambda, \Sigma-$hyperons etc. interactions with
nuclei~\cite{Oset:1981ih, Oset:1989ey,Carrasco:1989vq, FernandezdeCordoba:1991wf, FernandezdeCordoba:1992df, Nieves:1991ye,  FernandezdeCordoba:1992ky, Nieves:1992pm, Nieves:1993ev, Hirenzaki:1993jc, Carrasco:1992mg, Oset:1994vp, FernandezdeCordoba:1993az,  Oset:1994vp, GarciaRecio:1994cn, Hirenzaki:1995js, Gil:1997bm, Gil:1997jg, Albertus:2001pb,Albertus:2002kk}, and it was then extended to study charged-current (CC) \cite{Nieves:2004wx} and neutral-current (NC) \cite{Nieves:2005rq} (anti-)neutrino-nucleus interactions\footnote{For a recent review and compilation of results see Ref.~\cite{Nieves:2016yfc}.}. It aims to describe a wide range of nuclear processes (QE, muon and radiative pion captures, pion production, two-body processes) induced by electroweak probes. Being firstly compared with the existing data for inclusive electron scattering at intermediate energies~\cite{Gil:1997bm}, the model has proven to perform very well. 

 Besides the low-energy results mentioned above, we also present an analysis of intermediate energy QE neutrino scattering, in the range of energy transfers up to 400 MeV, for various targets of interest for oscillation experiments,  and both neutrino and antineutrino CC driven processes. We use full SFs, and improve also here the approach followed  in Ref.~\cite{Nieves:2004wx}, since we do  not neglect either at these energies  the width of the hole states. In all cases, we pay special attention to estimate the uncertainties affecting  the theoretical predictions, for which we use Monte Carlo simulations.  In particular, we show that errors on the $\sigma_\mu/\sigma_e$ ratio are much smaller than 5\%, and also  much smaller than the SF+RPA nuclear corrections,  which produce significant effects, not only in the individual cross sections, but also in their ratio for neutrino energies below 
400 MeV. These latter nuclear corrections, beyond Pauli blocking, turn out to be thus essential  to achieve a correct theoretical understanding of this ratio of cross sections of interest for appearance neutrino oscillation experiments. 

We will use the SFs derived in \cite{FernandezdeCordoba:1992df} to account for the modifications of  the dispersion relation of nucleons embedded in a nuclear medium, and their effects, with and without the inclusion of RPA corrections will play a central role in our discussions. Indeed, we will see how RPA (SF) effects in integrated decay rates or cross sections  become significantly smaller when SF (RPA) corrections are also taken into account. This interesting result was mentioned for the very first time in \cite{Nieves:2004wx}, and it is discussed in detail here. In particular at low energies, this interplay between both types of nuclear corrections becomes quite apparent, and it had not been addressed yet.  Modifications of  the differential-distributions shapes are, however,  always significant and relevant.

 Taking advantage of this work, we also redo some calculations presented in Ref.~\cite{Nieves:2004wx, Valverde:2006zn}, as they contain a small error in a form--factor used in the numerical computations. The error does not change the qualitative features (magnitude and
 behaviour of the RPA or SFs effects, size of the theoretical uncertainties, etc.) discussed in these references, since it only affects  the size of the elementary cross section on the nucleon. However, it produces some numerical effects (higher cross sections) which are around 20\% at most\footnote{The numerical results reported in  \cite{Nieves:2004wx, Nieves:2005rq, Valverde:2006zn} are inexact because of a mistake in the calculation of the magnetic form-factor $F_2^V$. To be more precise, the contribution to this form factor proportional to the neutron magnetic moment, $\mu_n$, was incorrectly taken as (see the notation of footnote 4 of Ref.~\cite{Nieves:2004wx})
\begin{equation}
-\frac12\frac{G_E^p}{1+\tau}   \frac{1+\lambda_n\tau+\tau}{1+\lambda_n}, \qquad \tau = -q^2/4M^2
\end{equation}
in the numerical computations. The correct expression is obtained from the above equation replacing  the denominator $(1+\lambda_n)$ by $(1+\lambda_n\tau)$, as can be seen for instance in Ref.~\cite{Nieves:2004wx}.
The mistake only affected to the QE cross sections and it was found in 2006, and all results published from 2007 on were obtained using a correct expression for this form-factor. The codes based in  \cite{Nieves:2004wx} that have been distributed  are free from this error as well.}. The updated results for neutrino scattering and muon capture can be found in this work.

In order to make this work self-explanatory, in Sec. \ref{sec:form} we start by sketching the formalism, which was presented in full details in \cite{Nieves:2004wx, Chiang:1989ni},  and in Sec. \ref{sec:muon_pion} we extend the scheme to study the inclusive muon and radiative pion captures in nuclei. Particle and hole  spectral functions ($S_p$ and $S_h$) and  long-range RPA correlations, in the context of an interacting local Fermi gas (LFG), are introduced and discussed in Sec. \ref{sec:nuc_cor}. A first brief analysis of the $S_{p,h}$ effects in the imaginary part of the particle-hole Green function is carried out in Sec. \ref{sec:analysis}.  These latter effects, together with those induced by the RPA re-summation, are fully discussed in Subsec.~\ref{sec:results_neutrino} for cross sections off argon, carbon and oxygen targets of interest for neutrino oscillation experiments, paying a special attention to  the $\sigma_\mu/\sigma_e$ ratio. We end up this subsection with a brief comparison of our 
predictions with those obtained 
within other representative approaches (Subsec.~\ref{sec:comp}). Total and differential decay rates for inclusive muon and 
radiative pion captures in nuclei are obtained in Subsecs.~\ref{sec:pioncapt} and  \ref{sec:muoncapt}, while the inclusive $^{12}{\rm C}(\nu_\mu,\mu^-)X$ and $^{12}{\rm C}(\nu_e,e^-)X$ reactions near threshold are analyzed in Subsec.~\ref{sec:lowcross}. Finally, we summarize the most important results of this work in Sec.~\ref{sec.concl}.

\section{Formalism and general considerations}\label{sec:form}
Let us consider the inclusive CC scattering of a neutrino\footnote{The generalization of the obtained expressions to antineutrino induced reactions, NC processes, or inclusive muon capture in nuclei is straightforward.} off a nucleus, $\nu_l+ A_Z \to l^-+X$.  The inclusive differential cross section in the laboratory frame for this process takes the form:
\begin{equation}\label{eq:1}
\frac{d^2\sigma}{d\Omega (\hat{k}') d E_l'} =\bigg(\frac{G_F}{2\pi}\bigg)^2 \frac{|\vec{k}|}{|\vec{k}'|} L_{\mu\nu} W^{\mu\nu}
\end{equation}
where $k^\mu$, $k'^\mu$ are the  incoming and outgoing lepton four-momenta, respectively. The Fermi constant $G_F=\sqrt{2}g^2/8M_W^2$ combines the gauge coupling constant ($g$)  and the mass ($M_W$)  of the $W$ gauge boson. The lepton tensor is given by ($\epsilon_{0123}=+1$):
\begin{equation}
L_{\mu\nu} = k'_{\mu} k_{\nu}+k_{\mu} k'_{\nu} - g_{\mu\nu} k\cdot k' +i\epsilon_{\mu\nu\alpha\beta}k'^{\alpha} k^{\beta}
\end{equation}
while the hadron tensor reads (see Ref.~\cite{Nieves:2004wx} for further details):
\begin{equation}\label{eq:2}
 W^{\mu\nu} = \frac{1}{2M_i} \overline{\sum}_f (2\pi)^3 \delta^4(P'_f-P-q) \langle f| j_{CC}^{\mu}(0)|i\rangle  \langle f| j_{CC}^{\nu}(0)|i\rangle ^*
\end{equation}
with $P^\mu$ the four-momentum of the initial nucleus $i$, $M_i=P^2$ the target nucleus mass, $P'_f$  the total four-momentum of the
hadronic final state $f$, and $q=k-k'$ the four-momentum transferred to the nucleus. The bar over the sum denotes the average
over initial spins and the CC is
\begin{equation}
j_{CC}^{\mu} = \bar{\Psi}_u\gamma^{\mu} (1-\gamma^5)(\cos\Theta_C \Psi_d + \sin\Theta_C \Psi_s)
\end{equation}
with $\Psi_{u,s,d}$ flavor quark fields and $\Theta_C$ the Cabibbo angle. The symmetry patterns governing  QCD  are 
assumed to work also for hadrons. This fact will allow us to construct in the next sections the hadron tensor for any of the reactions studied in this work.

The hadronic tensor has also symmetric and antisymmetric parts, the latter one being purely imaginary, which guaranties that the contraction with the lepton tensor gives a real value,  $W^{\mu\nu} = W^{\mu\nu}_s+iW^{\mu\nu}_a$. It can be expressed in terms of structure functions (Lorentz scalar real functions of $q^2$) using only the two four-vectors available, $q=k'-k$ and $P$:
\begin{equation}
\frac{W^{\mu\nu} }{2M_i}= -g^{\mu\nu} W_1 +\frac{P^{\mu}P^{\nu}}{M_i^2} W_2 + i\frac{\epsilon^{\mu\nu\delta\sigma}P_{\delta} q_{\sigma}}{2M_i^2}W_3+\frac{q^{\mu}q^{\nu}}{M_i^2}W_4 + \frac{P^{\mu}q^{\nu}+P^{\nu}q^{\mu}}{2M_i^2}W_5+i\frac{P^{\mu}q^{\nu}-P^{\nu}q^{\mu}}{2M_i^2}W_6
\end{equation}
Let us notice that it can be greatly simplified by choosing a natural reference frame of the nucleus at rest.

\subsection{Hadron tensor}\label{hadr-ten}
Even though we know the expression for the hadron tensor (Eq. (\ref{eq:2})), it should be evaluated for the nuclear configurations of interest, which makes the problem in general quite  demanding. To simplify the task, we will adopt the approximation of working in  nuclear matter and use the local density approximation (LDA) to obtain results in finite nuclei. 

As an introduction to the formalism used here, we show the relation between the hadron tensor and the self-energy of the gauge boson embedded in the nuclear medium. The following discussion (taken from  Refs.~\cite{Gil:1997bm,Nieves:2004wx}) can be performed for any gauge boson and incoming lepton. We present it for the CC neutrino-nucleus scattering, because this is a process which will be analyzed in Sec. \ref{sec:results_neutrino}. The approach consists of:

\begin{figure}[h]
 \includegraphics[scale=0.8]{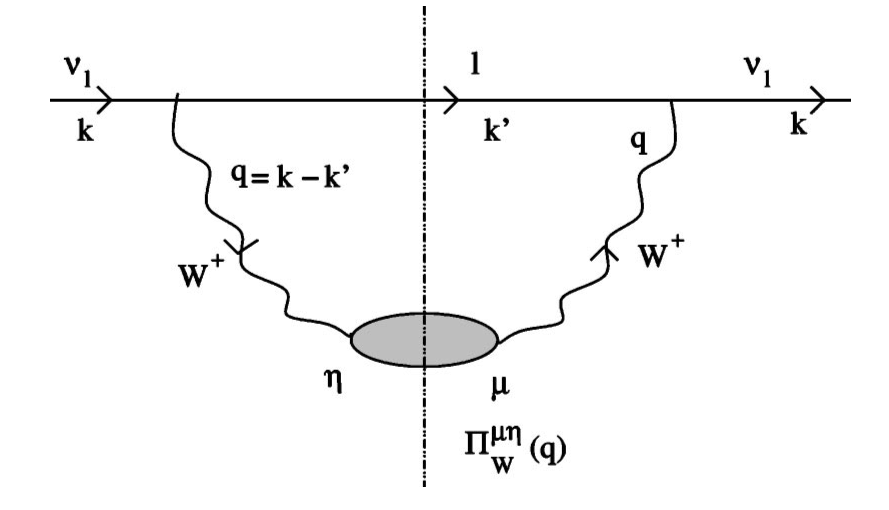}
\caption{Neutrino ($\nu_l$) self-energy in nuclear matter: in first approximation, the neutrino interacts via a $W^+$ boson, producing an intermediate lepton state $l$.}
\label{fig:1}
\end{figure}

\begin{enumerate}
\item Calculating the self-energy of the incoming lepton in the nuclear medium. It  depends on the self-energy of the gauge boson, denoted as $\Pi^{\mu\eta}_W$. 

The self-energy $\Sigma^r_{\nu}(k;\rho)$ of a neutrino of helicity $r$ and momentum $k$ in  nuclear matter of density $\rho$, 
at leading order in the Fermi constant, is  diagrammatically shown in Fig. \ref{fig:1}. This loop diagram is given by:
\begin{eqnarray}
\begin{split}
-i\Sigma^r_{\nu}(k;\rho) = \int \frac{d^4q}{(2\pi)^4} \bar{u}_r(k) & \Big[  -i\frac{g}{2\sqrt{2}}\gamma_L^{\mu}iD_{\mu\alpha}(q)(-i)\Pi_W^{\alpha\beta}(q;\rho)\\
& iD_{\beta\sigma}(q)i\frac{\slashed{k'}+m_l}{k'^2-m_l^2+i\epsilon} (-i\frac{g}{2\sqrt{2}})\gamma_L^{\sigma}  \Big] u_r(k)
\end{split}
\end{eqnarray}
where we have Dirac spinors $u_r(k)$ (with normalization $\bar u  u= 2m$) projected only to left-handed neutrinos by $\gamma_L^\mu= \gamma^\mu (1-\gamma_5)$, and the $W^{\pm}$ propagator, which for a low energy transfer becomes $D_{\mu\nu}(q) = -g_{\mu\nu}/M_W^2$ leading to a contact interaction. The sum over lepton spins produces a trace which results in the lepton tensor $L_{\mu\nu}$, 
\begin{equation}
\Sigma_{\nu}(k;\rho)=\frac{8iG_F}{\sqrt{2}M^2_W}\int \frac{d^4q}{(2\pi)^4}\frac{L_{\mu\nu}\Pi^{\nu\mu}_W(q;\rho)}{k'^2-m_l^2+i\epsilon}
\end{equation}
\item Relating the  lepton scattering cross section  with the imaginary part of its self-energy, which is computed by means of the Cutkosky cutting rules.
\\
The first step is to relate the imaginary part of the self-energy with the decay width of the particle,
\begin{equation}
\Gamma (k;\rho) = -\frac{1}{k^0} \text{Im}\Sigma_{\nu}(k;\rho)\label{eq:decay}
\end{equation}
To obtain $\text{Im}\Sigma_\nu$ we cut the loops of the Feynman diagram as shown in Fig. \ref{fig:1} by a vertical line putting on-shell the intermediate lepton ($l$) and the particles that are exchanged in the loops of the  $W$ self-energy (we have still not shown them explicitly). This allows to perform the integration over the energy,  and thus we get for $k^0 > 0$
\begin{equation}\label{eq:imsigma}
\text{Im}\Sigma_{\nu}(k;\rho)=\frac{8G_F}{\sqrt{2}M^2_W}\int \frac{d^3k'}{(2\pi)^3}\frac{\Theta(q^0)}{2E'_l} \text{Im}(L_{\nu\mu}\Pi^{\mu\nu}_W(q;\rho))
\end{equation}
Having this result, we next  relate the cross section with the decay
width of the particle. The probability of decay (interaction) is given
by $\Gamma dt$. The cross section measures the probability of interaction per unit of  area, $\sigma = \Gamma dt dS$, and since the 
integration over time may be related to an integration over space, $dt = v dx$ (where $v$ is a velocity of the particle), we obtain
\begin{equation}
d\sigma = \Gamma dt dS = \Gamma v dx dS = \Gamma v d^3r
\end{equation}
which leads to
\begin{equation}
\sigma = -\frac{1}{|\vec{k}|}\int \text{Im}\Sigma_\nu(k;\rho) d^3r
\end{equation}
Here we should make an important remark. The above derivation has been performed for nuclear matter of a constant density $\rho$. By means of the LDA, we can obtain results for finite nuclei.  At each point of the space, we calculate $\Sigma_\nu(k;\rho(r))$ in  infinite nuclear matter of  constant-density $\rho(r)$. Then we integrate over the volume (nucleus) taking into account that the density changes with the radius. This approximation is quite accurate for the study of inclusive responses to weak probes, which explore the whole nuclear volume, as shown in \cite{Carrasco:1989vq, Carrasco:1992mg}. 

Thus, the relation between the inclusive cross section and the gauge boson self-energy reads
\begin{equation}\label{eq:3}
\sigma = -\frac{1}{|\vec{k}|}\int \text{Im}\Sigma(k;\rho) d^3r = -\frac{1}{|\vec{k}|}\frac{8G_F}{\sqrt{2}M^2_W}\int d^3r\int \frac{d^3k'}{(2\pi)^3}\frac{\Theta(q^0)}{2E_l'} \text{Im}(L_{\mu\nu}\Pi^{\nu\mu}_W(q;\rho))
\end{equation}

\item Finally, the comparison of equations (\ref{eq:1}) and (\ref{eq:3})  allows to relate the hadron tensor to the gauge boson self-energy. Decomposing  the contraction\footnote{Note that lepton tensor splits also into a symmetric and antisymmetric part, i.e.,  $L_{\mu\nu} = L_{\mu\nu}^s+iL_{\mu\nu}^a$.} $L_{\nu\mu}\Pi^{\mu\nu}_W$ we get:
\begin{eqnarray}
\begin{split}
\sigma = -\frac{1}{|\vec{k}|}\frac{4G_F}{\sqrt{2}M^2_W}\int d^3r \int \frac{d^3k'}{(2\pi)^3}\frac{\Theta(q^0)}{2E_l'}& \bigg(L_{\mu\nu}^s \text{Im}(\Pi^{\mu\nu}_W(q;\rho)+\Pi^{\nu\mu}_W(q;\rho)) \\
&- L_{\mu\nu}^a \text{Re}(\Pi^{\mu\nu}_W(q;\rho)-\Pi^{\nu\mu}_W(q;\rho)) \bigg)
\end{split}
\end{eqnarray}
from where one obtains
\begin{eqnarray}
\begin{split}
W^{\mu\nu}_s = -\Theta(q^0)\bigg( \frac{2\sqrt{2}}{g}\bigg)^2\int \frac{d^3r}{2\pi} \text{Im}(\Pi^{\mu\nu}_W(q;\rho)+\Pi^{\nu\mu}_W(q;\rho))\\
W^{\mu\nu}_a = -\Theta(q^0)\bigg( \frac{2\sqrt{2}}{g}\bigg)^2\int \frac{d^3r}{2\pi} \text{Re}(\Pi^{\mu\nu}_W(q;\rho)-\Pi^{\nu\mu}_W(q;\rho))
\end{split}
\end{eqnarray}

\end{enumerate}
\begin{figure}[h]
\centering
\includegraphics[scale=0.8]{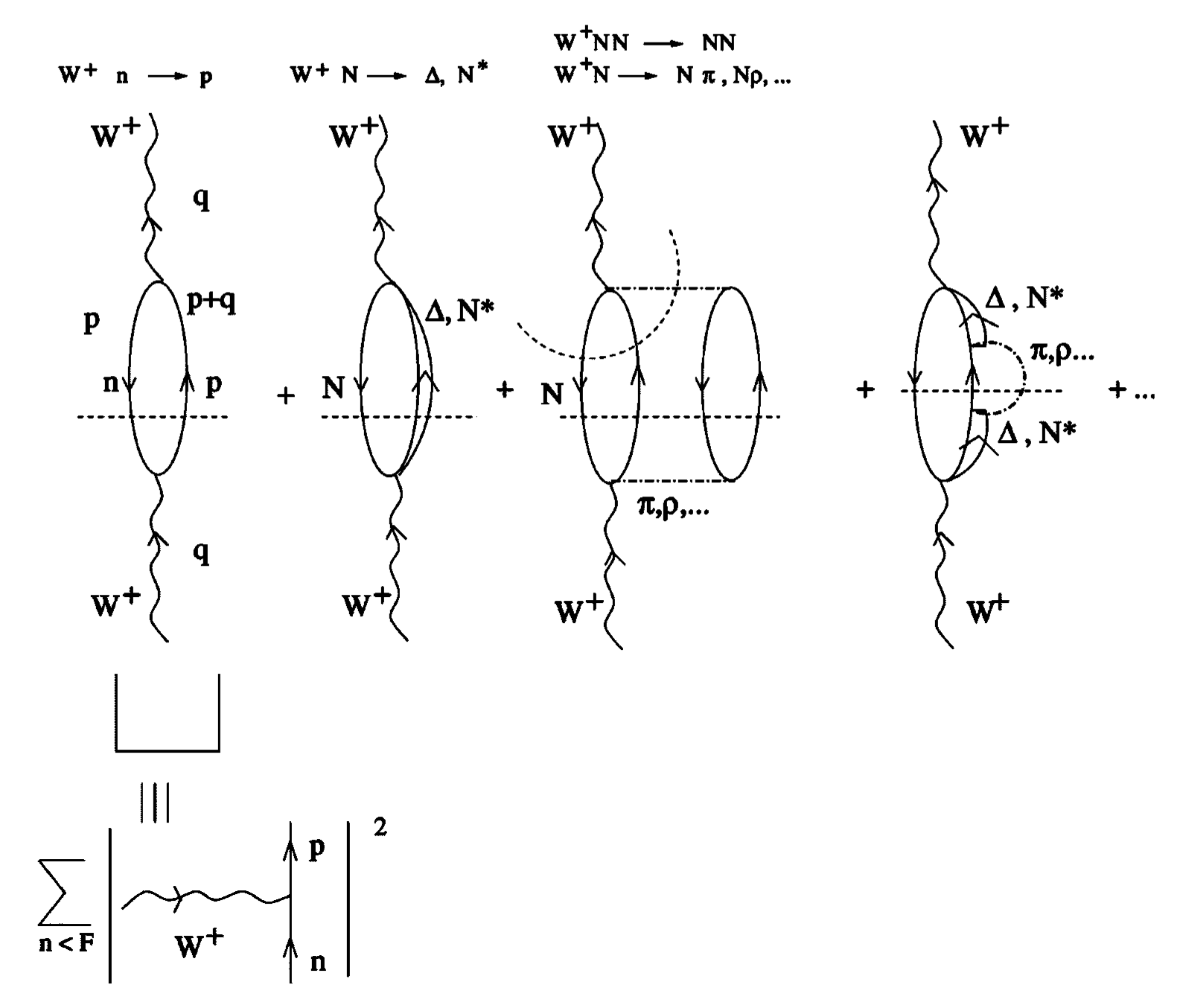}
\caption{Different contributions to the $W^+$ self-energy in nuclear matter.}
\label{fig:2}
\end{figure}
The self-energy of the gauge boson contains all possible modes of nuclear excitations: $1p1h$, $1p1h1\pi$, $2p2h$, $\Delta h$, etc., which are shown in Fig. \ref{fig:2}, where $ph$ ($\Delta h$) stands for the nuclear excitation of a particle--hole ($\Delta(1232)$--hole)) pair~\cite{fetterwalecka, Nieves:2016yfc}.
All these contributions were computed in a series of publications \cite{Carrasco:1989vq, Gil:1997bm, Nieves:2004wx, Nieves:2011pp, Nieves:2005rq} for real and virtual photons and CC and NC neutrino inclusive reactions. 

\subsection{An example: Charge current quasi-elastic (CCQE) reactions}\label{sec:cc}

In this section we will focus on one of the absorption modes listed in the previous subsection, $1p1h$, where the gauge boson interacts with just one nucleon of the nucleus, transferring to it the whole four-momenta $q^\mu$. 
It corresponds to the first diagram depicted in Fig. \ref{fig:2}.
\\
This loop diagram contains two basic ingredients: i) the interaction vertex of two nucleons and the gauge boson in the free space, and ii) the nucleon propagator inside of the nuclear medium.

The $W^+pn$ interaction vertex  has a vector-axial form:
\begin{equation}
< p; \vec{p}^{~\prime}=\vec{p}+\vec{q}~ | j^\alpha_{cc}(0) | n;
\vec{p}~> =
\bar{u}(\vec{p}{~^\prime})\Gamma^{\mu}u(p), \qquad \Gamma^{\mu}= V^{\mu}-A^{\mu}
\end{equation}
To construct the vertex, we consider Lorentz invariance, together with QCD symmetries and make use of  conservation and partial conservation of the vector and axial currents, CVC and PCAC, respectively. Both the vector and axial parts of the vertex can be expressed in terms of the form--factors that depend on $q^2$, the  only scalar at our disposal when dealing with on-shell nucleons,
\begin{eqnarray}
\begin{split}
&V^{\mu}=2\cos\theta_C\bigg(F_1^V(q^2)\gamma^{\mu}+i\mu_V\frac{F_2^V(q^2)}{2M}\sigma^{\mu\alpha}q_{\alpha}\bigg)\\
&A^{\mu}=\cos\theta_C G_A(q^2)\bigg(\gamma^{\mu}\gamma_5+\frac{2M}{m_{\pi}^2-q^2}q^{\mu}\gamma_5\bigg)
\end{split}
\end{eqnarray}
The structure of the vertex is the same for CC, NC and electromagnetic (EM) interactions (omitting the axial part in the latter case), and the difference lies in the form--factors, which are also related to each other by  isospin symmetry. The relations between vector CC and EM form-factors read~\cite{Nieves:2004wx}
\begin{eqnarray}
\begin{split}
&F_1^V(q^2) = \frac{1}{2}\big[F_1^p(q^2)-F_1^n(q^2)\big]\\
&\mu_V F_2^V(q^2) = \frac{1}{2}\big[\mu_pF_2^p(q^2)-\mu_nF_2^n(q^2)\big]
\end{split}
\end{eqnarray}
Among many existing form--factors' parameterizations,  we will use one by Galster et al.,~\cite{Galster:1971kv}, which specific details were compiled in \cite{Nieves:2004wx}.
The nucleon propagator in a free  local Fermi gas (LFG) has a form closely related to the free fermion propagator, except for the Pauli blocking factor, 
\begin{eqnarray}\label{eq:4}
\begin{split}
S(p;\rho) &= (\slashed{p}+M) G(p;\rho)\\
G(p;\rho) &= \frac{\Theta(k_F-|\vec{p}\,|)}{p^2-M^2-i\epsilon}+\frac{\Theta(|\vec{p}\,|-k_F)}{p^2-M^2+i\epsilon}\\
&=\frac{1}{p^0+E_p+i\epsilon}\,\frac{\Theta(k_F-|\vec{p}\,|)}{p^0-E_p-i\epsilon}+\frac{\Theta(|\vec{p}\,|-k_F)}{p^0-E_p+i\epsilon}\,\frac{1}{p^0+E_p-i\epsilon}
\end{split}
\end{eqnarray}
with $M$ the nucleon mass, $E_p=\sqrt{M^2+\vec{p}^{\,2}}$ and $k_F=(3\pi^2\rho/2)^{1/3}$, the Fermi momentum in  symmetric nuclear matter (the asymmetric matter would require different levels for protons and neutrons depending on their respective densities $\rho_p$, $\rho_n$). The first term of $G(p;\rho)$ represents a  hole state (nucleon below the Fermi level) and the second one  a particle state (nucleon above the Fermi level). The non-relativistic reduction of the nucleon propagator is obtained by approximating  $1/\left(p^0+E_p\pm i\epsilon\right)\sim 1/2M$ in Eq.~\eqref{eq:4}. With all these ingredients we can calculate the $1p1h$ contribution to the $W^+$ self-energy that reads~\cite{Nieves:2004wx}
\begin{eqnarray}
\begin{split}
-i\Pi^{\mu\nu}(q;\rho) &= -\left( \frac{g}{2\sqrt{2}} \right)^2 \int \frac{d^4 p}{(2\pi)^4} \text{Tr}[\Gamma^{\nu}(q) S(p;\rho)\bar\Gamma^{\mu}(q) S(p+q;\rho)]\\
&= -\cos^2\theta_C  \left( \frac{g}{2\sqrt{2}} \right)^2 \int \frac{d^4 p}{(2\pi)^4} A^{\mu\nu}(p,q) G(p;\rho) G(p+q;\rho)
\end{split}
\end{eqnarray}
where $\bar\Gamma^{\mu} = \gamma^0 \left(\Gamma^{\mu}\right)^\dagger \gamma^0$, and
\begin{equation}
 A^{\mu\nu}(p,q) = \frac{1}{\cos^2 \theta_C} \text{Tr} [\bar\Gamma^{\mu}(q) (\slashed{p}+\slashed{q}+M) \Gamma^{\nu}(q)(\slashed{p}+M) ]
\end{equation}
After integration over $p^0$ using Cauchy's theorem, we obtain 
for isospin symmetric nuclear matter
\begin{eqnarray}\label{eq:hadron_tensor}
W^{\mu\nu}(q) & = & -\frac{\cos^2\theta_C}{2M^2}\int_0^{\infty} dr r^2 \left\{ -\Theta(q^0) \int\frac{d^3p}{(2\pi)^2} \frac{M}{E_p}\frac{M}{E_{p+q}} \delta(q^0+E_p-E_{p+q}) \right. \nonumber\\
&  \times & \left. \Theta(k_F(r)-|\vec{p}\,|)\Theta(|\vec{p}+\vec{q}\,|-k_F(r))A^{\nu\mu}(p,q)\bigg|_{p^0=E_p} \right\} \label{eq:lindfree}
\end{eqnarray}
The generalization for asymmetric matter  is straightforward and we use it when studying nuclei where $\rho_p\ne \rho_n$. Since we are considering a neutrino CC  process, there will appear different Fermi levels for particle and hole states  $\Theta(k_F-|\vec{p}\,|), \Theta(|\vec{p}+\vec{q}\,|-k_F) \rightarrow \Theta(k_{F\text{hole}}-|\vec{p}\,|)\Theta(|\vec{p}+\vec{q}\,|-k_{F\text{particle}})$, where the corresponding Fermi levels will be determined by the neutron and proton densities. 

The hadron tensor, in this approximation, is determined by the  imaginary part of the $1p1h-$propagator. Using the  free nucleon propagator defined in Eq.~(\ref{eq:4}), we build the  $1p1h-$propagator - known also as the Lindhard function~\cite{fetterwalecka, Nieves:2016yfc}  that reads
\begin{equation}\label{eq:lindhard}
 {\bar U}(q;\rho) = -2 i \int \frac{d^4p}{(2\pi)^4}2MG(p;\rho)2MG(p+q;\rho) 
\end{equation}
The factor 2 comes from summing over spin. We do not sum over isospin (which would give another factor 2). In Sec. \ref{sec:pion_capture} we will introduce $U_N = 2 \bar U$ which is the nucleon Lindhard function summed over isospin. It is used in the denominator of the RPA response function,  which evaluation requires the sum over all possible intermediate $ph$ excitations. Integrating over $p^0$ we find
\begin{equation}\label{eq:lind}
{\bar U}(q;\rho) = 2\int \frac{d^3p}{(2\pi)^3}\frac{M}{E_p}\frac{M}{E_{p+q}} \frac{\Theta(k_F-|\vec{p}\,|)\Theta(|\vec{p}+\vec{q}\,|-k_F)}{q^0+E_p-E_{p+q}+i\epsilon}+ \cdots (q\rightarrow -q) 
\end{equation}
where some real terms for $q^2< 4M^2$, and suppressed in the non-relativistic limit, have been neglected.\footnote{The contribution of the free space loop function is also included in the definition given in Eq.~\eqref{eq:lindhard}. For $q^2 \ge 4M^2$,  the free space loop gets an imaginary part due to 
the  creation of a nucleon-antinucleon pair ($ph$ excitation of the Dirac  instead of the Fermi sea, using the terminology of Ref.~\cite{Nieves:2016yfc}), while its logarithmically divergent real part renormalizes the properties (mass and couplings) of the nucleon in the free space. Note that non-zero imaginary parts for $q^2<0$ are only produced by $ph$ excitations around the Fermi level.} 

The imaginary part of $\bar{U}(q;\rho)$ is easily obtained using the distribution identity 
\begin{equation}
\frac{1}{\omega\pm i\epsilon} = \mathcal{P}\left(\frac{1}{\omega}\right) \mp i\pi\delta(\omega)
\end{equation}
where $\mathcal{P}$ stands for the principal value. The second term ($q\rightarrow -q$) in Eq. (\ref{eq:lind}) describes a crossed term which does not contribute to the imaginary part when $q^0>0$, and thus we find
\begin{equation}
\text{Im}{\bar U}(q;\rho) = -\Theta(q^0)\int\frac{d^3p}{(2\pi)^2}\frac{M}{E_p}\frac{M}{E_{p+q}} \delta(q^0+E_p-E_{p+q}) \Theta(k_F-|\vec{p}\,|)\Theta(|\vec{p}+\vec{q}\,|-k_F)
\end{equation}
which appears between the curly brackets of the expression for the hadron tensor in Eq.~\eqref{eq:hadron_tensor}.
The integral above may be analytically calculated, even after introducing $A^{\mu\nu}$ as required to find the hadronic tensor for a non-interacting LFG. Expressions can be found in Appendix B of Ref.~\cite{Nieves:2004wx}. 

The non-relativistic reduction ($\text{Im} \bar U_{\rm NR}$) of $\text{Im}{\bar U}(q;\rho)$ is found by setting to one the factors $M/E_p$ and $M/E_{p+q}$ and using  non-relativistic nucleon dispersion relations to solve the energy-conserving delta function, 
\begin{equation}
\text{Im}{\bar U}_{\rm NR}(q;\rho) = -\Theta(q^0)\int\frac{d^3p}{(2\pi)^2}\delta\left(q^0+\vec{p}^{\,2}/2M-(\vec{p}+\vec{q}\,)^2/2M\right) \Theta(k_F-|\vec{p}\,|)\Theta(|\vec{p}+\vec{q}\,|-k_F)\label{eq:imNR}
\end{equation}
All the integrations involving the tensor $A^{\mu\nu}$ can also be  done analytically and are compiled  in the Appendix C of Ref.~\cite{Nieves:2004wx} for the non-relativistic case. 

We take density profiles from \cite{firestone, DeJager:1974liz, DeJager:1987qc}. Lighter nuclei are described by harmonic oscillator distributions, while heavier (above oxygen) by two-parameter Fermi profiles.  Additionally we take into account that nucleons are not point-like particles, and consider their finite size  by means of the prescription  discussed in Sec. II of Ref. \cite{GarciaRecio:1991wk} [see Eqs. (12-14) of this reference].

\subsection{Binding energy and Coulomb distortion effects}

These corrections are relevant at low energies. When a particle scatters off a nucleus and deposits energy, in the  $1p1h$ approximation, it is not  fully transferred  into the ejected nucleon's energy, but some part goes to compensate the binding energy of the hit bound nucleon. This is taken into account in the $1p1h$ contribution to the self-energy by considering that (we will discuss the situation for CC processes; the modifications for NC or EM ones are straightforward):
\begin{enumerate}
\item The initial and final nuclear configurations have different number of neutrons and protons. In that case, some energy $Q$ has to compensate the transition between the initial and final ground states.
\item In an isospin asymmetric nuclear-matter,  there is a gap between neutron and proton Fermi levels, so in the calculation of the hadron tensor (Eq.~(\ref{eq:hadron_tensor})) we get already a non-zero energy $Q_{gap}(r)$ which is the minimal energy needed for the process to occur within a LFG. It should be subtracted from the experimental $Q$ value  to enforce the correct (experimental) energy balance in the reaction.
\end{enumerate}
This means that in the calculation of the hadronic tensor, we  use a shifted value of $q^0$ (see Ref.~\cite{Nieves:2004wx} for more details),
\begin{equation}
q^0 \rightarrow q^0-(Q-Q_{gap}(r))
\end{equation}
The $Q$ and $Q_{gap}(r)$ values will be different for neutrino and antineutrino driven processes and in the latter case we will denote them as $\bar{Q}$ and $\bar{Q}_{gap}(r)$.

On the other hand, the charged lepton gets distorted by its electromagnetic interaction with the nucleus which produces a change of its propagation in the nuclear medium. We will implement this effect using the semi-classical approach proposed in Refs.~\cite{Singh:1992dc, Kosmas:1996fh, Singh:1998md}, where the self-energy acquired by the charged lepton is taken into account. In a good approximation, this self-energy is proportional to the Coulomb potential created by the nucleus:
\begin{equation}
\Sigma = 2k'^0 V(r)
\end{equation}
where $V(r)$ depends on the charge distribution of the nucleus, $\rho_{\rm ch}(r)$,
\begin{equation}
 V(r) = -4\pi\alpha \bigg[\frac{1}{r}\int_0^r r'^2 \rho_{\rm ch}(r') dr' + \int_r^{\infty}r' \rho_{\rm ch}(r') dr'\bigg]
\end{equation}
and $\alpha\sim 1/137$ is the fine structure constant. This self-energy will affect both energy and momentum of the lepton, making them local functions depending on $r$, $E_{k'}(r)$ and $k'(r)$. Asymptotically for $r\rightarrow \infty$ we have  $E_{k'}(r)\rightarrow E_{k'}$ and $k'(r)\rightarrow k'$, so that the energy and momentum are conserved in the reaction. From the conservation of energy we have
\begin{equation}
 V(r)+E_{k'}(r) = E_{k'}
\end{equation}
and then $|\vec{k}'(r)| = \sqrt{(E_{k'}-V(r))^2-m_l^2}$. This affects also the momentum transfer $\vec{q}(r) = \vec{k}- \vec{k}'(r)$ and should be taken into account in the integration over $d^3k'$ in Eq. (\ref{eq:imsigma}). 

Including these effects we get a modified CCQE hadron tensor that now reads
\begin{eqnarray}
\begin{split}
W^{\mu\nu}(q) = & -\frac{\cos^2\theta_C}{2M^2}\int_0^{\infty} dr r^2 \frac{|\vec{k}'(r)| E_{k'}(r)}{|\vec{k}'| E_{k'}} \Theta(E_{k'}(r)-m_l) \left(-\Theta(q'^0)\right)\int\frac{d^3p}{(2\pi)^2}\frac{M}{E_p}\frac{M}{E_{p+q'}} \\ 
& \delta(q'^0+E_p-E_{p+q'}) \Theta(k_F-|\vec{p}\,|)\Theta(|\vec{p}+\vec{q'}|-k_F)A^{\nu\mu}(p,q')\bigg|_{p^0=E_p}
\end{split}
\end{eqnarray}
where $q'^0 = q^0-(Q-Q_{gap}(r))$ and $\vec{q}\,' = \vec{k}-\vec{k}'(r)$.

Coulomb distortion is rather a small effect for light nuclei, getting only sizable for heavier ones and low energy outgoing charged leptons, when $V(r)$ is of the same order as $E_{k'}$.

\section{Further nuclear corrections}\label{sec:nuc_cor}

In what follows we  incorporate additional and relevant nuclear effects into the simple model presented in the previous section.

\subsection{Nucleon self-energy and spectral functions}\label{sec:fsi}
Nucleons in nuclear matter are not free particles, they interact with each other. These collisions introduce a change in the energy-momentum dispersion relation and a collision broadening. In other words, each nucleon propagator would be dressed with a self-energy depending on its energy, momentum and nuclear density, $\Sigma(\omega,q;\rho)$. Thus, the free nucleon propagator $G(p;\rho)$  of Eq.~\eqref{eq:4} should be replaced by a dressed one that in the non-relativistic limit reads
\begin{equation}
G_{\rm dressed}(p;\rho) = \frac{1}{2M}  \frac{1}{p^{0}-M -\vec{p}^{\,2}/2M-\Sigma(p^{0},
  \vec{p}\,;\rho)} 
\end{equation}
The real part of the self-energy modifies the nucleon dispersion relation in the nuclear medium, while the imaginary part accounts for some many-body decay channels, $NN\to NN$. As an observable effect, the QE peak deduced in a non-interacting LFG picture of the nucleus would be shifted and would get wider, spreading its strength. 

Here we will use  a semi-phenomenological model for the nucleon self-energy derived in \cite{FernandezdeCordoba:1991wf}. It implements the low-density theorems and  the used effective $NN$ potential in the medium  is obtained from the experimental\footnote{This allows to account for some short-range correlation effects in the model.} elastic $NN$ scattering cross section incorporating some medium polarization (RPA) corrections. The approach is non-relativistic and it is derived for isospin symmetric nuclear matter. The resulting spectral functions stay in a good agreement with microscopic calculations \cite{Fantoni:1983zz, Fantoni:1984zz, Ramos:1989hqs, Muther:1995bk, Mahaux:1985zz}. The use of non-relativistic kinematics is sufficiently  accurate for the hole, but its applicability to the ejected nucleon limits the range of energy and momentum transferred to the nucleus.

\begin{figure}[h]
\centering
\includegraphics[scale=1]{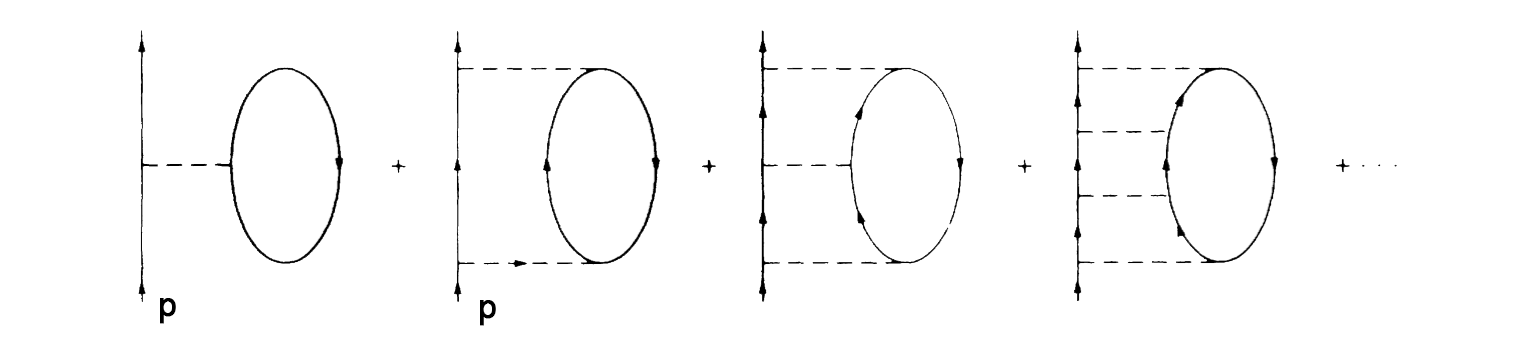}
\caption{Series of diagrams that are summed up to obtain the nucleon self-energy.}
\label{fig:6}
\end{figure}

The self-energy consists of a ladder sum of nuclear corrections generated by the series of diagrams depicted in Fig. \ref{fig:6}. The dashed lines stand for the effective in-medium $NN$ potential (see Ref.~\cite{FernandezdeCordoba:1991wf} for details).
There are few additional nuclear effects and approximations implemented in the model of Ref.~\cite{FernandezdeCordoba:1991wf}, eg. prescription on how to extrapolate the experimental $NN$ cross section to off-shell nucleons or   the inclusion of polarization effects, which take into  account both the $ph$ and the $\Delta h$ excitations, etc. As the first step of the evaluation, the imaginary part of the self-energy is obtained. It accounts  for collisional broadening effects and the results found  for ${\rm Im}\Sigma$ are 
quite close to those obtained in the elaborate many-body calculations of Refs.~\cite{Fantoni:1983zz, Fantoni:1984zz}.
The real part of the self-energy is calculated using a dispersion relation, summing an additional Fock diagram which provides a purely real contribution. Only pieces of
the Hartree type, which should be independent of the momentum, are missing in the model. Hence, up to an unknown momentum independent term 
in the self-energy, the rest of the nucleon properties in the medium can be calculated,
like effective masses, nucleon momentum distributions, etc., which are also in good agreement with
sophisticated many--body calculations~\cite{Ramos:1989hqs, Mahaux:1985zz}.

\begin{figure}[h]
\centering
\includegraphics[scale=0.26]{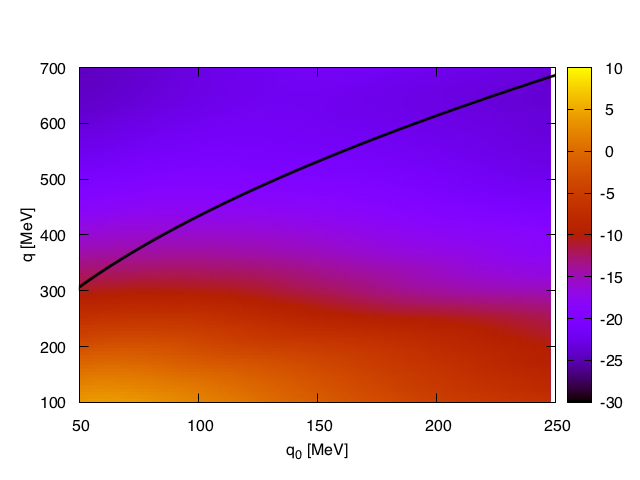}
\includegraphics[scale=0.26]{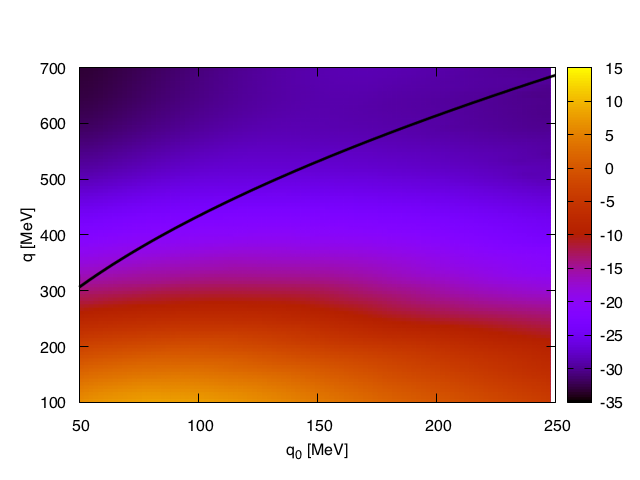}
\includegraphics[scale=0.26]{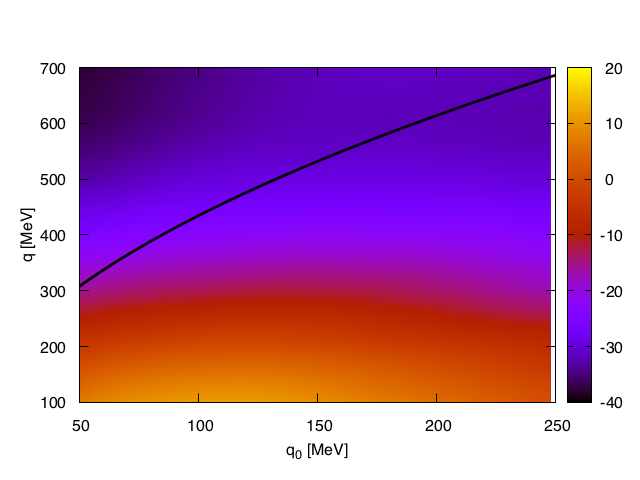}
\caption{Difference between the real parts of hole and particle self-energies, i.e.,  $\text{Re}\Sigma(E,\vec{p};\rho)-\text{Re}\Sigma(E+q^0, \vec{p}+\vec{q}; \rho)$, with $(E,p)= (\frac{3}{5} \frac{k_F^2}{2M},\sqrt{\frac{3}{5}} k_F)$ and $\vec{p}\, \bot\, \vec{q}$. Panels correspond to three different densities: $\rho=0.05\, \text{fm}^{-3}$ (left), $\rho=0.1\, \text{fm}^{-3}$ (center) and $\rho=0.15\, \text{fm}^{-3}$ (right). Results are displayed in [MeV] units. The contour shows the approximate position, $q^0=\vec{q\,}^2/2M$, of the QE peak.}
\label{fig:diff-realself}
\end{figure}

From the above discussion,  it is clear that the obtained result for the real part of the self-energy should not be treated as an absolute value. However, in our case we do not deal with a nucleon propagator but with that of a $ph$ excitation (see Eq. (\ref{eq:lindhard})), where only differences between two nucleon self-energies appear. Thus, the  constant terms  of the hole and particle self-energies cancel in the computation of the imaginary part of the Lindhard function.
In Fig.~\ref{fig:diff-realself} we show  for three different nuclear densities, the difference between the real parts of the self-energies of two nucleons of four-momenta $(E,\vec{p}\,)$ and $(E+q^0,\vec{p}+\vec{q}\,)$, respectively,  as a function of $q^0$ and $|\vec{q}\,|$. We see how, for a fixed momentum transfer, dressing the particle-hole lines moves the QE peak towards larger energy 
transfers. This is  because the difference of the real parts of hole and particle self-energies is negative in the vicinity of the naive (free) position of the QE peak.

Nevertheless, one can estimate appropriate (absolute) values for the real part of the nucleon-hole self-energy by looking at the binding energy per nucleon. We  follow Ref.~\cite{Marco:1995vb}, where the EMC effect was studied using the nucleon self-energy derived in \cite{FernandezdeCordoba:1991wf}, and
include phenomenologically a constant term $C\rho$ in  $\text{Re}\Sigma$ and demand  the binding energy per nucleon, $|\epsilon_A|$, to be the
experimental one. Thus for example, the parameter $C$ in carbon turns out to be around 0.8 fm$^2$, which provides around 25-30 MeV
repulsion at $\rho=0.17$ fm$^3$ and leads to  $|\epsilon_A| = 7.8$ MeV (see Table I of Ref.~\cite{Marco:1995vb}). 

Energy-dependent  Dirac optical model potentials for several nuclei were determined in  \cite{Cooper:1993nx} by fitting proton-nucleus elastic scattering
data in the energy range 20-1040 MeV. This approach has been widely employed in analyses of electron-induced proton knockout~\cite{Woo:1998zz}. It uses scalar (S)  and vector (V) complex potentials in the Dirac equation, 
and the dependences of these potentials on the kinetic energy, $t_{\rm kin}$, and radial coordinate, $r$, are found by fitting the scattering solutions to the measured elastic cross section, analyzing power, and spin rotation function.  Schr\"odinger equivalent (SE) potentials, constructed out of the scalar and vector potentials,  are also given in \cite{Cooper:1993nx}.

\begin{figure}[h]
\centering
\includegraphics[scale=0.35]{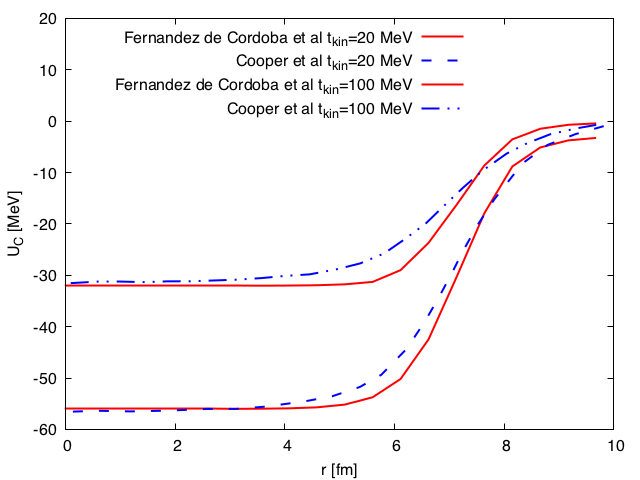}\hspace{0.5cm}
\includegraphics[scale=0.35]{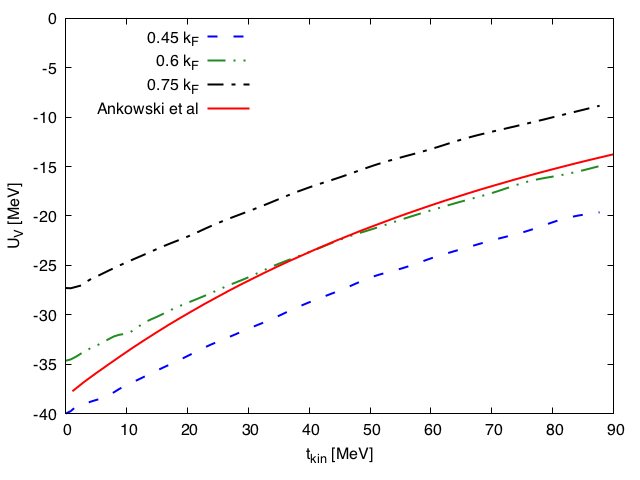}
\caption{Left: The dashed and dashed dotted lines stand for the Schr\"odinger equivalent central potentials, taken from the top panel of Fig. 6  of Ref.~\cite{Cooper:1993nx}, for $t_{\rm kin}=20$ MeV and 100 MeV in $^{208}$Pb. The solid lines show ${\rm Re}\Sigma(t_{\rm kin}, \vec{q}\, ;\rho)$  obtained with the model of Ref.~\cite{FernandezdeCordoba:1991wf}, with $t_{\rm kin}=  \vec{q}^{\,2}/2M$. Right: The solid line (red) shows the real part of the carbon optical potential for proton, obtained from the Dirac phenomenological fit of Ref.~\cite{Cooper:1993nx} and taken from the Fig. 1 of Ankowski et al.,~\cite{Ankowski:2014yfa}, as a function of the nucleon kinetic energy. Also in the right panel, the dashed lines show the results obtained from Eq.~\eqref{eq:UV} for three different values of the modulus of the hole (local) momentum.  In all cases we take $\vec{p}\, \bot\, \vec{q}$.
 Both in the left and right panels,  we add a constant term $C\rho$, with $C= 0.8$ fm$^2$, to the real part of the nucleon self-energy,}
\label{fig:UV}
\end{figure}

In the left panel of Fig.~\ref{fig:UV}, we compare  the SE $^{208}$Pb central potentials displayed in 
the top panel of Fig. 6  of Ref.~\cite{Cooper:1993nx} for $t_{\rm kin}=20$ MeV and 100 MeV  with ${\rm Re}\Sigma(t_{\rm kin}=  \vec{q}^{\,2}/2M, \vec{q}\, ;\rho)$, as a function of $r$. We reproduce quite well the Wood-Saxon form of the potentials, which is not surprising since the model of Ref.~\cite{FernandezdeCordoba:1991wf} satisfies the low densities theorems, and describe simultaneously the results for both kinetic energies. The overall scale (depth) is determined by the phenomenological, kinetic-energy independent, term $C\rho$, for which we take  $C= 0.8$ fm$^2$ as in carbon.

Next and to further test the energy dependence of the real part of the nucleon self-energy, we follow Ref.~\cite{Ankowski:2014yfa}. 
In the presence of the scalar and vector potentials of Ref.~\cite{Cooper:1993nx}, the total energy of a proton, $E'_{\rm tot}$, is
\begin{equation}
 E'_{\rm tot} = \sqrt{(M+S)^2+\vec{p}^{\,'2}}+ V
\end{equation}
From this in-medium energy, and considering only the real parts,  in Ref.~\cite{Ankowski:2014yfa}  it is defined a kinetic-energy dependent potential $U_V$ as
\begin{equation}
 U_V(t_{\rm kin}) = \int d^3 r \rho(r) {\rm Re}(E'_{\rm tot})-\sqrt{M^2+\vec{p}^{\,'2}}, \qquad  t_{\rm kin}=\sqrt{M^2+\vec{p}^{\,'2}}-M
\end{equation}
and it is depicted in Fig. 1 of this reference for carbon. This potential is used in ~\cite{Ankowski:2014yfa} to 
modify the energy spectrum of the final-state nucleon taking  
\begin{equation}
t_{\rm kin} = \frac{|\vec{k}\,|^2 (1-\cos\theta')}{M+ |\vec{k}\,| (1-\cos\theta')},
\end{equation}
where $|\vec{k}\,|$ and $\theta'$ denote the energy of the beam particle (massless) and the angle of the outgoing lepton, respectively\footnote{In ~\cite{Ankowski:2014yfa}, it is shown  
that in the low-$t_{\rm kin}$ region, particularly relevant to QE scattering, interactions
with the spectator system lead to a sizable modification
to the struck protons's spectrum.}. Such definition corresponds to assume that $t_{\rm kin} = q^0$ and  $q^0=-q^2/2M$, with $q^\mu$ the lepton momentum transfer. Since the three-momentum of the final-state nucleon in our formalism is $\vec{p}+\vec{q}$, with $\vec{p}$ the momentum of the hole state, we could estimate the above potential for moderate kinetic energies as
\begin{equation}
 U_V(t_{\rm kin}) \sim \int d^3 r \rho(r) \bar{E}(\vec{p}+\vec{q}\,) - \frac{\vec{q}^{\,2}}{2M}, \quad  t_{\rm kin}=\frac{\vec{q}^{\, 2}}{2M} \label{eq:UV}
\end{equation}
with $\bar{E}$ a self-consistent solution of 
\begin{equation}
\bar{E}(\vec{p}\,) =M + \frac{\vec{p}^{\,2}}{2M} + \text{Re}\Sigma(\bar{E}(\vec{p}\,),\vec{p}; \rho) \label{eq:10}
\end{equation}
In the right panel of Fig.~\ref{fig:UV}, we show results obtained with the  model of Ref.~\cite{FernandezdeCordoba:1991wf} for ${\rm Re}\Sigma$, supplemented with a constant term $C\rho$, for three different values of the hole (local) momentum ($p\propto k_F(r)$) and compare with the potential obtained from the fits carried out in Ref.~\cite{Cooper:1993nx}.  We find a reasonable agreement with similar dependences on $t_{\rm kin}$ and  differences in the magnitude of the potential of the order of 5-10 MeV at most, which could be partially re-absorbed either by modifying the parameter $C$ or using an appropriate average momentum.

The comparisons in Fig.~\ref{fig:UV} are sensitive to the absolute values of ${\rm Re}\Sigma$. However, we should stress again that cross sections depend on the difference between the real parts of hole and particle self-energies, where Hartree-type constant terms cancel.

\subsubsection{$2p2h$ contribution}
Concerning the interest in the $2p2h$ excitations \cite{Martini:2009uj,Martini:2010ex, Nieves:2011pp, Nieves:2011yp, Martini:2011wp,  Nieves:2012yz, Nieves:2013fr, Martini:2013sha,Gran:2013kda, Amaro:2011aa, Simo:2014esa, Megias:2014qva, RuizSimo:2016ikw, Katori:2013eoa, Benhar:2015ula, Barbaro:2016hrt, Pastore:2012rp} we want to stress that there is one contribution of this type  taken into account in the nucleon self-energy  (although it is only a  part of the  $2p2h$ calculation performed in \cite{Nieves:2011pp}). 
\begin{figure}[h]
\centering
\includegraphics[scale=1]{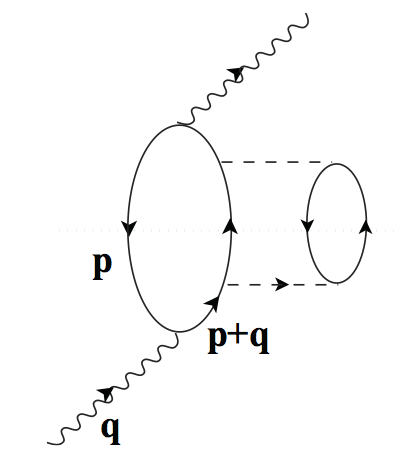}
\caption{The $2p2h$ contribution included in the nucleon self-energy. By cutting the diagram with a horizontal line (Cutkosky cut) we put two hole  and two particle states on-shell.}
\label{fig:2p2h}
\end{figure}
This is depicted in Fig.~\ref{fig:2p2h}, where the nucleon particle propagator is dressed up by a $ph$ excitation. The real part of the total nucleon self-energy, obtained from the imaginary part, also contains information about this $2p2h$ excitation. Even in the approximation where the imaginary part of the nucleon self-energy is neglected in the calculation of the SFs, this $2p2h$ contribution would be partially taken into account.

\subsubsection{Non-free Lindhard function}
Knowing the nucleon self-energy $\Sigma(\omega,q;\rho)$, one can use it to get the SFs. The Lehmann representation of the dressed nucleon propagator in the nuclear medium reads:
\begin{equation}
2M G_{\rm dressed}(\omega, q;\rho) = \int_{\mu}^{\infty} \frac{S_p(\omega',q)}{\omega-\omega'+i\epsilon} d\omega'+\int^{\mu}_{-\infty} \frac{S_h(\omega',q)}{\omega-\omega'-i\epsilon}d\omega'
\end{equation}
where the particle and hole SFs $S_{p,h}$ are determined by the nucleon self-energy. Thus, in the non-relativistic limit, we have
\begin{equation}
S_{p,h}(\omega,\vec{q}\,) = \mp \frac{1}{\pi} \frac{\text{Im}\Sigma(\omega,\vec{q}\,)}{\big(\omega - M-\vec{q}^{\,2}/2M - \text{Re}\Sigma(\omega,\vec{q}\,) \big)^2 + \text{Im}\Sigma(\omega,\vec{q}\,)^2} \label{eq:defSpSh}
\end{equation}
with $\omega \le \mu$ or $\omega \ge \mu$ for $S_h$ and $S_p$ respectively, and the chemical potential $\mu$ is defined by:
\begin{equation}
\mu(k_F) = M + \frac{k_F^2}{2M} + \text{Re}\Sigma (\mu(k_F) ,k_F) \label{eq:defchm}
\end{equation}
We have omitted the $\rho$ dependence in the SFs and $\Sigma$ to shorten the notation. Obviously the SFs depend on the density through the nucleon self-energy. 

In what follows, we take into account the nucleon self-energy,  and in this manner we obtain the $ph-$propagator in a LFG of interacting nucleons. This modified propagator  plays the role of the Lindhard function in this case. As mentioned above, only the  imaginary part of this new Lindhard function is needed to compute the hadron tensor. It is obtained by using Cauchy's residue theorem and it reads~\cite{Nieves:2004wx},
\begin{equation}
\text{Im} {\bar U}_{SF}(q,\rho) = -\frac{\Theta(q^0)}{4\pi^2}\int d^3p\int_{\mu-q^0}^{\mu} d\omega S_h(\omega,\vec{p}\,) S_p(q^0+\omega,\vec{p}+\vec{q}\,)\label{eq:linSF}
\end{equation}
It means for example that for CCQE scattering,  one can  account for the nucleon self-energy effects  in an isospin
symmetric nuclear medium of density $\rho$ by substituting in Eq.~(\ref{eq:lindfree})
\begin{equation}
-\Theta(q^0) \int\frac{d^3p}{(2\pi)^2} \frac{M}{E_p}\frac{M}{E_{p+q}} \delta(q^0+E_p-E_{p+q}) \Theta(k_F(r)-|\vec{p}\,|)\Theta(|\vec{p}+\vec{q}\,|-k_F(r))A^{\nu\mu}(p,q)\bigg|_{p^0=E_p} 
\end{equation}
by
\begin{equation}
 -\frac{\Theta(q^0)}{4\pi^2}\int d^3p\int_{\mu-q^0}^{\mu} d\omega S_h(\omega,\vec{p}\,) S_p(q^0+\omega,\vec{p}+\vec{q}\,)A^{\nu\mu}(p,q)\bigg|_{p^0=E_p}
\end{equation}
\subsubsection{Asymmetric case}
The spectral functions were derived in \cite{FernandezdeCordoba:1991wf} for  symmetric nuclear matter. However, one can generalize them to the asymmetric case, introducing separate chemical potentials  for protons and neutrons, 
and referring the self-energies to these two different Fermi levels. Thus, for instance, the imaginary part of the Lindhard function, when the hole state is a proton and the particle state is a neutron, takes the form:
\begin{equation}
\text{Im} {\bar U}_{SF}(q,\rho) = -\frac{\Theta(q^0)}{4\pi^2}\int d^3p\int_{\mu_n-q^0}^{\mu_p} d\omega\Theta(\mu_p+q^0-\mu_n)  S_h^{(p)}(\omega,\vec{p}\,) S_p^{(n)}(q^0+\omega,\vec{p}+\vec{q}\,) \label{eq:USF}
\end{equation}
where $\mu_n$ and $\mu_p$ are chemical potentials for neutrons and protons respectively.  Because the model of Ref.~\cite{FernandezdeCordoba:1991wf} was developed in  symmetric nuclear matter, here we should necessarily take $\rho=\rho_p +\rho_n$ to evaluate the nucleon self-energy, which would be the same for protons and neutrons. We could use, however,  
$\rho_p$ or $\rho_n$ to obtain the chemical potentials from Eq.~\eqref{eq:defchm} as needed. 
\subsubsection{Possible approximations}
The result in Eq.~\eqref{eq:linSF} has a simple form, however it is not easy from the computational point of view. The spectral functions have forms of  narrow peaks (see Figs. 14,16 of Ref. \cite{FernandezdeCordoba:1991wf}), especially for energies close to the Fermi level (where $\text{Im}\Sigma(q^0,\vec{q}\,) \rightarrow 0$). Moreover, $\text{Re}\Sigma(q^0,\vec{q}\,)$ obtained from dispersion relations is   result of yet another integration, which is also quite time consuming. Because of the large computational time needed to evaluate the imaginary part of the non-free Lindhard function, it is advisable  to introduce approximations that work well in some situations. As mentioned, the spectral functions have form of peaks. Their width depends on the distance to the Fermi level and in some cases they become nearly delta functions (low/high energies for particle/hole SFs; see Fig. 10 of Ref.~\cite{FernandezdeCordoba:1991wf}).
Thus, for energy transfers $q^0$ high enough, the width of the particle SF is much broader than that of the hole SF (see analysis in Sec. \ref{sec:analysis}). 
In this region, one could explore the validity of approximating  $S_h$ by a delta function:
\begin{equation}
S_h(\omega,\vec{p}\,) = \delta\left(\omega-\bar{E}(\vec{p}\,)\right)\Theta\left(\mu - \bar{E}(\vec{p}\,)\right) \label{eq:approxSh}
\end{equation}
with $\bar{E}(\vec{p}\,)$ defined in Eq.~\eqref{eq:10}. This simplification, used in Ref.~\cite{Nieves:2004wx}, saves one integration and then we are  left with: 
\begin{equation}
\text{Im} \bar{U}_{SF{\rm approx}}(q,\rho) = -\frac{\Theta(q^0)}{4\pi^2}\int d^3p S_p\left(q^0+\bar{E}(\vec{p}\,),\vec{p}+\vec{q}\,\right) \Theta(\mu - \bar{E}(\vec{p}\,)) \label{eq:imsfappox}
\end{equation}
The reliability of this approximation will be discussed in detail in Sect.~\ref{sec:analysis}. There, we will see that it is reasonable at intermediate energies, where it leads to cross sections around 5-10\% larger than those obtained with the correct expression for $\text{Im} \bar{U}_{SF}(q,\rho)$. Nevertheless, here we will not adopt this approximation, and we will present results from the many body model derived in Ref.~\cite{Nieves:2004wx} using for the very first time full SFs for both particle and hole nucleon lines.
%
\subsection{RPA corrections}
\label{sec:rpa}
\begin{figure*}[h]
\centering
\includegraphics[scale=0.9]{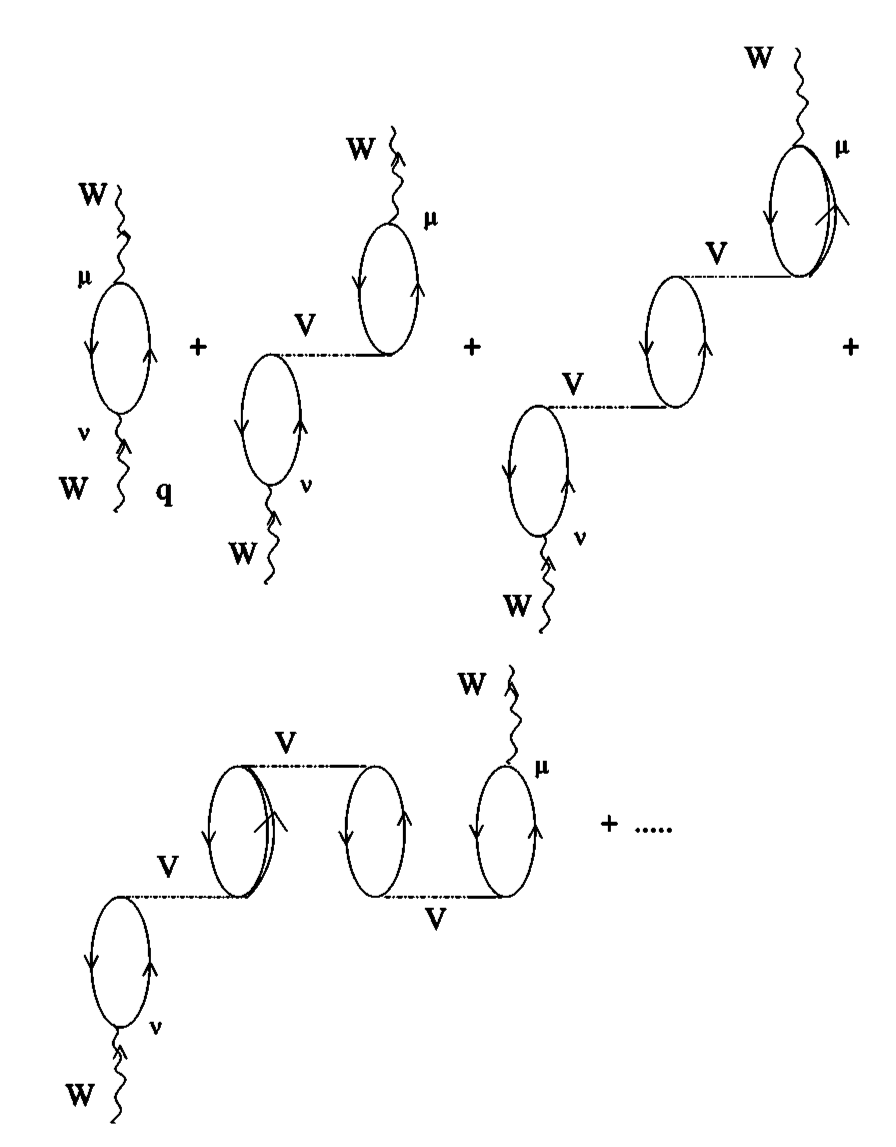}
\caption{RPA series of $ph$ and  $\Delta h$ excitations.}
\label{fig:rpa}
\end{figure*}
RPA  correlations account for some nuclear medium polarization effects sensitive to the collective degrees of freedom of the nucleus. These corrections bear some resemblance with the polarization experienced by a probe charge inside of an electron gas~\cite{Nieves:2016yfc}. Within the model employed in \cite{Gil:1997bm, Nieves:2004wx, Nieves:2005rq}, a series of $ph$ and $\Delta h$ excitations (Fig.~\ref{fig:rpa}),  which interact via an effective spin-isospin non-relativistic potential, is summed up~\cite{Nieves:2016yfc}.(Also here we are limited to moderate energy and momentum transfers because of the use of non-relativistic approximations.) This effective interaction includes a contact Landau-Migdal potential,
\begin{equation}\label{eq:11}
V = c_0\Big\{ f_0(\rho) + f_0'(\rho)\vec{\tau}_1\cdot\vec{\tau}_2 + g_0(\rho) \vec{\sigma}_1\cdot \vec{\sigma}_2+g_0'(\rho)\left(\vec{\sigma}_1\cdot\vec{\sigma}_2\right)\left(\vec{\tau}_1\cdot\vec{\tau}_2\right)\Big\}
\end{equation}
The constants in Eq. (\ref{eq:11}) were determined  from (low energy) calculations of
nuclear electric and magnetic moments, transition probabilities,
and giant electric and magnetic multipole resonances~\cite{Speth:1980kw, rpa_par2},
\begin{equation}
f_i(\rho(r)) =\frac{\rho(r)}{\rho(0)}f_i^{(in)}+\bigg(1-\frac{\rho(r)}{\rho(0)}\bigg)f_i^{(ex)}
\end{equation}
with 
\[
f_0^{(in)}=0.07 \ \ \ \ f_0^{(ex)}=-2.15 \ \ \ \ {f'}_0^{(in)}=0.33\ \ \ \ {f'}_0^{(ex)}=0.45
\]
and $c_0=380$ MeV fm$^3$, $g_0=0.575$ and ${g'}_0=0.725$.

In the $S=T=1$ sector, we improve the interaction and include explicitly pion and $\rho$ meson exchanges, which separate the non-relativistic potential into transverse and longitudinal channels,
\begin{equation}
c_0 {g'}_0(\vec{\sigma}_1\cdot\vec{\sigma}_2)(\vec{\tau}_1\cdot\vec{\tau}_2) \rightarrow \vec{\tau}_1\cdot \vec{\tau}_2 \sum_{i,j} \sigma^i\sigma^j V_{ij}^{\sigma\tau}
\end{equation}
\begin{equation}
V_{ij}^{\sigma\tau} = \hat{q}_i \hat{q}_j V_l(q) + (\delta_{ij}-\hat{q}_i \hat{q}_j )V_t(q)
\end{equation}
with $ \hat{q} = \vec{q}/|\vec{q}\,|$ and the longitudinal and transverse potentials given by,
\begin{equation}
V_l(q) = \frac{f^2}{m_{\pi}^2} \bigg\{  \bigg( \frac{ \Lambda_{\pi}^2-m_{\pi}^2}{ \Lambda_{\pi}^2-q^2} \bigg)^2\frac{\vec{q}^{\,\,2}}{q^2-m_{\pi}^2}+g' \bigg\},  \ \ \ \ \ \ \ \ f^2/4\pi =0.08,  \ \ \ \ \ \Lambda_{\pi}=1200\, \text{MeV}
\end{equation}
\begin{equation}
V_t(q) = \frac{f^2}{m_{\pi}^2} \bigg\{ C_{\rho} \bigg( \frac{ \Lambda_{\rho}^2-m_{\rho}^2}{ \Lambda_{\rho}^2-q^2} \bigg)^2\frac{\vec{q}^{\,\,2}}{q^2-m_{\rho}^2}+g' \bigg\},  \ \ \ \ \ \ \ \ C_{\rho}=2,  \ \ \ \ \ \Lambda_{\rho}=2500\, \text{MeV}
\end{equation}
and $g'=0.63$, as used in \cite{Gil:1997bm, Nieves:2004wx, Nieves:2005rq}. Moreover $\Delta(1232)$ degrees of freedom in the nuclear medium are also considered, which opens the possibility of taking into account 
$\Delta h$ excitations in the RPA series, as mentioned above. It affects only the $S=T=1$ sector and the interaction $ph$-$\Delta h$ and $\Delta h$-$\Delta h$ is taken from \cite{Oset:1981ih} (see also \cite{Nieves:2016yfc} for details).  The RPA sum leads to  substitutions in some terms  of the hadron tensor obtained within the $1p1h$ approximation (see Appendix A of Ref.~\cite{Nieves:2004wx}). For instance, the $(S=T=1)-$RPA sum produces, in a schematic way and for a free LFG, a replacement of the type
\begin{equation}
{\rm Im} {\bar U}(q;\rho) \left[a \hat q_i  \hat q_j + b \left(\delta_{ij}- \hat q_i  \hat q_j \right) \right] \to  {\rm Im} {\bar U}(q;\rho) \left[a \frac{\hat q_i  \hat q_j}{|1-U(q;\rho)V_l(q)|^2} + b \frac{\delta_{ij}- \hat q_i  \hat q_j}{|1-U(q; \rho)V_t(q)|^2}  \right] \label{eq:rpa-example}
\end{equation}
where $U(q;\rho)= U_N+U_\Delta$ takes into account the $ph$ and the $\Delta h$ excitations, with $U_N=2\bar{U}$ (the factor of 2 accounts for a sum over isospin, not explicitly carried out in the definition given in Eq.~(\ref{eq:lind})) in a symmetric medium. For positive values of $q^0$, the backward propagating $ph$ excitation has no imaginary part, and for QE kinematics the $\Delta(1232)$ Lindhard function $U_\Delta$ is also real\footnote{Analytical expressions for $U_\Delta$ can be found for example in Ref.~\cite{Nieves:2016yfc}, while expressions for the  real part of the relativistic Lindhard function $U_N$ can be found in Ref.~\cite{Barbaro:2005pq}. The corresponding non-relativistic counterparts, obtained by setting to one the factors $M/E_p$ and $M/E_{p+q}$ and using  non-relativistic nucleon dispersion relations in Eq.~\eqref{eq:lind}, can be found in Refs.~\cite{fetterwalecka, Nieves:2016yfc}.  }. Nevertheless, we refer the reader to \cite{Nieves:2004wx} for a detailed description of the RPA re-
summation within this  formalism. 

We should mention that the interaction used to compute the RPA corrections is
in principle unrelated to the semi-phenomenological one employed in \cite{FernandezdeCordoba:1991wf} to evaluate the nucleon self-energies.

Here we would like to focus on the situation when  RPA and SF effects are included together. As sketched above, polarization effects are computed by summing up  an infinite  series of $ph$  and $\Delta h$  excitations. In principle to be fully consistent, one should include also the nucleon self-energy into all of them, which means that in the denominator of each RPA correction we should have $\bar{U}_{SF}$ instead of $\bar{U}$ (both imaginary and real parts). Moreover one should consider the $\Delta$ spectral function in the nuclear medium. All these refinements would introduce further corrections in the density expansion implicitly assumed in the model. However, one should be cautious. The RPA coefficients that appear in the $ph$($\Delta$h)--$ph$($\Delta$h) effective interaction   were long time ago fitted to data, using a model of non-interacting nucleons~\cite{Speth:1980kw, rpa_par2, Oset:1981ih, Nieves:1993ev}, and since then, they have been successfully used in several nuclear calculations at 
intermediate 
energies, as mentioned in the introduction. Note that the imaginary part of the $ph-$propagator (the Lindhard function) appears both in the numerators and denominators of Eq.~\eqref{eq:rpa-example}. Its contribution to the latter ones is in general small because in most of the available phase space, the denominators of the RPA series are being dominated by the real parts, which start by 1 in addition to the $({\rm Re}\,U\,V_{l,t})$ contribution. However, the role of the imaginary part of the $ph-$propagator in the numerators is essential, because it determines the allowed $(q^0,|\vec{q}\,|)$ regions, together with their relative weight into the final response. These allowed regions are obviously different when an interacting LFG or a free LFG of nucleons is being considered. Even in this latter case and for moderate energy and momentum transfers, allowed  $(q^0,|\vec{q}\,|)$ regions depend on whether relativistic or non-relativistic nucleon kinematics is being used. Because our treatment of the RPA and the 
SF effects is non-relativistic, this will be an important source of systematic uncertainties affecting our predictions.  Later we will come back to this point in more detail. 

Thus, we consider $\text{Im} {\bar U}_{SF}$ in the numerators of the RPA series, and to avoid having to re-tune the  RPA parameters which affect the real part of the denominators,
we have adopted the following  strategy. We leave the real part of the Lindhard function in the RPA denominators unchanged, which for consistency with the $ph(\Delta h)-ph(\Delta h)$ force is computed in the non-relativistic limit, while we also use SFs to compute the  imaginary parts in the denominators. In this manner we remove unphysical peaks, that would be generated  when in the denominator $\text{Im} {\bar U} = 0$ and in the numerator $\text{Im} {\bar U}_{SF} \neq 0$. Next and  to estimate the theoretical uncertainties,  we follow the work of Ref.~\cite{Valverde:2006zn} and 
we  take uncorrelated Gaussian distributions with relative errors of 10\%, for all the parameters that enter into the  effective interaction employed in the construction of the RPA series. In the case of CC-driven processes, these are $f_{0}^{\prime (in)}$, $f_{0}^{\prime (ex)}$, $f$, $f^*$, $\Lambda_\pi$,$C_\rho$, $\Lambda_\rho$ and $g^\prime$, since the isoscalar terms of the effective interaction do not contribute to CC induced reactions. Finally, by means of a
Monte Carlo (MC) simulation, we find for any observable predicted by the model its probability distribution. 
Theoretical errors and uncertainty bands on the derived quantities will be always obtained by discarding the highest and lowest 16\%
of the sample values, to leave a 68\% confidence level (CL) interval. 

The CC hadron tensor with inclusion of Coulomb distortion, binding energy, RPA and SF effects has a form:
\begin{eqnarray}\label{eq:hadron_full}
\begin{split}
W^{\mu\nu}(q) = \frac{\cos^2\theta_C}{2M^2}\int_0^{\infty} & dr r^2 \frac{|\vec{k}'(r)| E_{k'}(r)}{|\vec{k}'| E_{k'}}  \Theta(E_{k'}(r)-m_l)\\
&\Theta(q'^0) \int \frac{d^3p}{(2\pi)^2}\int_{\mu-q'^0}^{\mu} d\omega S_h(\omega,\vec{p}) S_p(q'^0+\omega,\vec{p}+\vec{q}\,')  A_{RPA}^{\nu\mu}(p,q')\bigg|_{p^0=E_p} \label{eq:SF+RPA}
\end{split}
\end{eqnarray}
with $q'^0 = q^0-(Q-Q_{gap}(r))$ and $\vec{q}\,' = \vec{k}-\vec{k}'(r)$, as discussed above and $ A_{RPA}^{\nu\mu}$ given in Appendix A of Ref.~\cite{Nieves:2004wx}, with the real part of the RPA denominators computed using the non-relativistic reduction of $\bar U(q;\rho)$.  We recall here that the SFs depend on $r$ through the dependence of the particle and hole self-energies on the local density.

\section{Inclusive muon and radiative pion capture in nuclei}\label{sec:muon_pion}
In this section we will shortly describe the capture of a bound pion or muon by the nucleus. In particular, we will study
\begin{equation}\label{eq:90}
(A_Z-\mu^-)^{1s}_{\rm bound} \rightarrow \bar{\nu} + X
\end{equation}
\begin{equation}\label{eq:91}
(A_Z -\pi^-)_{\rm bound} \rightarrow \gamma + X
\end{equation}
Both $\mu^-$ and $\pi^-$ are electromagnetically bound to the nucleus, but since their masses are of the order of 200-300 heavier than that of the electron, their wave functions significantly overlap with the density distribution of the nucleus.  This is the reason why they do not form stable atoms and the strong interaction produces (complex) corrections to the electromagnetic energy levels in the case of pionic atoms.   We analyze these low energetic\footnote{Note that the energy transferred to the nuclear system is at most the mass ($m$) of the muon or the pion, and in practice, it is significantly smaller since the QE peak is located in the vicinity of $m^2/2M$.} processes because in this energy range,  the nuclear effects are essential and clearly visible,  while they play a lesser role at intermediate energies. Muon capture dynamics is governed by  CC interactions and hence  the formalism presented in Sec.~\ref{hadr-ten} can be employed. Radiative pion capture is on the other hand  governed by a 
different dynamics,
 which will be shortly presented in the next subsection. The general argumentation from Sec.~\ref{hadr-ten} holds, but the self-energies of the pion and the muon in the nuclear medium are strongly dominated, because of  kinematical reasons, by the QE reaction mechanism (i.e.,  $1p1h$ excitation).
\\
 The decay width is computed (schematically) in the following way within the LDA:
\begin{enumerate}
\item We calculate the width $\hat\Gamma(q,\rho_n(r),\rho_p(r))$ for proton and neutron nuclear matter densities using a formalism derived from that outlined in Sec.~\ref{sec:form}. 
\item For the considered nucleus, we obtain the $\mu^-$ or $\pi^-$ wave functions, $\phi(r)$,  and the energy levels by 
solving the Schr\"odinger or Klein-Gordon  equations, respectively. In this latter case (pionic atoms), besides the electromagnetic potential\footnote{Both for the muon and pion cases, finite size and vacuum polarization corrections are taken into account in the derivation of this part of the potential.} ,  a  pion-nucleus optical (strong) potential is  additionally taken into account. This potential has been developed microscopically and it is exposed in detail in Ref.~\cite{Nieves:1993ev}.
\item Finally, we evaluate
\begin{equation}
\Gamma = \int d^3r |\phi(r)|^2 \hat\Gamma(q,\rho_n(r),\rho_p(r)) \label{eq:wfunc}
\end{equation}
to obtain the decay width in finite nuclei.
\end{enumerate}
The idea behind  the above approximation is the following: At every point of the nuclear matter, there is ''a piece'' of $\mu^-$ ($\pi^-$) given by $|\phi(r)|^2d^3r$, which  has a decay width $ \hat\Gamma(q,\rho_n(r),\rho_p(r))$. Integration over the whole volume leads to the total width. We make the additional kinematical assumption that the bound $\mu^-$ or $\pi^-$ is at rest. 
\newpage
\subsection{Radiative pion capture}\label{sec:pion_capture}
\begin{wrapfigure}{r}{0.25\textwidth}
\centering
\includegraphics[scale=1]{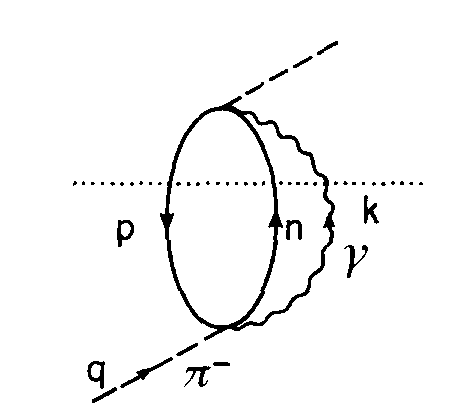}
\caption{Pion self-energy related to the $\pi N \to \gamma N$ process.}
\label{fig:pion}
\end{wrapfigure}

In the case of radiative pion capture, we follow the formalism derived in Ref.~\cite{Chiang:1989ni},  its $1p1h$ self-energy (see Fig. \ref{fig:pion}) is given by
\begin{equation}
-i\Pi(q;\rho) = -\sum_{s,\lambda} \int \frac{d^4k}{(2\pi)^4} \int \frac{d^4p}{(2\pi)^4} i G(p;\rho) i G(q-k+p;\rho) i D_0(k)(-i)T_\lambda(-i)T^{\dagger}_\lambda
\end{equation}
where  a sum over the spin $s$ of the nucleons and the photon polarization $\lambda$ is performed. On the other hand, $D_0(k) = 1/((k^0)^2-\vec{k}^2+i\epsilon)$ is the photon propagator and $T$ is the amplitude for the process $\pi^- p\rightarrow n \gamma$. For low $\pi$ momentum (the pion is bound), the contact (Kroll-Ruderman) term gives by far the largest contribution, which with recoil corrections reads
\begin{equation}
T_\lambda = i 2M e\sqrt{2}\frac{f}{m_{\pi}}(1+\frac{m_{\pi}}{2M})\vec{\sigma}\vec{\epsilon}(\lambda)
\end{equation}
where $e$ is the proton charge ($e^2/4\pi= \alpha \sim 1/137$) and $\vec{\epsilon}$ is the photon polarization vector. 
Let us notice that there is no dependence on  momenta in the vertex, so the integration over $p$ gives us the Lindhard function $\bar U(q-k)$. After summing over spins and polarizations we get
\begin{equation}
\Pi(q;\rho) =  i\int \frac{d^4k}{(2\pi)^4} \bar U(q-k;\rho) D_0(k) 16\pi\alpha \frac{f^2}{m_{\pi}^2}\bigg(1+\frac{m_{\pi}}{2M}\bigg)^2
\end{equation}
Next we use the Cutkosky's rules to calculate the imaginary part of this self-energy diagram (putting the particles cut by the dotted line in Fig.~\ref{fig:pion} on-shell), and assuming a static pion $q^0=m_{\pi}$, $\vec{q}=0$, justified  to study the capture from bound states, and thus we find
\begin{equation}
\text{Im}\Pi(q;\rho) = \int_0^{m_{\pi}} \frac{d |\vec{k}||\vec{k}|}{(2\pi)^2 }\text{Im} \bar U (m_{\pi}-|\vec{k}|, \vec{k};\rho) 16\pi\alpha\frac{f^2}{m_{\pi}^2}\bigg(1+\frac{m_{\pi}}{2M}\bigg)^2
\end{equation}
Recalling Eq. (\ref{eq:decay}), we find
\begin{equation}
\frac{d\hat\Gamma(\rho)}{d |\vec{k}|}= -\frac{\Theta(m_\pi- |\vec{k}|)}{m_\pi}\frac{4\alpha|\vec{k}|}{\pi }\text{Im} \bar U(m_{\pi}-|\vec{k}|, \vec{k};\rho) \frac{f^2}{m_{\pi}^2}\bigg(1+\frac{m_{\pi}}{2M}\bigg)^2
\end{equation}
The final result in finite nuclei is obtained by folding the above expression with the pion bound wave function as indicated in Eq.~(\ref{eq:wfunc}). We will also  enforce the correct energy balance in the decay, which changes  the argument of the Lindhard function (energy that is transferred into the final nuclear system).
\begin{equation}
\text{Im} \bar U(m_{\pi}-|\vec{k}|, \vec{k};\rho) \rightarrow \text{Im} \bar U\left(m_{\pi}-|\vec{k}|-(Q-Q_{gap}(r)), \vec{k};\rho\right)
\end{equation}
Taking into account the RPA effects is also much less complicated in this decay than in the case of lepton scattering because of the simplicity of the vertex. We have only one RPA series to sum up (driven by the transverse effective interaction in the medium), where we include both the $ph$ and the $\Delta h$ excitations~\cite{Chiang:1989ni}:
\begin{equation}
\text{Im} \bar U \rightarrow \frac{\text{Im} \bar U }{|1-(U_N+U_{\Delta})V_t|^2}
\end{equation}
In addition, the consideration of the particle and hole SFs affects only the imaginary part of the Lindhard function, and  
considering all effects together,
\begin{equation}\label{eq:pion_dec}
\frac{d\hat\Gamma(\rho)}{d|\vec{k}|} = -\frac{4\alpha}{\pi}\frac{f^2}{m_{\pi}^2}\bigg(1+\frac{m_{\pi}}{2M}\bigg)^2 \Theta\left(\widehat m_\pi(r)- |\vec{k}|\right)|\vec{k}|\,\frac{\text{Im} \bar U_{SF}\left(\widehat m_\pi(r)-|\vec{k}|, \vec{k};\rho\right) }{|1-(U'_{N}+U_{\Delta})V_t|^2}
\end{equation}
with $\widehat m_\pi(r)=m_{\pi}-(Q-Q_{gap}(r))$, and we use the notation $U'_{N}$ to recall that its imaginary part is computed using SFs to avoid fictitious singularities 

\subsection{Muon capture}
Muon capture is studied in full analogy to pion capture. A major difference is that the  outgoing particle is a neutrino $\nu_{\mu}$ instead of a $\gamma$, 
which implies that this process is  driven by  CC interactions. We have shown in Sec.~\ref{sec:form} that the neutrino self-energy is determined by the $W^+$ spectral properties. The inclusive decay width of a bound muon absorbed by the nucleus  is obtained from the imaginary part of its self-energy (spin-averaged) in the nuclear medium, which in turn is determined by the $W^-$ self-energy, $\bar\Pi^{\mu\nu}_W(q;\rho_p,\rho_n)=\Pi^{\mu\nu}_W(q;\rho_n,\rho_p)$, in this case. The latter quantity is computed following the steps outlined in  Sec. \ref{sec:form} for the $W^+$ case. Thus one easily gets~\cite{Nieves:2004wx}
\begin{eqnarray}
\begin{split}
\hat\Gamma(\rho_p,\rho_n) &= -\frac{1}{m_{\mu}}\frac{4G_F}{\sqrt{2}M_W^2}\int \frac{d^3k}{(2\pi)^3}\frac{\Theta(q^0)}{2|\vec{k}|} \text{Im}\left[L_{\mu\nu}\bar\Pi^{\mu\nu}_W(q;\rho_p,\rho_n)\right]\\
&=\frac{G_F^2 \cos^2\theta_C}{m_{\mu}}\int \frac{d^3k}{(2\pi)^3}\frac{1}{2|\vec{k}|} L_{\mu\nu}T^{\mu\nu}(q;\rho_p,\rho_n)
\end{split}
\end{eqnarray}
where we have assumed that the muon  is at rest, which simplifies the kinematics and the computation of the hadronic tensor ($T^{\mu\nu}$) that is clearly dominated by the excitation of a $ph$ nuclear component (QE mechanism). The muon binding energy,  $B^{1s}$, is also taken into account, however its value for the considered (light) nuclei in this work is at most $1$ MeV - see Table 1 in \cite{Nieves:2004wx}. We have also enforced the correct energy balance: $q^0 \rightarrow q^0 - [\bar{Q}-\bar{Q}_{gap}] = m_{\mu} - B^{1s} - |\bar{k}|- [\bar{Q}-\bar{Q}_{gap}]$, considering that the muon is captured from the $1s$ orbit. The $1p1h$ hadron tensor, after including  SF and  RPA corrections reads
\begin{equation}
T^{\mu\nu}(q; \rho_p,\rho_n)= \frac{\Theta(q^0)}{4M^2}\int \frac{d^3p}{(2\pi)^2} \int_{\mu_n-q^0}^{\mu_p} d\omega S_h(\omega,\vec{p})S_p(\omega+q^0,\vec{p}+\vec{q}\,)A_{RPA}^{\mu\nu}(p,q)\bigg|_{p^0=E(\vec{p})}
\label{eq:muon-cpt}
\end{equation}
As in the case of radiative pion capture, the final result in finite nuclei is obtained by folding with the muon bound wave function as indicated in Eq.~(\ref{eq:wfunc}).
\section{Analysis of SF effects}\label{sec:analysis}
As we have shown in Eqs.~(\ref{eq:hadron_full}),  (\ref{eq:pion_dec})  and (\ref{eq:muon-cpt}),  the inclusive neutrino-nucleus cross section and the muon and radiative pion captures in nuclei depend on the imaginary part of the Lindhard function\footnote{For the sake of clarity, in this section we will omit the arguments of the Lindhard function when possible.} $\text{Im} \bar U_{SF}$. In the case of pion capture this dependence is direct, while for a CC process the situation is more complicated because the interaction vertex gives rise to the $L_{\mu\nu} W^{\mu\nu}$  contraction, inducing a dependence of  the $A^{\mu\nu}_{RPA}(p,q)$ tensor on the hole momentum $p$.   Thus, we will present first a short analysis of the SF effects on the imaginary part of Lindhard function for  two different energy regimes. Both,  real and imaginary parts of the particle and hole self-energies enter into the evaluation of $\text{Im} \bar U_{SF}$. As mentioned above, the real part modifies the dispersion relation of the 
nucleon embedded in the nuclear medium, while the imaginary part accounts for some many-body decay channels. 

In Ref.~\cite{Nieves:2004wx}, the 
imaginary part of the hole self-energy was neglected  (see Eq.~(\ref{eq:approxSh})) to save computational time. We will discuss below that, though this approximation could be reasonable for intermediate neutrino energies, it is not  appropriate for low nuclear excitation energies. Moreover, for intermediate energies, we will show the approximation of Eq.~(\ref{eq:approxSh}) overestimates the cross sections by around 5-10\%. Given that highly accurate theoretical  predictions are essential to conduct the analysis of neutrino properties, here we will improve on this and in Subsec~\ref{sec:results_neutrino}, neutrino and antineutrino cross sections for argon, carbon and oxygen targets will be obtained using full particle and hole SFs. 

\subsection{Low energy transfers}
\begin{figure}[h]
\begin{center}
\includegraphics[scale=0.4]{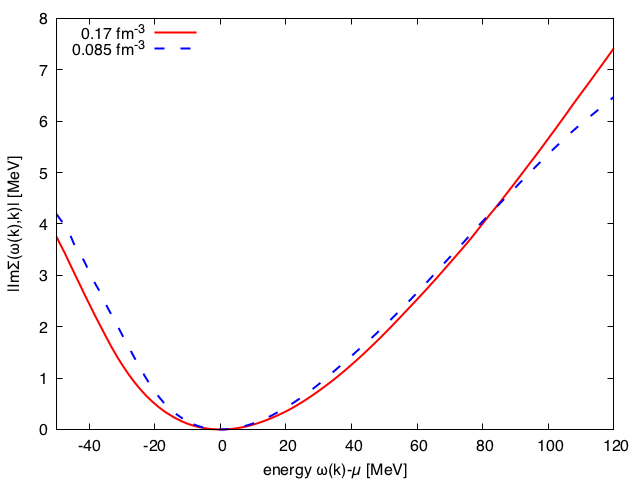}
\end{center}
\caption{$\text{Im}\Sigma(\omega(k),k)$ as a function of $\omega(k)$ (self-consistent solution of Eq. (\ref{eq:10})) calculated for two different nuclear matter densities using the model of Ref.~\cite{FernandezdeCordoba:1991wf}.}
\label{fig:7}
\end{figure}

\begin{figure}[h]
\centering
\includegraphics[scale=0.35]{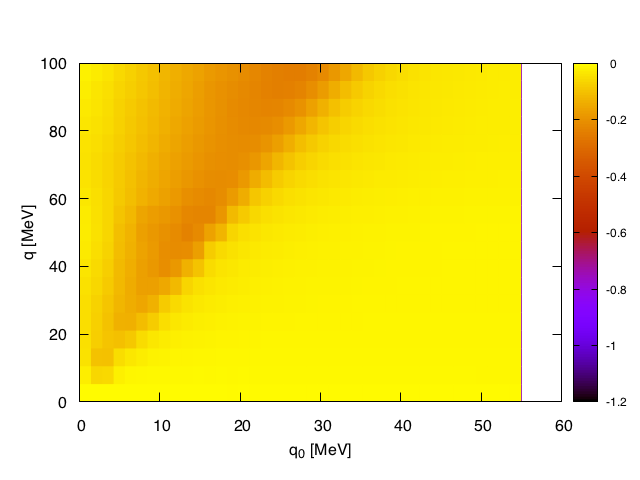}
\includegraphics[scale=0.35]{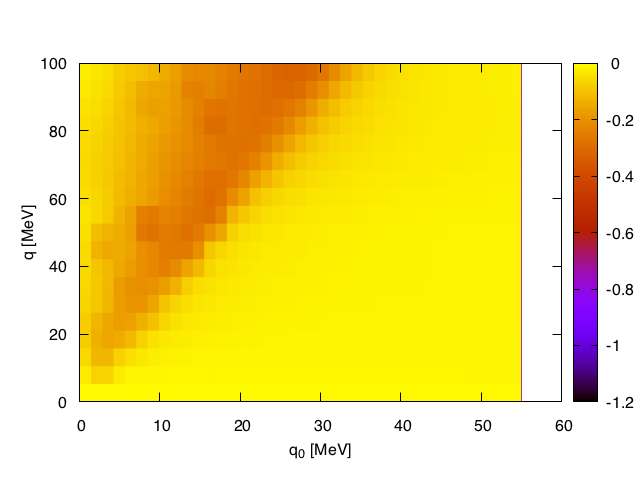}
\caption{Comparison of $\text{Im}\bar U_{SF}(q^0,q;\rho)$  computed from Eq.~(\ref{eq:USF}) keeping the width of both particle and hole lines (left) and the approximated  $\text{Im}\bar U_{SF{\rm approx}}$ (Eq.~\eqref{eq:imsfappox}) obtained by neglecting the  imaginary part of the hole self-energy (right). The density employed is $\rho=0.09$ fm$^{-3}$ and the Lindhard functions are displayed in [${\rm fm}^{-2}$] units. }
\label{fig:8}
\end{figure}

\begin{figure}[h]
\centering
\includegraphics[scale=0.35]{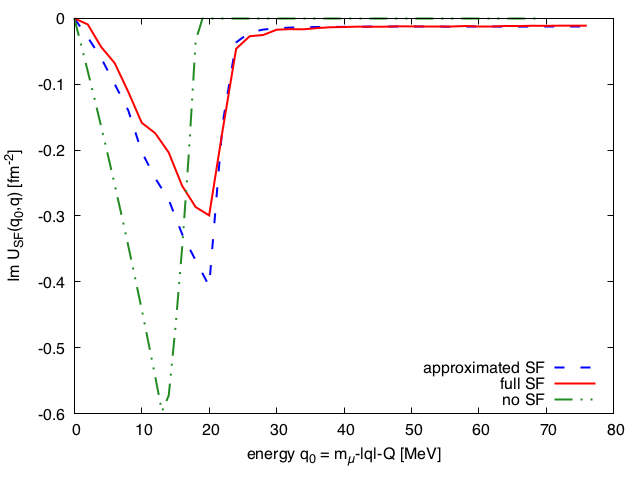}
\includegraphics[scale=0.35]{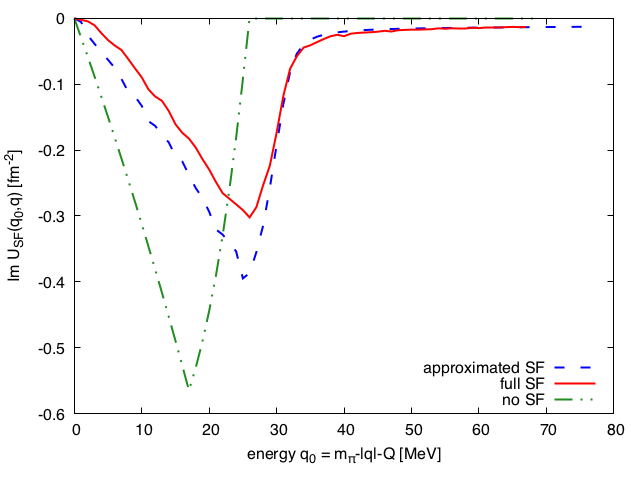}
\caption{Comparison of different approaches to $\text{Im}\bar U(q^0,|\vec{q}\,|;\rho)$ as a function of $q^0$ in $^{12}$C and $\rho=0.074$ fm$^{-3}$. Inclusive muon/pion capture kinematics is used and hence 
$|\vec{q}\,| = m_{\mu}-q^0 -{\cal Q}_{\mu}$ (left) and $|\vec{q}\,| = m_{\pi}-q^0-{\cal Q}_{\pi}$ (right). The  dashed-double dotted green curves stand for the imaginary part of the Lindhard function computed in a free LFG using non-relativistic kinematics.}
\label{fig:profile}
\end{figure}

For low energy transfers, $q^0$, we should take into account the width of the hole state (imaginary part of the nucleon self-energy) going beyond the approximation of  Eq.~(\ref{eq:approxSh}). The reason  can be understood from the results of Fig.~\ref{fig:7}. There, we show the imaginary part of the self-energy $\text{Im}\Sigma(\omega(k),\vec{k}; \rho)$ as a function of the energy $\omega(k)$, with $\omega$ being the solution of Eq.~(\ref{eq:10}), and two different nuclear matter densities.  We have adopted the model derived in Ref.~\cite{FernandezdeCordoba:1991wf}. Naturally, there is a lower limit for $\omega$, when the momentum is equal to 0, and an upper limit to be consistent with the non-relativistic approximations. There exists a minimum at the Fermi surface ($\omega(k_F)=\mu$), and in its vicinity, both the hole and particle state widths are of the same magnitude, while for higher energies the imaginary part of the particle self-energy grows and it becomes in modulus significantly larger than the 
typical values taken by that of the hole state.  Hence, it is not 
justified to neglect the hole width in the low excitation-energies regime, while for higher energy transfers keeping it is much less important.
In Fig. \ref{fig:8} we show both $\text{Im}\bar U_{SF}$ and the  approximated   $\text{Im}\bar U_{SF{\rm approx}}$ (Eq.~\eqref{eq:imsfappox}), obtained from Eq.~(\ref{eq:approxSh}) when $S_h$ is replaced by a  delta function (see Eq.~\eqref{eq:approxSh}). The full calculation 
leads to smaller values (in modulus), which can be even better appreciated if we compare a profile of this 3D plot. For this, we  use the energy-momentum dependence from muon and pion capture kinematics, i.e., $\text{Im}\Sigma(q^0,|\vec{q}\,|;\rho) = \text{Im}\Sigma(q^0=m_{\mu / \pi}-|\vec{q}\,|-{\cal Q}_{\mu / \pi}, |\vec{q}\,|;\rho)$, for $\rho=0.074$ fm$^{-3}$ in $^{12}$C (${\cal Q}_{\mu / \pi}$ accounts for the binding energy effects and the existing difference between the experimental $Q$ values and those deduced from the isospin-asymmetric LFG picture of the nucleus). Results are shown in Fig. \ref{fig:profile}, where we can see that the difference induced by keeping the imaginary part of the nucleon self-energy in the hole state could reach $30\%$ at the peak.
Let us remind here  that the integration over a function which contains two delta-like peaks is highly demanding from the computational point of view. Fig.~\ref{fig:7} shows that as the excitation energy approaches the Fermi surface,  the widths of both, particle and hole, SFs are getting smaller, making both SFs  similar to  delta functions. This is why for very low energy transfers, of the order of few MeV, the calculation may show some numerical instabilities, as can be appreciated in Fig. \ref{fig:profile}. 

\begin{figure}[h]
\centering
\includegraphics[scale=0.4]{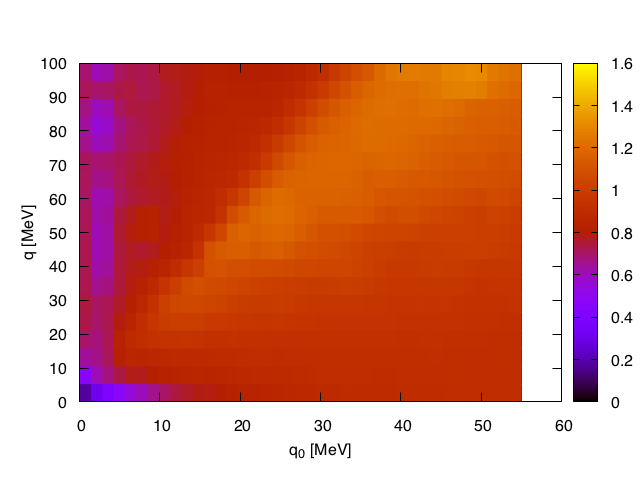}
\caption{Ratio of $\text{Im}\bar U_{SF}/\text{Im}\bar U_{SF{\rm approx}}$ for $\rho=0.09$ fm$^{-3}$}
\label{fig:fsi_comp}
\end{figure}

Finally, in Fig. \ref{fig:fsi_comp} we show the ratio $\text{Im}\bar U_{SF}/\text{Im}\bar U_{SF{\rm approx}}$ for low energy and momentum transfers, where the error induced by  neglecting the hole width can be better appreciated. 
\subsection{Intermediate energy transfers}

\begin{figure}[h]
\centering
\includegraphics[scale=0.35]{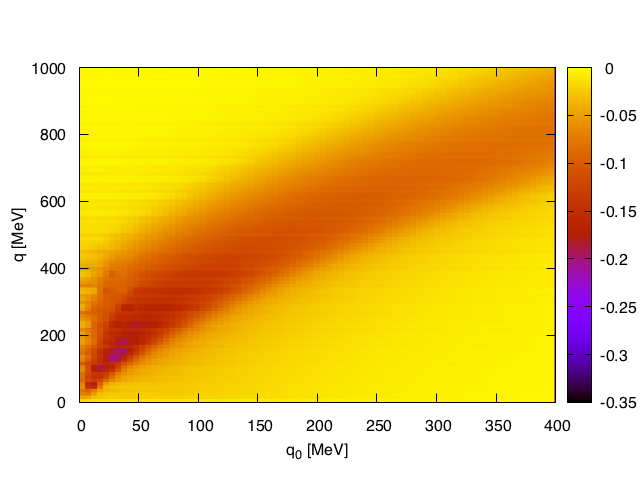}\hspace{1cm}
\includegraphics[scale=0.35]{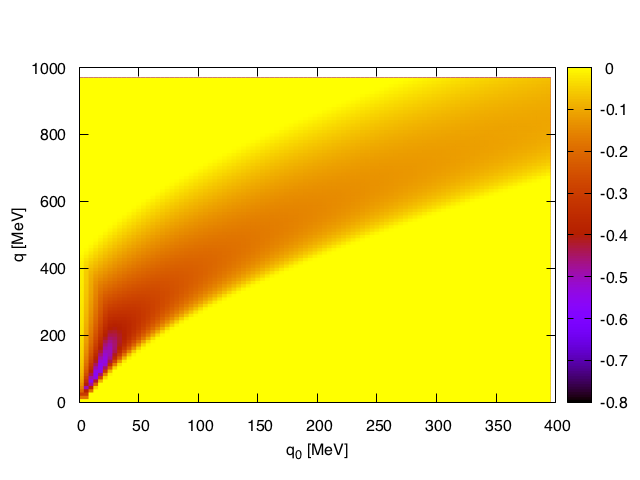}
\caption{Imaginary parts  of the full SF and the free LFG Lindhard functions in fm$^{-2}$ units and for density $\rho=0.09$ fm$^{-3}$. On the left, we show results for $\text{Im} \bar U_{\rm SF}$,  while on the right panel the non-interacting LFG $\text{Im} \bar U_{\rm NR}$ is depicted.  }
\label{fig:9}
\end{figure}

\begin{figure}[h]
\centering
\includegraphics[scale=0.4]{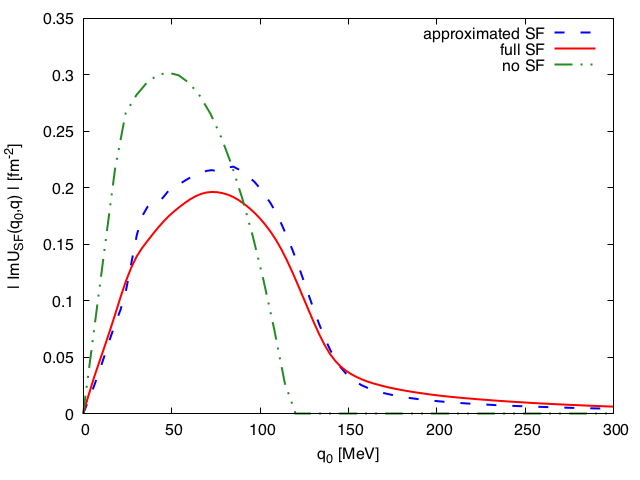}
\caption{$|\text{Im} \bar U_{\rm SF}|$ (solid),  $|\text{Im} \bar U_{\rm SF\rm{approx}}|$ (dashed)  and $|\text{Im}\bar U_{\rm NR}(q)|$ (dashed-dotted)  for $|\vec{q}\,|=$300 MeV and density $\rho=0.09$ fm$^{-3}$, as a function of the energy transfer.}
\label{fig:10}
\end{figure}

In Fig.~\ref{fig:9}, we show  $\text{Im}\bar U_{SF}$  and the imaginary part of the free LFG non-relativistic Lindhard function, $\text{Im} \bar U_{\rm NR}$ (Eq.~\eqref{eq:imNR}), in a wider energy region. We clearly observe\footnote{Note that  in the high energy and momentum transfer region, there will be relativistic effects not considered in the plots of Fig.~\ref{fig:9}.}  that $\text{Im}\bar U_{SF}$ (on the left) takes non-zero values in a much wider part of the $(q^0,|\vec{q}\,|)$ available phase-space. In the case of $\text{Im} \bar U_{\rm NR}$ (on the right),  there is a very well marked band of nonzero values. On the other hand,  $\text{Im} \bar U_{\rm NR}$ takes values generally lower (larger in absolute value) than $\text{Im}\bar U_{SF}$.   These effects  are clearly  visible  in Fig.~\ref{fig:10}, where  $\text{Im}\bar U_{SF}$ and $\text{Im} \bar U_{\rm NR}$ for $|\vec{q}\,|=300$ MeV and density $\rho=0.09$ fm$^{-3}$ are displayed. Although this plot cannot be 
directly compared 
with the cross section for neutrino scattering, one may expect that the SF corrections would move the position of the QE peak (the dispersion relation of a nucleon embedded in the nuclear medium is different because the effects of $\text{
Re}  \Sigma(q^0,|\vec{q}\,|;\rho)$; see also Fig.~\ref{fig:diff-realself}) and this peak would be generally lower, with a partial, but sizable, spreading of its strength. 

In Fig.~\ref{fig:10}, we also show results for $\text{Im}\bar U_{SF{\rm approx}}$, as a function of the energy transfer. We see  that though, the approximation of Eq.~\eqref{eq:approxSh} used in Eq.~(\ref{eq:imsfappox}) works better than for low energies, it produces values of $\text{Im}\bar U_{SF}$ (in modulus) around the QE peak systematically larger ($\sim 7$\%) than those obtained when the width of the hole state is maintained. The largest part of this enhancement is produced for having neglected in Eq.~\eqref{eq:approxSh} the inverse of the Jacobian determinant
\begin{equation}
\left|1 -\frac{\partial\text{Re}\Sigma(\omega,\vec{p}\,)}{d\omega}\right|^{-1}_{\omega=\bar E(\vec{p}\,)},\label{eq:Jacobian}
 \end{equation}
that appears in the reduction of $S_h(\omega,\vec{p}\,)$ to $\delta\left(\omega-\bar{E}(\vec{p}\,)\right)$, when the $\text{Im}\Sigma(\omega,\vec{p}\,) \to 0$ limit is taken. The above factor is the 
quasi-particle strength and it is related to the inverse of the effective $\omega-$mass~\cite{Mahaux:1985zz}.

Computing the partial 
derivative of $\text{Re}\Sigma(\omega,\vec{p}\,)$ is also numerically involved, and since accurate theoretical cross sections are important to conduct 
neutrino oscillation analyses, we improve in this work the predictions presented in Ref.~\cite{Nieves:2004wx}, by considering  
full SF effects also at intermediate  energies. Thus in the next section, we will show results obtained using fully dressed particle and hole propagators, maintaining both real and imaginary parts of the in-medium nucleon self-energies. 

\section{Results}\label{sec:results}
The use of non-relativistic kinematics is sufficiently  accurate for the computation of  hole SF, but its applicability to the ejected nucleon limits the range of energy ($q^0$) and momentum ($|\vec{q}\,|$) transfers 
to regions where, at least, $|\vec{q}\,|< 500-600$ MeV.  The energy  of the projectile is an issue for totally integrated
cross sections because if it is large, there will be phase space regions where the $q^0$  and $|\vec{q}\,|$ will be 
too large to accept the accurateness of our non-relativistic description of the particle SF. For
differential cross sections, however, we could address large projectile energies at forward angles to keep  $|\vec{q}\,|$ sufficiently
small. On the other hand, RPA effects decrease as $-q^2$ increases, and become necessarily small when the associated wave-length of the electro-weak probe is much shorter than the nuclear size. As the energy of the
projectile increases, the available phase-space includes larger regions where one might expect that RPA effects are small. However,
one should admit larger uncertainties in the RPA corrections at these large values of $-q^2$, because their calculation probes $NN$,
$N\Delta$ and $\Delta\Delta$ interactions at high virtualities. The model used here includes some exchanges of virtual mesons, and it has been shown to work well at
intermediate energies in different hadronic processes, as pointed out in the Introduction. Thus, with some precautions, the idea is that we could realistically
compute RPA corrections up to a region of $-q^2$ values where they become quite small and hence, the possible existence of some systematic errors on their
computation will have little effect in the final observables. Indeed, the present model for RPA corrections has been successfully applied to describe 
MiniBooNE~\cite{Nieves:2011yp} (see Fig.~\ref{fig:MiniBooNE} below) and MINER$\nu$A~\cite{Gran:2013kda} CCQE integrated cross sections. 
\subsection{Neutrino scattering at intermediate energies}\label{sec:results_neutrino}
\begin{table*}[h]
\centering
\begin{tabular}{|c c| c c c|}
  \hline
& & \multicolumn{3}{c|}{$\sigma(\nu_{\mu}+^{16}$O $\rightarrow \mu^-+X)\, [10^{-40} {\rm cm}^2]$}\\
\hline
 & & Non-relativistic & Relativistic & SF   \\
$500$ MeV & Pauli & 625 & 580 &  494\\
 & RPA &$520\pm 40$  & $470\pm 40$ & $445\pm 27$\\ [5pt]
$375$ MeV & Pauli  & 443 & 418 & 328\\
 & RPA & $329\pm 24$  & $308\pm 22$ & $274\pm 14$ \\  [5pt]
$250$ MeV & Pauli  & 199 & 192 & 132\\
 & RPA & $123 \pm 7$ & $ 118 \pm 7$ & $101\pm 5$ \\ [5pt]
\hline
& & \multicolumn{3}{c|}{$\sigma(\bar{\nu}_{\mu}+^{16}$O $\rightarrow \mu^{+}+X)\, [10^{-40} {\rm cm}^2]$}\\
\hline
 & & Non-relativistic & Relativistic & SF   \\
$500$ MeV  & Pauli  &143.8 & 134.4 & 118.9\\
 & RPA & $106.3 \pm 1.9$ & $98.5 \pm 1.9$ & $105.6\pm 1.5$ \\ [5pt]
$375$ MeV  & Pauli &99.8 & 94.1 & 78.2 \\
 & RPA &$71.6\pm 1.4$ & $66.9\pm1.3$ & $68.6 \pm 1.2$\\ [5pt]
$250$ MeV  & Pauli &51.5 & 49.0 & 37.6 \\
 & RPA &$34.3\pm 0.8$  & $32.5\pm 0.8$ & $31.0\pm 0.7$  \\
 \hline
\end{tabular}
\caption{Muon  neutrino  and antineutrino  inclusive QE integrated cross sections from oxygen.
We present results for relativistic  and non-relativistic nucleon kinematics. In this latter case, we present results with  and without SFs effects. Results, denoted as RPA and Pauli have been obtained with and without including RPA  and  Coulomb corrections, respectively. SF results have been computed using a complex self-energy to dress  both, particle and hole nucleon lines.  Theoretical errors on the RPA predictions show MC 68\% CL intervals derived from the uncertainties in the $ph$($\Delta$h)--$ph$($\Delta$h) effective interaction, as detailed  in Subsec.~\ref{sec:rpa}.}
\label{table:3a}
\end{table*}
\begin{table*}[h]
\centering
\begin{tabular}{|c c| c c c|}
  \hline
& & \multicolumn{3}{c|}{$\sigma(\nu_e+^{16}$O$\rightarrow e^-+X)\, [10^{-40} {\rm cm}^2]$}\\
\hline
 & & Non-relativistic & Relativistic & SF   \\
$310$ MeV  & Pauli & 370 & 350 & 271\\
 & RPA & $259\pm 18$ & $244\pm 16 $ & $219 \pm 11$\\  [5pt]
$220$ MeV  & Pauli &191 & 183 & 131\\
 & RPA & $117\pm 7$ & $112\pm 6$ & $101\pm 5$ \\  [5pt]
$130$ MeV  & Pauli &44.6 & 43.1 & 28.3\\
 & RPA & $25.6 \pm 1.2$ & $24.8\pm 1.1$ & $23.2\pm 0.8$ \\ [5pt]
\hline
& & \multicolumn{3}{c|}{$\sigma(\bar{\nu}_e+^{16}$O $\rightarrow e^{+}+X)\, [10^{-40} {\rm cm}^2]$}\\
\hline
 & & Non-relativistic & Relativistic & SF   \\
$310$ MeV  & Pauli & 81.6 & 77.3 &63.1 \\
 & RPA & $57.9\pm 1.1$ & $54.2\pm 1.1$ & $55.6\pm 0.9$\\ [5pt]
$220$ MeV & Pauli  & 49.2 & 47.0 & 36.2\\
 & RPA & $32.3\pm 0.8$ & $30.8\pm 0.8$ & $30.4 \pm 0.7$ \\ [5pt]
$130$ MeV & Pauli  & 17.9 & 17.3 & 12.2 \\
 & RPA &$10.3\pm 0.3$ & $9.8\pm 0.3$ & $ 9.6 \pm 0.3$  \\
  \hline
 \hline
\end{tabular}
\caption{As in Table~\ref{table:3a} but for electron  neutrino  and antineutrino  inclusive QE scattering. }
\label{table:3b}
\end{table*}
\begin{figure*}[h]
\centering
\includegraphics[scale=0.35]{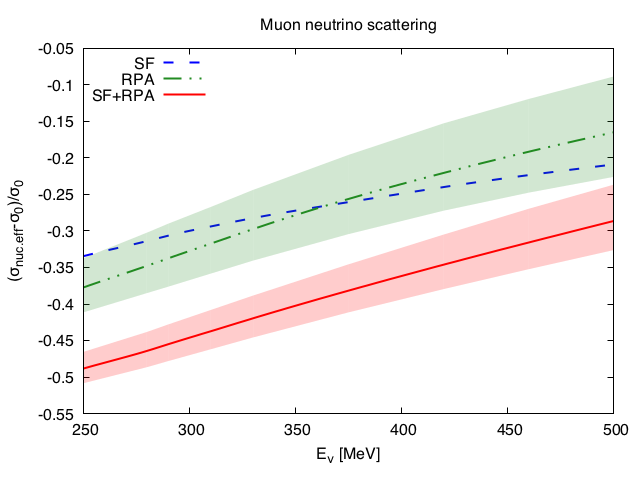}
\includegraphics[scale=0.35]{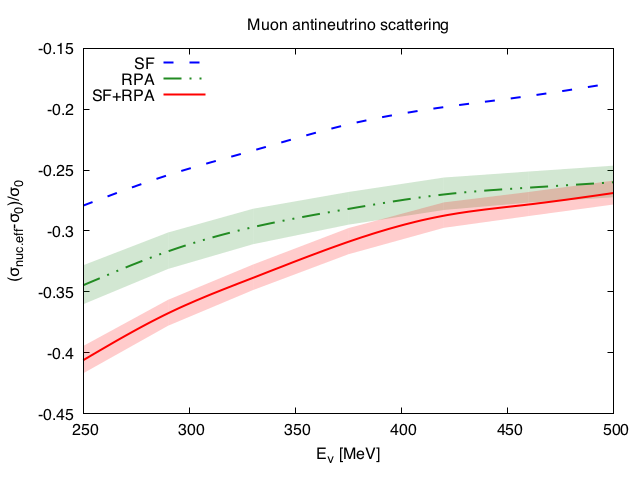}
\hspace{0mm}
\includegraphics[scale=0.35]{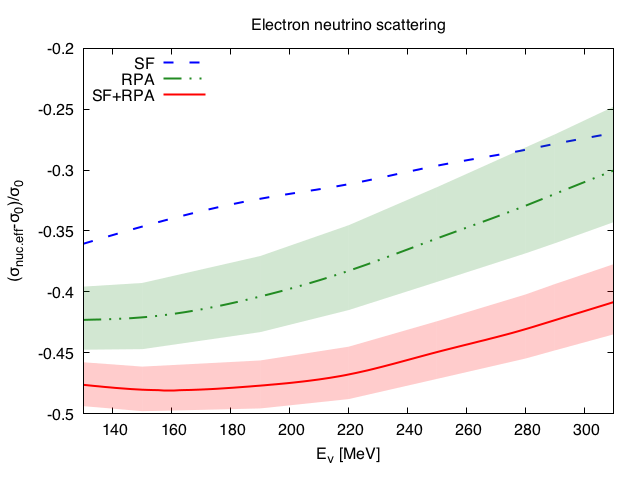}
\includegraphics[scale=0.35]{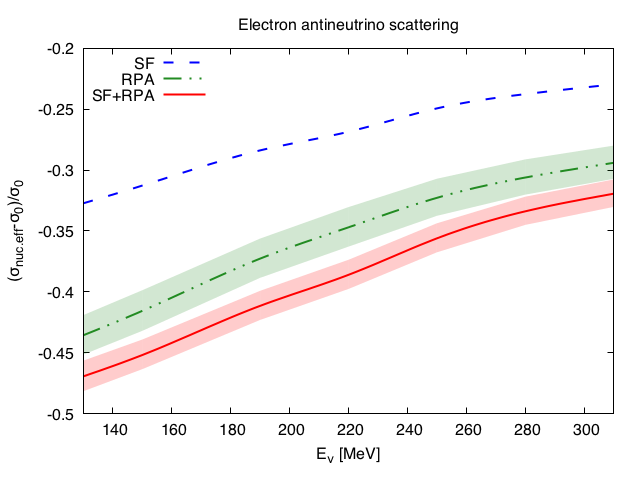}
\caption{Importance of nuclear effects compared to the non-relativistic free LFG cross section ($\sigma_0$). We display $(\sigma_{\text{nuc eff}}-\sigma_0)/\sigma_0$ where \textit{nuc eff} stands for a  nuclear effect (RPA, SF or SF+RPA). The bands show 68\% CL intervals derived from the uncertainties on the $ph$($\Delta$h)--$ph$($\Delta$h) effective interaction.}
\label{fig:14}
\end{figure*}

\begin{figure*}[h]
\centering
\makebox[0pt]{\includegraphics[scale=0.3]{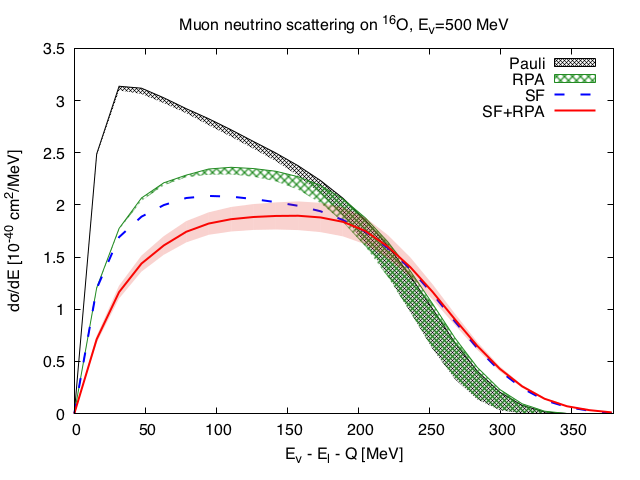}\hspace{0.5cm}\includegraphics[scale=0.3]{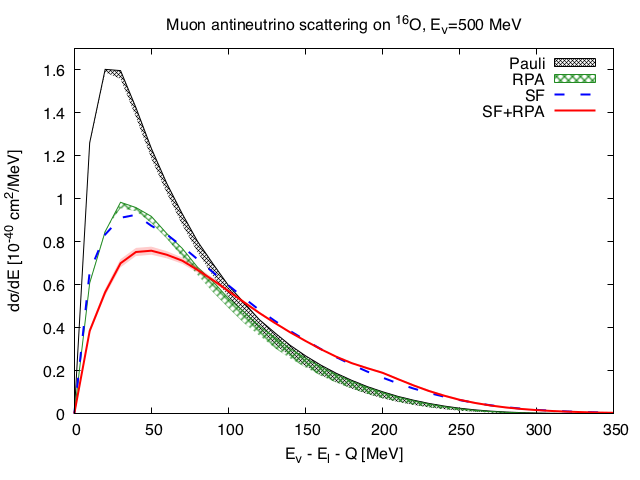}}\\
\makebox[0pt]{\includegraphics[scale=0.3]{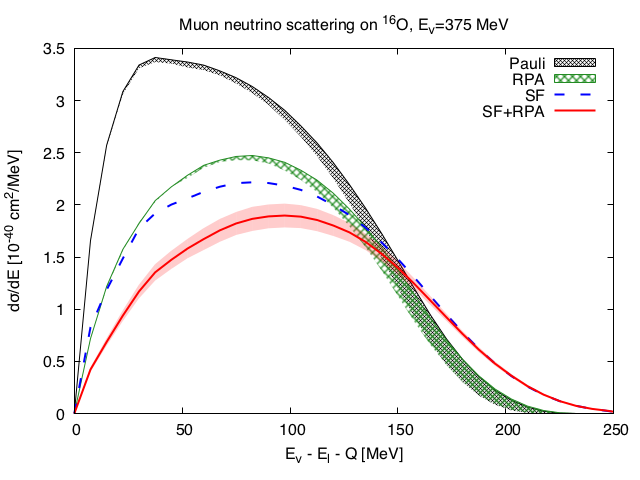}\hspace{0.5cm}\includegraphics[scale=0.3]{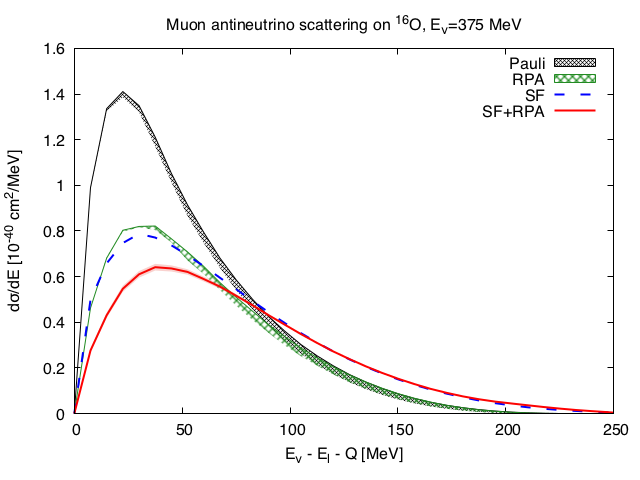}}\\
\makebox[0pt]{\includegraphics[scale=0.3]{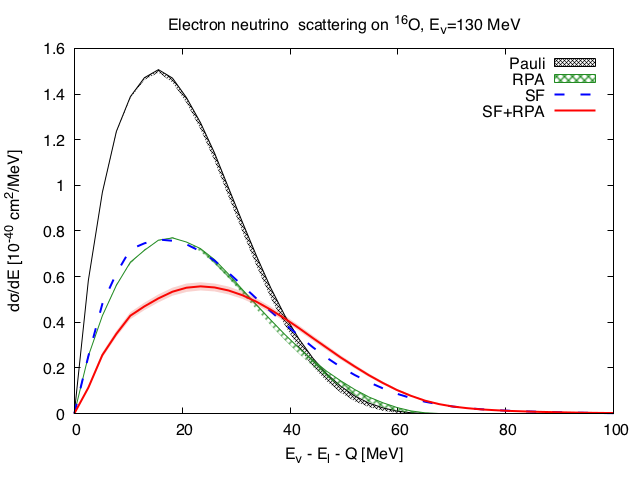}\hspace{0.5cm}\includegraphics[scale=0.3]{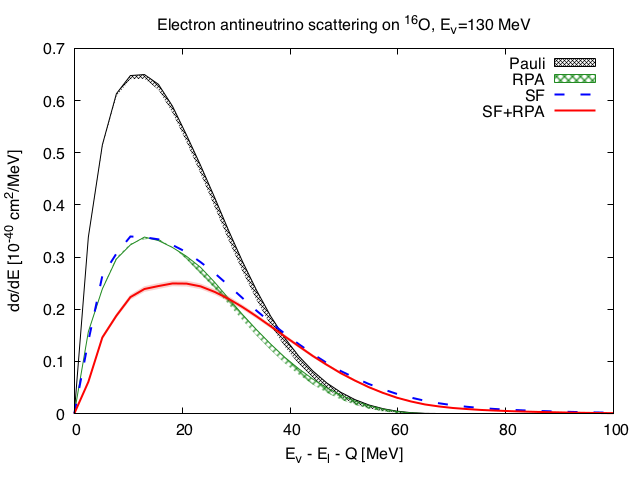}}
\caption{Neutrino and antineutrino differential cross sections from $^{16}$O at various energies. ''Pauli'' and ''RPA'' curves were calculated with non-relativistic kinematics. The use of relativistic kinematics causes a decrease of the cross section shown as stripped pattern bands below those curves.  SF results have been computed using a complex self-energy to dress  both, particle and hole nucleon lines.  Theoretical errors on the SF+RPA predictions show MC 68\% CL intervals (red bands).  }
\label{fig:15}
\end{figure*}
In Tables~\ref{table:3a} and \ref{table:3b}, we present  results in oxygen for  inclusive electron and muon (anti-)neutrino-nucleus scattering and energy transfers up to $400$ MeV. We examine RPA and the SF corrections and their dependence on the energy. First, we observe the differences stemming from the use of non-relativistic and relativistic Lindhard functions. (As mentioned, in the case of non-relativistic kinematics, we use the non-relativistic nucleon dispersion relations and set to one the factors $M/E_p$ and $M/E_{p+q}$ in Eq.~\eqref{eq:lindfree}.)  For the highest considered energies ($E_{\nu_\mu}=500$ MeV), relativistic effects are approximately $7-10\%$ and decrease down to $3-4\%$ for $E_{\nu_\mu}=250$ MeV. We should be aware of this fact when considering SF+RPA corrections because they have been  computed using  non-relativistic kinematics.  

Next, we pay attention to both RPA and SF corrections that suppress the total cross sections. Results are graphically shown in Fig. \ref{fig:14}. For a free LFG, the RPA effects\footnote{For both the non-relativistic and SF set of results, the real part of the $ph-$Lindhard function that appear in the RPA denominators has been computed  using its non-relativistic expression derived in a free LFG.} are especially significant at lower energies,  where we find a very drastic reduction of about $35-40\%$, the corrections being still large (of the order of 20--25\%) for the higher energies examined in  Table~\ref{table:3a}. SF effects change importantly  both, the integrated and the shape of
the differential cross sections, as we will see. When medium polarization (RPA) effects are not considered, the SFs provide significant reductions (20--35\%) of the
neutrino cross sections, and somewhat smaller effects in the case of antineutrinos\footnote{The SF effects reported in Ref.~\cite{Nieves:2004wx} were smaller because in that work, the imaginary part of the hole self-energy was neglected. }. The SF corrections decrease as the (anti-)neutrino energy increases. However, when  RPA correlations are included, the reductions  become more moderate, around 15\% for neutrino reactions, and much smaller for antineutrinos. Indeed, in this latter case and for the higher energies examined in  the Tables~\ref{table:3a} and ~\ref{table:3b}, the integrated cross sections remain practically unchanged.  SF effects are responsible for a certain quenching of the QE peak and a redistribution of its strength as can be seen in Fig.~\ref{fig:15}, where (anti-)neutrino differential cross sections from $^{16}$O at various energies are shown. The use of non-free SFs produces a tail which goes to higher energies inducing in general a significant change of the ($q^0,|\vec{q}\,|$)-region 
accessible in the process.  It does not change the strength of the interaction between the gauge boson and the nucleons (the form--factors), which is how the RPA effect is included in  our  formalism. 

As mentioned  above, when we take into account  RPA corrections, the differences between  SF and non-relativistic LFG total cross sections are small, and in general mostly covered 
by the theoretical errors of the RPA predictions (see Fig. \ref{fig:14}), derived from the uncertainties on the $ph$($\Delta$h)--$ph$($\Delta$h) effective interaction.  This is because the SFs diminish the height of the QE peak and increase the cross section for the high energy transfers. But for
nuclear excitation energies higher than those around the QE peak, the RPA corrections are certainly less important than in the peak region. Hence, the RPA suppression of the SF distribution is significantly smaller than the RPA reduction of the distributions determined by the ordinary Lindhard function. In  Fig.~\ref{fig:15}, we also observe that antineutrino distributions are narrower than neutrino ones and more significantly peaked towards lower energy transfers. Also in these plots, we can see (stripped pattern bands) the size of the relativistic effects. These introduce a systematic error in our predictions in the higher energy transfer region of the differential cross sections, because SF+RPA corrections have been  computed within a  non-relativistic scheme.  

In Fig. \ref{fig:14} we present how the size of the nuclear effects depends on the energy of the incoming (anti)neutrino. We appreciate some differences between neutrino and antineutrino reactions. Both  SF and RPA effects suppress the cross section and as already mentioned, these two combined effects  yield results  similar to those obtained when  only RPA correlations are considered. On the other hand, for antineutrinos, the use of non-free SFs leads to smaller effects.

\begin{figure*}[h]
\centering
\makebox[0pt]{\includegraphics[scale=0.27]{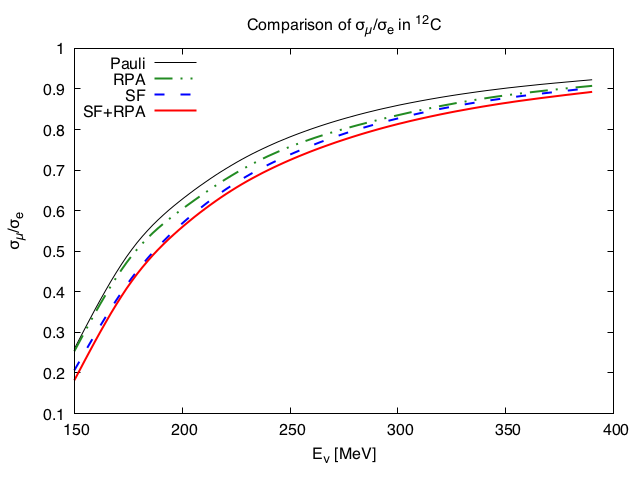}\hspace{0.05cm}\includegraphics[scale=0.27]{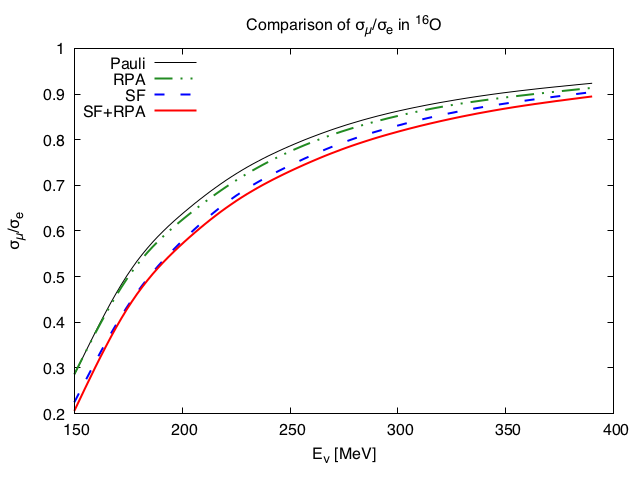}\hspace{0.05cm}\includegraphics[scale=0.27]{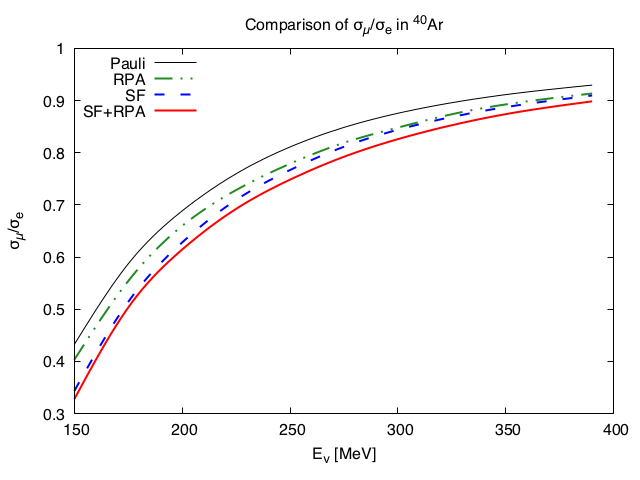}}\\
\makebox[0pt]{\includegraphics[scale=0.27]{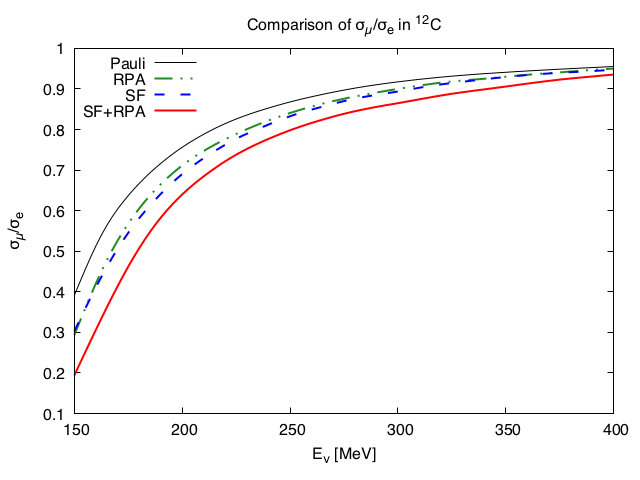}\hspace{0.05cm}\includegraphics[scale=0.27]{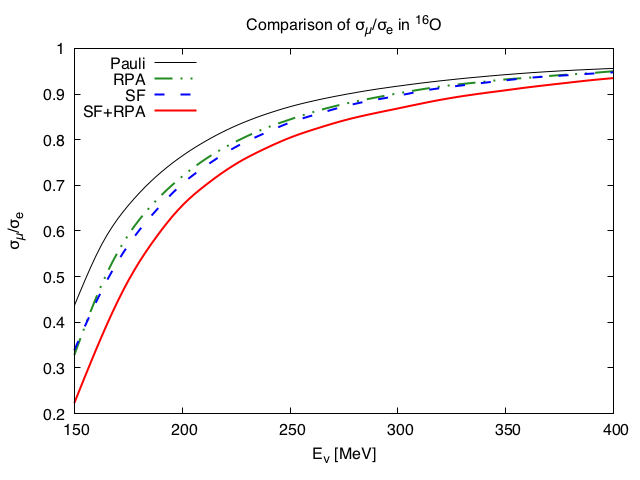}\hspace{0.05cm}\includegraphics[scale=0.27]{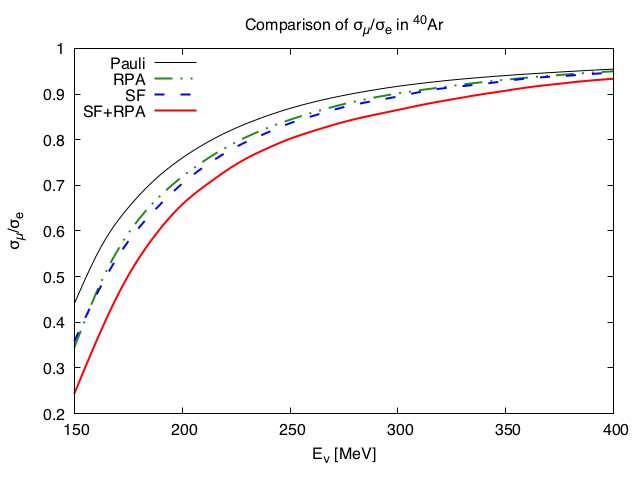}}\\
\makebox[0pt]{\includegraphics[scale=0.3]{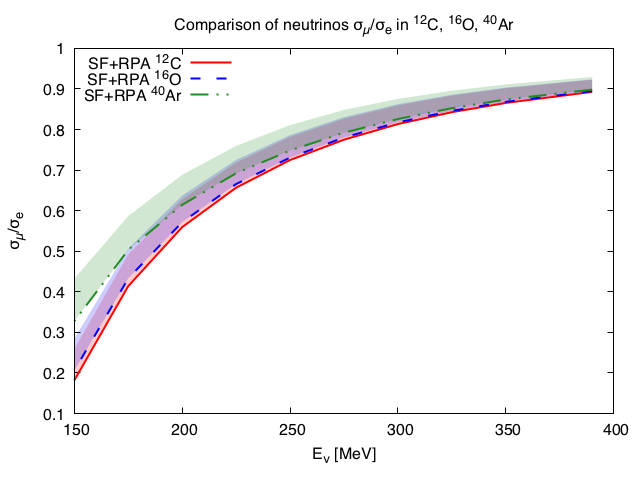}\hspace{0.05cm}\includegraphics[scale=0.3]{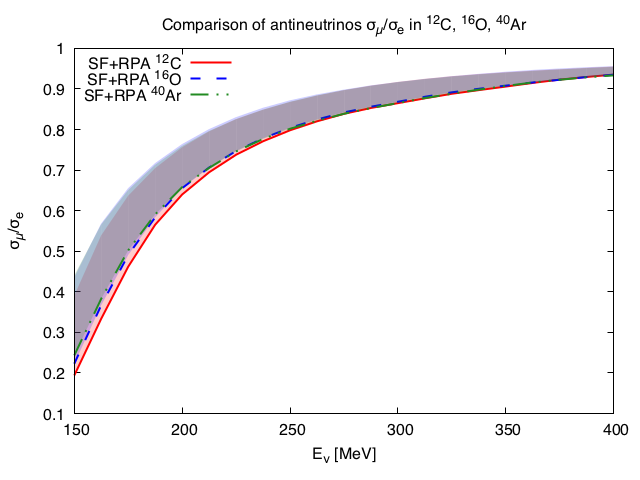}}
\caption{Ratio $\sigma(\mu)/\sigma(e)$ of inclusive neutrino (first row) and antineutrino (second row) QE cross sections 
   for carbon, oxygen and argon, as a  function of the incoming (anti-)neutrino energy. We show  non-relativistic free LFG (labeled as {\it Pauli} in the plots), RPA, SF and SF+RPA results. In the two bottom plots, we show bands (red  for carbon, blue for oxygen and green for  argon) whose upper and lower limits are given by the {\it Pauli} and SF+RPA predictions, respectively. }
\label{fig:ratios}
\end{figure*}

Theoretical errors practically cancel out in the ratio
$\sigma(\mu)/\sigma(e) \equiv \sigma(\nu_\mu+ ^AZ \to \mu^- + X)/\sigma(\nu_e+ ^AZ \to e^- + X) $, and in the equivalent one constructed for antineutrinos. These ratios are depicted in Fig.~\ref{fig:ratios} for carbon, oxygen and argon. Theoretical uncertainties on these ratios turn out to
be much smaller than 1\% and are hardly visible in the plots. On the other hand, predictions for these ratios
obtained from a simple Lindhard function\footnote{It is to say from a local Fermi gas model of non-interacting nucleons.}
incorporating a correct energy balance in the reaction (lines denoted
as ``Pauli'' in the plots) differ from the most realistic ones obtained including also SF+RPA effects  at the level of 5-10\% for neutrino energies above 300 MeV, in sharp contrast
with the situation found for each of the the individual 
$\sigma(\nu_\mu+ ^AZ \to \mu^- + X)$, $\sigma(\nu_e+ ^AZ \to e^- +
X)$, $\sigma(\bar\nu_\mu+ ^AZ \to \mu^+ + X)$ and $\sigma(\bar\nu_e+ ^AZ \to e^+ + X)$ cross sections (see Fig.~\ref{fig:14}). However, these differences are much larger at low energies, especially for the antineutrino ratios. Note that RPA corrections greatly cancel out, especially in carbon and oxygen, in the neutrino ratios calculated with full SFs. For antineutrino ratios,  though,  RPA effects are clearly visible when SFs are used. 
Besides, we should note that  in the ratio
$\sigma(\mu)/\sigma(e)$,  relativistic nucleon kinematics effects are
quite small, being always smaller than 1\% in the whole 
energy interval studied in this work,  as it was pointed out in Ref.~\cite{Valverde:2006zn} (see Fig.~6 of that reference).
\subsubsection{Comparison with other approaches}
\label{sec:comp}
Here, we briefly discuss predictions obtained within other approaches. There is an abundant literature in the field, and we do not aim at performing an exhaustive comparison, but we will rather focus in some representative works, where RPA or SF effects have been examined. 

\begin{itemize}
\item We will begin with the continuum RPA (CRPA) scheme examined in Ref.~\cite{Kolbe:2003ys}. As explained in this latter reference, the main difference between RPA and CRPA approaches lies in  the treatment of the excited states. In the case of RPA, all of them are treated as bound
states, leading to a discrete excitation spectrum, while within a CRPA scheme, the final states asymptotically have the appropriate scattering wave-function  
for energies above the nucleon-emission thresholds; consequently the excitation spectrum
in the CRPA is continuous. In this sense, it is clear that the approach followed here (see Subsec.~\ref{sec:rpa}) should be understood as a CRPA one.  

In Ref.~\cite{Kolbe:2003ys}, it is argued that the RPA or CRPA are the methods 
of choice at intermediate neutrino energies. The CRPA calculations carried out in this reference used
a finite range residual force based on the Bonn potential, and all multipole operators with $J\le 9$ and both parities were included. Free nucleon form factors were used in \cite{Kolbe:2003ys}, with no
quenching, and thus this RPA approach provided a realistic description of collective nuclear excitations due to one-particle one-hole excitations of the correlated ground state.  However, neither short range nucleon-nucleon correlation effects included in realistic SFs, nor the excitation of $\Delta h$ components in the RPA responses are taken into account in the scheme of Ref.~\cite{Kolbe:2003ys}. 

In Fig.~\ref{fig:CRPA}, we compare our RPA predictions for 
$d\sigma/d(\cos\theta')$ with those obtained in \cite{Kolbe:2003ys} for oxygen and  two different electron--neutrino energies. We find a reasonable agreement, which is substantially improved when $\Delta h$ excitations are not allowed in our approach (black dashed curves). (The role played by the inclusion of $\Delta h$ components in the RPA series at intermediate energies was already mentioned in Ref.~\cite{Nieves:2004wx}.) There exist some discrepancies  for $E_\nu= 500$ MeV and $\theta'> 90^{\text 0}$. In this region, the momentum transfers are larger than those for which our non-relativistic RPA treatment is adequate. Nevertheless, we clearly see that in both approaches, RPA corrections lower the cross section at forward angles, but raise it at more backwards angles. This is also seen for $E_\nu= 300$ MeV.  
\begin{figure}[h]
\centering
\makebox[0pt]{\includegraphics[scale=0.35]{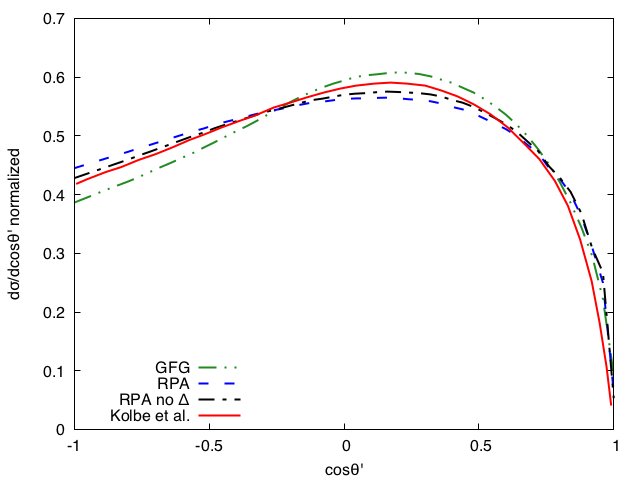}\hspace{0.05cm}\includegraphics[scale=0.35]{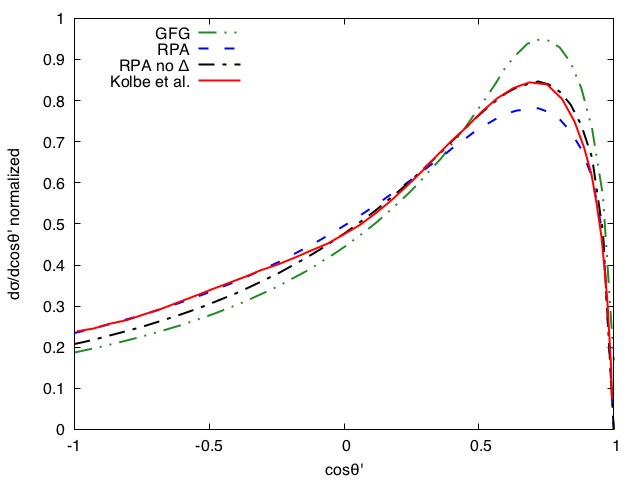}}
\caption{Angular distributions of the emitted electron in the $\nu_e+^{16}$O$\rightarrow e^-+X$ inclusive reaction for $E_\nu=300$ MeV (left) and 500 MeV (right). The curves labeled by GFG and Kolbe et al. are taken from the bottom panel of Fig.~3 of Ref.~\cite{Kolbe:2003ys}, and stand for  the relativistic global Fermi gas model  and the CRPA calculations presented in that work, respectively. In addition, we also show our full RPA predictions and the distributions obtained when the excitation of $\Delta h$ components in the RPA responses are not taken into account (this amounts to setting $U_\Delta$ to zero  in the denominators of Eq.~\eqref{eq:rpa-example}). 
Relativistic free LFG (non-interacting) SFs have been used in all cases.}
\label{fig:CRPA}
\end{figure}

\item The double differential neutrino-carbon quasielastic cross sections measured by the MiniBooNE collaboration triggered an enormous theoretical activity, since a large value of the axial nucleon mass, $M_A$, is needed to describe the data when RPA and $2p2h$ nuclear effects are not considered~\cite{AguilarArevalo:2010zc}. The solution to this puzzle came from the consideration of these nuclear corrections, which were computed by two  different groups: Lyon~\cite{Martini:2011wp} and  IFIC~\cite{Nieves:2011yp}.  The latter one included RPA corrections using the many-body scheme described in this work, while the Lyon group  accounted for
RPA effects as described in Ref.~\cite{Martini:2009uj}.  In Fig.~\ref{fig:MiniBooNE}, we show results~\cite{Nieves:2011yp},  calculated with the model used in this work,  for the  QE contribution
  to the CC quasielastic $\nu_\mu-
  ^{12}$C double differential cross section convoluted with the
  MiniBooNE flux. There, we also display results from the Lyon model taken from
  \cite{Martini:2011wp}.  Both sets of predictions  for this
  genuine QE contribution, with and without RPA effects, turn out to
  be in an excellent  agreement, despite the large corrections produced by the RPA re-summation.  Note that the comparison in Fig.~\ref{fig:MiniBooNE} is quite appropriate, not only for the repercussion of the $M_A$ puzzle, but also because the MiniBooNE flux peaks at muon-neutrino energies around 600 MeV~\cite{AguilarArevalo:2008yp}, below 1 GeV that is the energy used to show predictions in Ref.~\cite{Martini:2009uj}. Our RPA treatment is non-relativistic and it should be used with some caution, as discussed at the beginning of Subsect.~\ref{sec:results_neutrino},  for neutrino energies well above those compiled in Table~\ref{table:3a}. We understand that some relativistic corrections could also limit the validity of the RPA predictions of Ref.~\cite{Martini:2009uj}. 
\begin{figure}[h]
\centering
\includegraphics[scale=0.75]{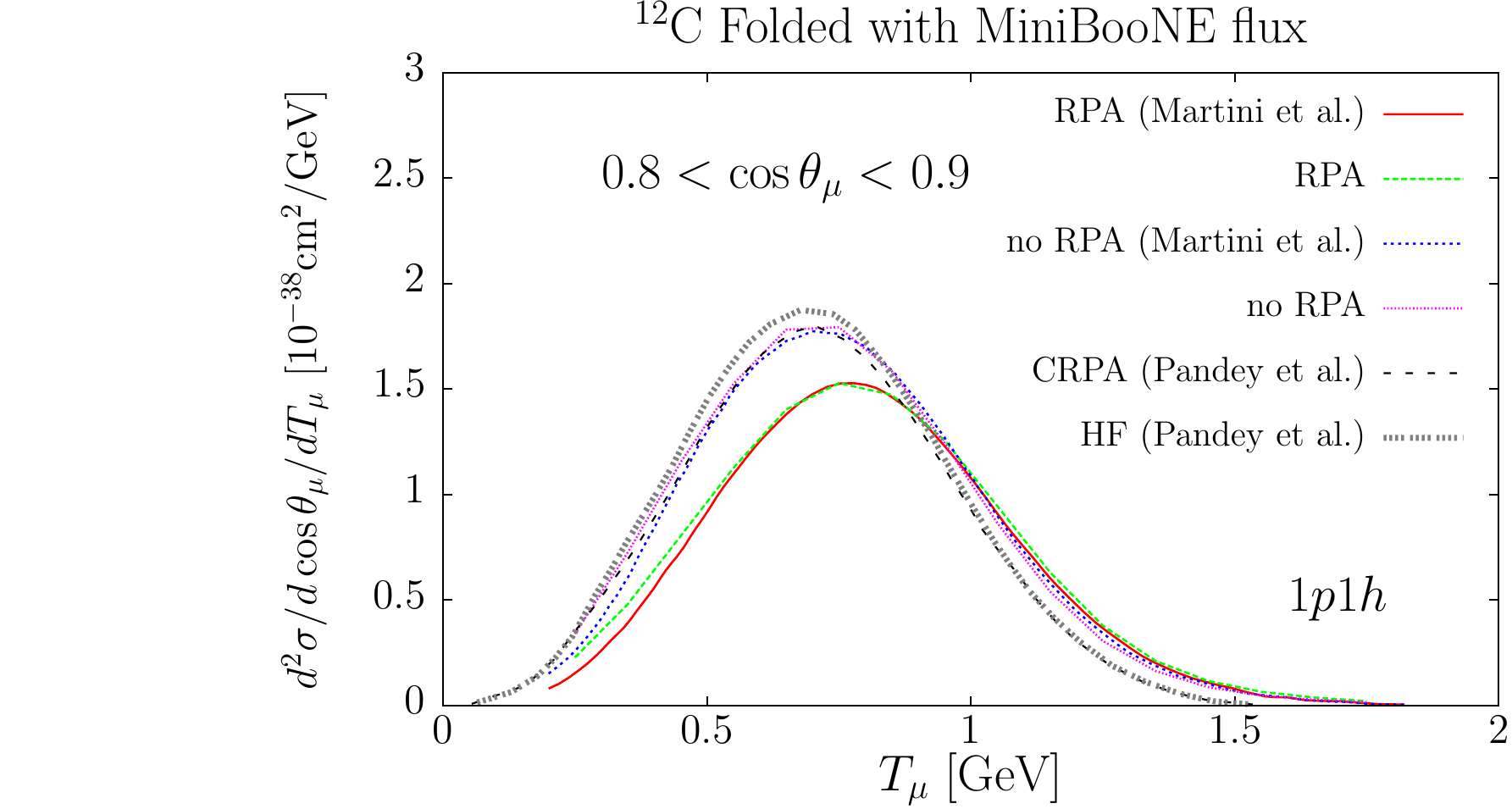}
\caption{RPA effects on the QE contribution to the MiniBooNE flux--averaged  $\nu_\mu- ^{12}$C double differential cross section per
  neutron for $0.8 < \cos\theta_\mu < 0.9$, as a function of the outgoing muon
  kinetic energy. The curves labeled by Martini et al. and Pandey et al. are taken from Fig. 6 of Ref.~\cite{Martini:2011wp} and Fig. 4 of Ref.~\cite{Pandey:2016jju}, respectively, while the other two curves have been calculated using the model presented in this work, and they were first showed in Fig. 3 of Ref.~\cite{Nieves:2011yp}. Relativistic free LFG (non-interacting) SFs have been used in our predictions. }
\label{fig:MiniBooNE}
\end{figure}

\item There exist other RPA or CRPA approaches available in the literature. Thus,  for instance a detailed study of a CRPA approach to QE electron-nucleus
and neutrino-nucleus scattering has been recently presented in Ref.~\cite{Pandey:2014tza}. There, a special attention to low-energy
excitations is paid, together with an exhaustive comparison of the $^{12}$C$(e,e')$ and $^{16}$O$(e,e')$ experimental double differential cross sections with   CRPA  and Hartree-Fock (HF) predictions. The work of Ref.~\cite{Pandey:2014tza} is in principle self-consistent, because the same interaction is used in both the HF and CRPA calculations. This is however not completely true, since the parameters of the momentum-dependent nucleon-nucleon force used in \cite{Pandey:2014tza} were
optimized against ground-state and low-excitation energy properties~\cite{Waroquier:1986mj}, and this force tends to be unrealistically strong at large $Q^2=-q^2$ values. This is corrected in \cite{Pandey:2014tza} by introducing a phenomenological dipole hadronic form factor at the nucleon-nucleon interaction vertices. Qualitative features reported in  \cite{Pandey:2014tza} agree with those  found in this work. To be more specific, let us focus  in the $^{12}$C$(e,e')$ cross sections showed  for different kinematics in Fig.~5 of this reference. There, we see that  being a collective effect, RPA corrections decrease as the associated wave-length of the virtual photon becomes significantly shorter than the typical size of the nucleus~\cite{Gran:2013kda}. Thus, RPA effects become little relevant for the highest $Q^2-$panels showed in that figure, which in general correspond to incoming electron energies above 1 GeV or in the case of smaller energies to large scattering angles. However, large RPA corrections are 
clearly visible for the lowest electron energies (first seven panels of the figure), where in addition $Q^2< 0.1$ GeV$^2$. Indeed, in most of these panels, where $Q^2$ is even smaller than 0.025 GeV$^2$,  we see how the consideration of RPA correlations lead to the appearance of peaks in some regions. In the next subsection (Subsec.~\ref{sec:low}), where the predictions of our model for low energies are discussed,  we will see how something similar also occurs within our model,  and in some regions we find clear enhancements  of the   SF+RPA distributions as compared to those obtained without including RPA corrections. 

In general, and besides the extremely low $Q^2-$panels, we conclude from Fig. 5 of Ref.~\cite{Pandey:2014tza} that RPA effects on top of the HF results are moderately small. This is 
in good agreement with our observation that RPA corrections are smaller when realistic 
SFs are taken into account. (Note that within a HF scheme, the nucleons acquire a real self-energy, and thus somehow this would be equivalent to use SFs obtained neglecting the particle and hole widths). The less important role played by RPA corrections, at sufficiently high $Q^2$ values when some realistic mean field potentials are used, could provide some understanding 
of the success of the SuSA~\cite{Amaro:2004bs, Amaro:2006pr, Amaro:2010sd, Amaro:2011qb, GonzalezJimenez:2012bz, Gonzalez-Jimenez:2014eqa} 
or the bound local FG model (used in the GiBUU--Giessen Boltzmann-Uehling--Uhlenbeck- transport approach~\cite{Gallmeister:2016dnq}), in predicting neutrino cross sections despite  not incorporating RPA effects.

Nevertheless, the approach of Ref.~\cite{Pandey:2014tza}  has some limitations mostly because zero-range Skyrme interactions
do not properly describe processes involving momentum
transfers, significantly larger than $m_\pi$. Indeed, though a zero-range Skyrme force
might be adequate for a microscopic description of both ground- and
excited-state properties of nuclei, it might not be well suited to describe the dynamics of the ejected nucleon ($S_p$) or to compute
RPA corrections for large momentum and energy  transfers (let us say above 150 MeV). In this
latter case, including $\Delta(1232)$ degrees of freedom (as we have shown
in the discussion of the results of Ref.~\cite{Kolbe:2003ys}), or 
considering explicitly pion exchange contributions to the interaction
produce significant effects. This has been shown in multitude of works~\cite{Oset:1981ih, Oset:1989ey,Carrasco:1989vq, FernandezdeCordoba:1991wf, FernandezdeCordoba:1992df, Nieves:1991ye,  FernandezdeCordoba:1992ky, Nieves:1992pm, Nieves:1993ev, Hirenzaki:1993jc, Carrasco:1992mg, Oset:1994vp, FernandezdeCordoba:1993az,  Oset:1994vp, GarciaRecio:1994cn, Hirenzaki:1995js, Gil:1997bm, Gil:1997jg, Albertus:2001pb,Albertus:2002kk}, where photon,
electron, pion, kaon, $\Lambda, \Sigma-$hyperons etc. interactions with nuclei have been described within the many-body framework used here. 
Thus, as an example, in Fig.~\ref{fig:Ghent}, we compare with data the predictions of the approach of Ref.~\cite{Pandey:2014tza} for inclusive QE cross section for scattering of electrons on carbon at 560 MeV and 60$^o$ ($|\vec{q}\,|= -0.508$ GeV). We observe that the results of  Ref.~\cite{Pandey:2014tza} describe already the data in the region of QE peak, leaving almost no room for $2p2h$ contributions, which according  to the 
empirical fit to electron-nucleus scattering data carried out in \cite{Bosted:2012qc} provide a significant cross section in that region. Note that our SF+RPA QE predictions (RPA effects are moderately small, as one can expect for this value of $q^2$) lie below the data, and one might expect that some $2p2h$ contributions would improve notably the agreement with data\footnote{We cannot simply add up the $2p2h$ contribution displayed in Fig.~\ref{fig:Ghent} to our predicted cross section. This is because our model for the SFs contains contributions from the $2p2h$ diagram depicted in Fig.~\ref{fig:2p2h},  and then the addition of the rest of $2p2h$ contributions will include some interference terms which sign is not defined. Moreover and according to Ref.~\cite{Gallmeister:2016dnq}, the  $2p2h$ curve also accounts for some short-range and RPA effects.}.  In addition, one should bear in mind that  our  results for the energy transfers larger than that of the QE peak are affected from relativistic corrections, 
which will make the distribution narrower, as can be inferred from the reddish-shaded region shown in Fig.~\ref{fig:Ghent}. Indeed, the position of the QE peak is also affected and a relativistic calculation will shift its position around 10 MeV towards lower energy transfers.

We should note that the GiBUU 2016 QE plus $2p2h$ cross sections, supplemented by  $\Delta(1232)-$driven mechanisms and some non-resonant pion background  terms provide a fairly good description of the data for all energy transfers shown in Fig.~\ref{fig:Ghent}, as can be seen in the original Fig.~3 of Ref.~\cite{Gallmeister:2016dnq}.
\begin{figure}[h]
\centering
\includegraphics[scale=0.4]{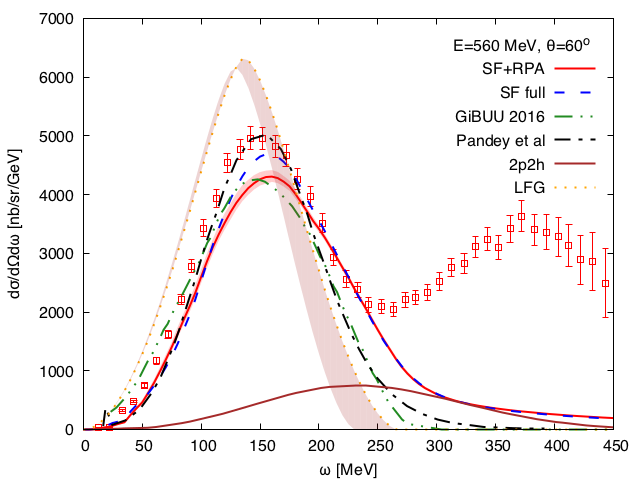}
\caption{Inclusive QE cross section for scattering of electrons on
carbon at 560 MeV and 60$^o$ ($q^2= -0.242$ GeV$^2$ at the QE peak). 
Besides the SF and SF+RPA results obtained within the many-body framework used in this work (RPA corrections are included 
as in the case of the vector contributions to the neutrino-induced inclusive
QE reaction driven by the electroweak NC  studied in Ref.~\cite{Nieves:2005rq}; see also \cite{Gil:1997bm}),  predictions from Refs.~\cite{Pandey:2014tza} (Pandey et al.) and \cite{Gallmeister:2016dnq} (GiBUU 2016) are also shown. These latter curves are taken from panel $j$ of Fig.~5 of \cite{Pandey:2014tza} and Fig.~3 of \cite{Gallmeister:2016dnq}, respectively. The $2p2h$ curve, taken also from Fig.~3 of  \cite{Gallmeister:2016dnq}, stands for contributions of meson exchange currents (genuine $2p2h$), and it might include also short-range  and RPA effects~\cite{Gallmeister:2016dnq}. It is obtained from an empirical fit to electron-nucleus scattering data carried out in \cite{Bosted:2012qc}. Finally, the reddish-shaded region shows the difference between relativistic and non-relativistic non-interacting LFG predictions. Data taken from Ref.~\cite{Benhar:2006er}.}
\label{fig:Ghent}
\end{figure}

The HF and CRPA approaches  of Ref.~\cite{Pandey:2014tza}  were used in Ref.~\cite{Pandey:2016jju} to evaluate the  QE contribution
  to the CC quasielastic $\nu_\mu-  ^{12}$C double differential cross section convoluted with the MiniBooNE flux. These latter results for $0.8 < \cos\theta_\mu < 0.9$, as a function of the outgoing muon
  kinetic energy, are also displayed in Fig.~\ref{fig:MiniBooNE}. The size of the quenching is smaller in the CRPA model of Refs.~\cite{Pandey:2014tza,Pandey:2016jju}, resulting
in a larger predicted cross section for the QE process, than in the approaches of Refs.~\cite{Nieves:2011yp} and \cite{Martini:2011wp}. We expect here a situation similar to that discussed in Fig.~\ref{fig:Ghent} for electron scattering, since all the available estimates~\cite{Nieves:2011yp, Martini:2011wp, Megias:2016fjk, Gallmeister:2016dnq} for the
  $2p2h$ contribution to the CCQE-like cross section measured by MiniBooNE will lead to the  CRPA or HF models used in  \cite{Pandey:2016jju} to
  overestimate the data\footnote{See for instance the results for the QE and 2p2h
  cross sections given in Fig. 5 of Ref.~\cite{Gallmeister:2016dnq}, which sum describes fairly well the MiniBooNE data. }.

\item  Next we pay attention to schemes involving realistic SFs. We begin with the formalism, based on factorization and a state-of-the-art model of the nuclear SFs, used in Refs.~\cite{Benhar:2005dj, Benhar:2006nr, Benhar:2009wi, Benhar:2010nx, Vagnoni:2017hll} to describe neutrino-nucleus interactions.  Such scheme has been extensively and  successfully tested in electro-nuclear reactions at relatively large energies. We  
first compare in Fig.~\ref{fig:SFs+fact} our results with the most recent QE neutrino predictions reported in Ref.~\cite{Vagnoni:2017hll}. 
\begin{figure}[h]
\centering
\includegraphics[scale=0.5]{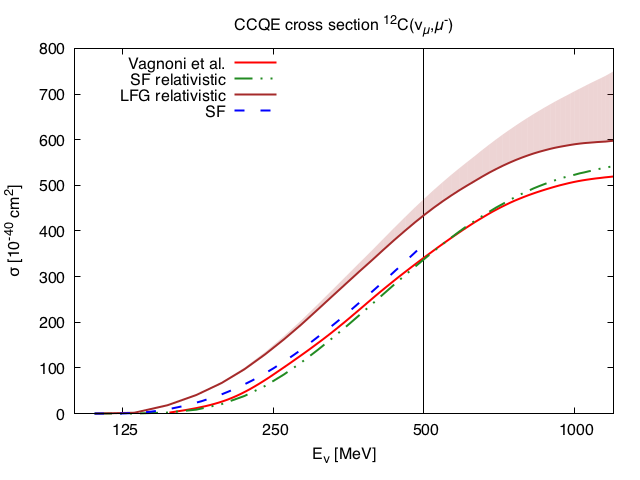}
\caption{CCQE cross section of the reaction $\sigma(\nu_\mu+ ^{12}{\rm C} \to \mu^- + X)$
as a function of neutrino energy. Besides the results taken from the bottom panel of Fig.~3 of Ref.~\cite{Vagnoni:2017hll}, and labeled as Vagnoni et al., we also display (blue dashed line) our full SF predictions up to 500 MeV, and relativistic and non-relativistic free LFG (upper limit of the reddish band) cross sections for the entire neutrino energy range. We also show results (SF relativistic) obtained keeping the full hole SF but replacing the particle spectral function $S_p(q^0+\omega,\vec{p}+\vec{q}\,)$ in Eq.~\eqref{eq:linSF} by $M \delta(q^0+\omega-E_{p+q})/E_{p+q}$, with $E_{p+q}$ the relativistic free energy of the outgoing nucleon.   }
\label{fig:SFs+fact}
\end{figure}
The calculation of  Ref.~\cite{Vagnoni:2017hll} considers a fully dressed nucleon-hole, but uses a free particle SF, i.e., it employs a plane wave for the outgoing nucleon, satisfying a free relativistic energy-momentum dispersion relation. In the terminology of this reference, FSI effects are not taken into account. In spite of this, we see that our results, obtained dressing both particle and hole nucleon lines with a complex self-energy, agree  quite well with the predictions given in Ref.~\cite{Vagnoni:2017hll} up to $E_\nu=500$ MeV, where relativistic corrections could start being relevant. This confirms the validity of the approximation, some times used by this group, of neglecting FSI nuclear effects when studying inclusive total cross sections\footnote{FSI effects on inclusive integrated cross sections are mostly produced by the consideration of the real part of the 
self-energy in the energy conservation equation, and are in general small.}. To extend the comparison to higher energies, we have adopted the same approximation  as in \cite{Vagnoni:2017hll},   and replaced $S_p$ in Eq.~\eqref{eq:linSF} by an energy conserving delta function\footnote{As discussed in Subsec.~\ref{sec:fsi}, the real part of the nucleon self-energy is evaluated in \cite{FernandezdeCordoba:1991wf} up to  momentum independent pieces that appear both in the hole and particle self-energies and  that cancel in the computation of the imaginary part of the fully dressed Lindhard function. However, to obtain results using a dressed hole and an undressed particle, an absolute value for the real part of the nucleon-hole self-energy is needed. Here, we  include phenomenologically a constant term $C\rho$ in the nucleon self-energy, with $C=0.8$ fm$^2$ for carbon, 
fixed to  a binding energy per nucleon $|\epsilon_A| = 7.8$ MeV (see Subsec.~\ref{sec:fsi}).}, including also the $M/E_{p+q}$ factor that appears in the evaluation of the Lindhard function when relativistic kinematics is used. The green dash-dotted curve, labeled as SF relativistic,
in Fig.~\ref{fig:SFs+fact} shows the results of this new calculation. The agreement with the predictions of Ref.~\cite{Vagnoni:2017hll} is remarkable for the entire neutrino energy range displayed in the figure, even above 1 GeV. 

Note  that the use of a realistic hole SF produces significant corrections, which clearly need to be accounted for to achieve an accurate description of the cross section.

FSI effects (use of a non-trivial particle SF) in the scheme of Refs.~\cite{Benhar:2005dj, Benhar:2006nr, Benhar:2009wi, Benhar:2010nx} are taken into account by means of a convolution~\cite{Benhar:2013dq, Ankowski:2014yfa}, which involves the real part of a nucleon-nucleus optical potential--responsible
for a certain shift in the QE peak position--, the nuclear transparency, and  the in-medium $NN$ scattering cross section. The imaginary part of the Lindhard function calculated using the SFs of Ref.~\cite{FernandezdeCordoba:1991wf} also nicely agrees with that deduced within the scheme of Refs.~\cite{Benhar:2005dj, Benhar:2006nr, Benhar:2009wi, Benhar:2010nx} when FSI effects are taken into account. This is work in progress that will be presented elsewhere~\cite{new}. Some preliminary results can be found in \cite{talk-noemi}, where the scaling function~\cite{Alberico:1988bv, Barbaro:1998gu, Donnelly:1999sw} is computed and compared in both approaches (the scaling function is essentially, up to a factor $|\vec{q}\,|$ and some other constants,  the imaginary part of the Lindhard function~\cite{Rocco:2017hmh}).

In Fig.~\ref{fig:Benhar500}, we have also compared the results of our approach in the QE region for several $e+^{12}{\rm C}\to e' X$ double differential distributions at different  scattering angles and incoming electron energies with data and with the predictions of Ref.~\cite{Ankowski:2014yfa}. The approach of Ref.~\cite{Ankowski:2014yfa}, in addition to the use of a realistic hole spectral function,   takes also into account the effects of FSI (non-trivial particle SF) between the struck nucleon and the residual nucleus through the convolution mentioned above. Our full SF results agree reasonably well with the predictions of Ref.~\cite{Ankowski:2014yfa} for all  examined kinematics. Nevertheless in the bottom panels, for which  $|\vec{q}\,| > 365$ MeV, our distributions are wider than those obtained within the approach of Ref.~\cite{Ankowski:2014yfa}, showing clear differences above the QE peak. Relativistic corrections will make our distributions narrower, as can be inferred from the reddish-shaded 
regions in Fig.~\ref{fig:Benhar500}.
\begin{figure}[h]
\centering
\includegraphics[scale=0.25]{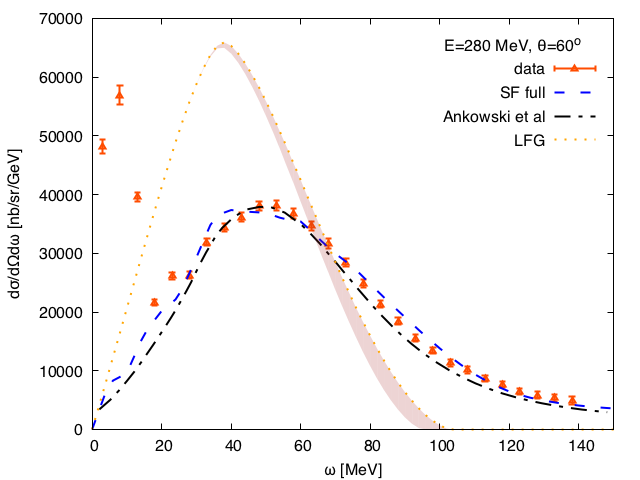}
\includegraphics[scale=0.25]{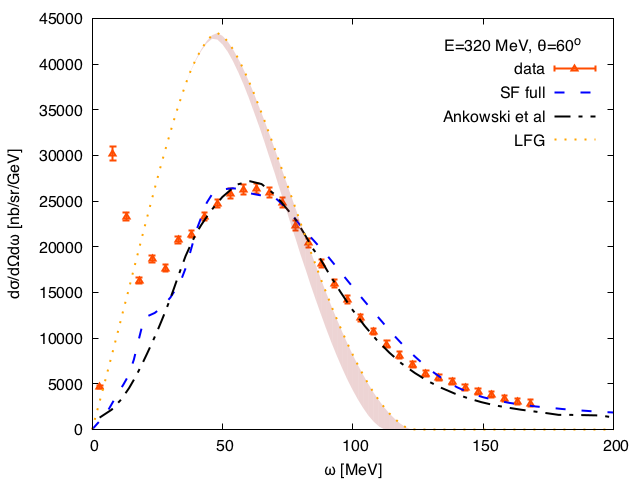}
\includegraphics[scale=0.25]{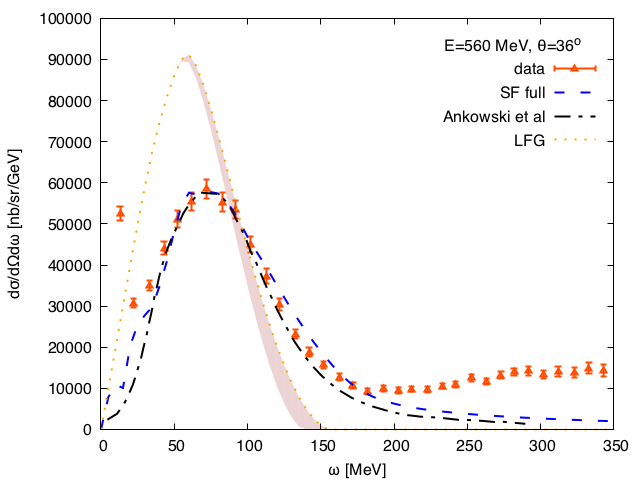}
\includegraphics[scale=0.25]{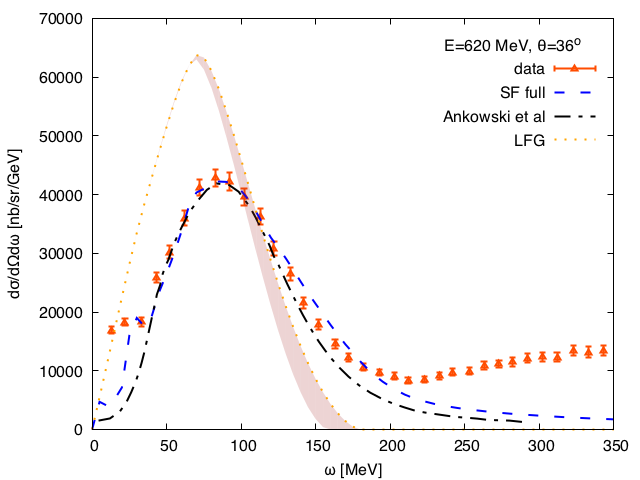}
\includegraphics[scale=0.25]{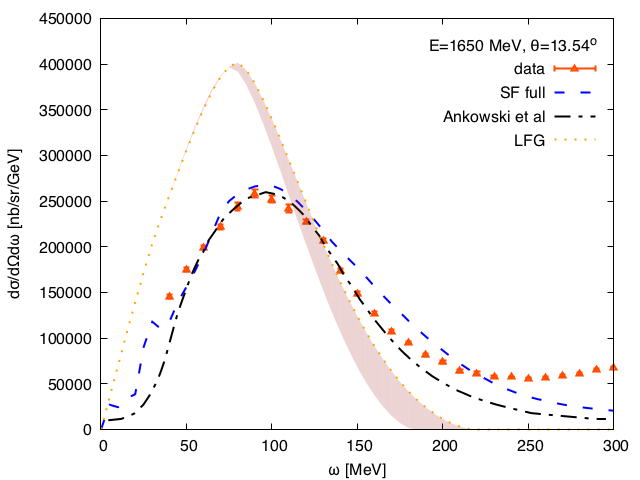}
\includegraphics[scale=0.25]{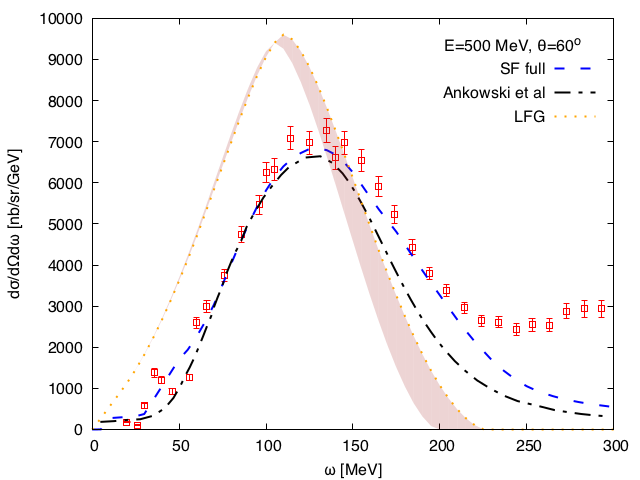}
\caption{Inclusive QE cross sections for scattering of electrons on
carbon at different  scattering angles and incoming electron energies. 
Besides the SF results obtained within the many-body framework used in this work,  predictions (Ankowski et al.) taken from panels (d)--(i) of Fig.~2 of Ref.~\cite{Ankowski:2014yfa} are also shown. At the QE peaks, the momentum transfers $|\vec{q}\,|$ are 259, 295, 331, 366, 390 and 450 MeV, respectively.  Data taken from Refs.~\cite{Barreau:1983ht,Baran:1988tw,  Whitney:1974hr}.  As in Fig.~\ref{fig:Ghent}, the reddish-shaded regions show the difference between relativistic and non-relativistic non-interacting LFG predictions.}
\label{fig:Benhar500}
\end{figure}

\item We finish these comparisons discussing the similarities of our approach with the GiBUU framework  used in Ref.~\cite{Leitner:2008ue} to make predictions for
CC and NC inclusive scattering of oxygen at beam energies ranging from 0.5 to 1.5 GeV. The scheme takes into account various nuclear effects: the LDA 
for the nuclear ground state, mean-field potentials, and in-medium spectral functions. For the spectral function of the initial state nucleon, it was considered only the real part of the self-energy generated by  a mean-field potential and neglected the imaginary part\footnote{Neglecting the hole width is a priori a reasonable approximation, as can be inferred from Fig.~\ref{fig:7}, and it was also used in Ref.~\cite{Nieves:2004wx}. Note however that in this latter work, the Jacobian determinant discussed in Eq.~\eqref{eq:Jacobian} was further approximated to one.}. All these  in-medium modifications were tested by comparing the predictions of the model with  electron scattering results.

In Fig.~\ref{fig:GiBUU}, we show the CCQE predictions for oxygen at $E_\nu=0.5$ GeV given in  the top panel of Fig.13 of Ref.~\cite{Leitner:2008ue} (orange-dotted curve labeled as Leitner et al in Fig.~\ref{fig:GiBUU}), together with our  SF and free LFG results. The agreement is not as good as in the previous cases, and our full SF distribution at the QE peak 
is smaller (around 30\%) than that obtained in Ref.~\cite{Leitner:2008ue}, and it is also significantly wider. The agreement improves when the results of Ref.~\cite{Leitner:2008ue} are compared with the differential cross section obtained within our model neglecting the imaginary part of the hole self-energy, 
as in ~\cite{Leitner:2008ue}. 

We should note that the GiBUU framework used in \cite{Leitner:2008ue} overestimates the similar $(e,e')$ differential QE cross sections  for incoming electron energies and outgoing scattering angles close to those examined in Fig.~\ref{fig:GiBUU}. This can be seen in Figs.~10 and 11 of the same reference \cite{Leitner:2008ue}. Indeed looking at the top panels of these two figures, one can appreciate deviations from data of around 20-25\% at the QE peak. Moreover, the discrepancies seem to increase when both, the incoming electron energy and scattering angle decrease. (Note that in the top panels of Figs.~10 and 11, 
both  scattering angle and energy are larger than those examined in Fig.~\ref{fig:GiBUU}.) One certainly expects that the approximate SF-treatment used in \cite{Leitner:2008ue} should work much better and be quite accurate 
for angular-integrated cross sections. 

A new release of GiBUU became available in 2016~\cite{Gallmeister:2016dnq}, among other improvements, a better preparation of the nuclear ground state and its momentum distribution are implemented. This corresponds in our language to use  more accurate SFs. The new result is also shown (blue-dashed curve, labeled as GiBUU 2016) in Fig.~\ref{fig:GiBUU}, where we could see the agreement with the two versions of the present work is now quite good. 

Moreover, the update GiBUU version provides an excellent description of electron data, not only for  QE scattering   discussed in this work, but also in the dip and $\Delta-$peak regions~\cite{Gallmeister:2016dnq, talk-mosel}.
\begin{figure}[h]
\centering
\includegraphics[scale=0.5]{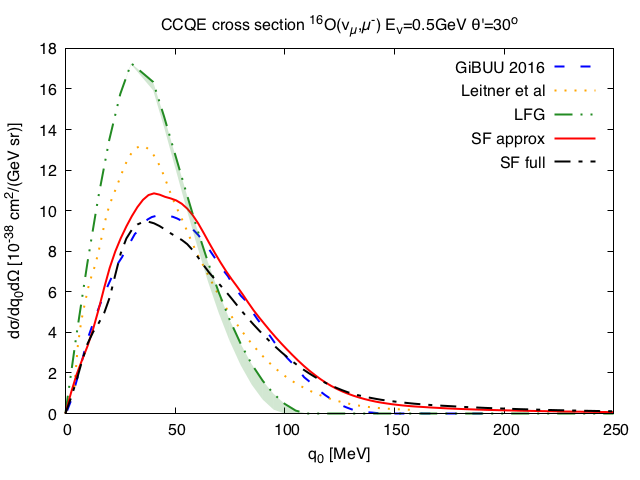}
\caption{CCQE neutrino double differential cross section $d^2\sigma/d\Omega (\hat{k}') d E'$  on $^{16}$O as a function of the energy transfer  at $E_\nu= 0.5$ GeV and a scattering angle of $\theta'=30^{\rm o}$. The orange-dotted curve, labeled by Leitner et al., stands for 
the full calculation of Ref.~\cite{Leitner:2008ue} (full in-med. SF curve of top panel of Fig.~13 of this reference). We also show relativistic and non-relativistic free LFG, and the full SF and approximated SF (neglecting the hole width)  sets of predictions calculated within the many-body framework discussed in this work.  Finally the blue-dashed curve, labeled as GiBUU 2016, has been obtained with the 2016 updated version of the GiBUU code~\cite{Gallmeister:2016dnq}.  }
\label{fig:GiBUU}
\end{figure}

\end{itemize}
\subsection{Low energy results: muon and radiative pion captures and neutrino scattering near threshold}
\label{sec:low}
Even though one may expect that in the low energy regime (excitations below 50 MeV) the LFG based formalism should break down, it was argued in \cite{Amaro:1997ed, Amaro:2004cm} that  the current scheme gives reasonable  estimates for inclusive (integrated) quantities, for instance inclusive muon and radiative pion capture widths in nuclei. Thus, the shapes of the differential cross sections or widths which will be presented in this subsection do not pretend to recover the physical spectra (which might contain discrete transitions and/or resonances in this energy range). They rather illustrate the general trend of the SF and the RPA effects, and their areas might provide reasonable predictions for integrated observables.

In addition, the studies carried out in Refs.~\cite{Chiang:1989ni} and \cite{Nieves:2004wx} of the inclusive muon and radiative pion captures in nuclei, and the  LSND~\cite{Albert:1994xs, Athanassopoulos:1997rn, Auerbach:2002iy, Athanassopoulos:1997rm}, LAMPF~\cite{Krakauer:1991rf} and KARMEN~\cite{Bodmann:1994py}   
near threshold $^{12}$C $(\nu_\mu,\mu^-)X$ and  $(\nu_e,e^-)X$ reactions did not take into account SF effects. In this section we would mainly focus on this aspect of our model. We will present results from the full SF calculation, where both particle and hole nucleon lines have been dressed with a complex self-energy. In this energy region, this full SF treatment leads to results around $30\%$ lower  at the peak than those obtained with the approximated SF, where the width of the nucleon-hole is neglected (see Fig.~\ref{fig:profile}). This sizable difference becomes more moderate when we include  RPA corrections, however it still is of the order of $10-20\%$. We will neglect relativistic effects in all the results presented in this subsection.

\subsubsection{Inclusive radiative pion capture}\label{sec:pioncapt}
\begin{figure}[h!]
\centering
\includegraphics[scale=0.35]{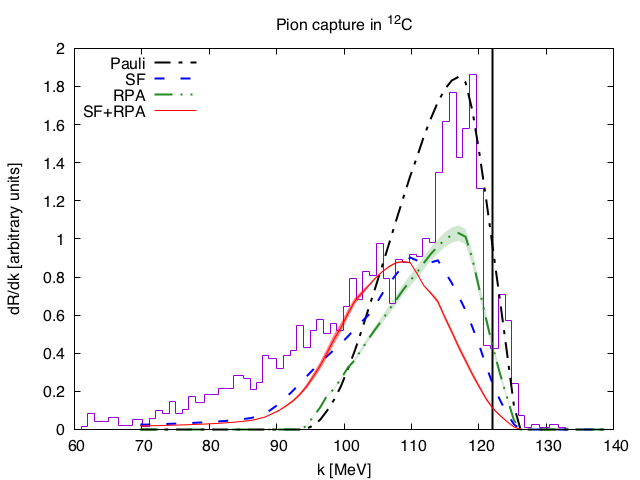}
\includegraphics[scale=0.35]{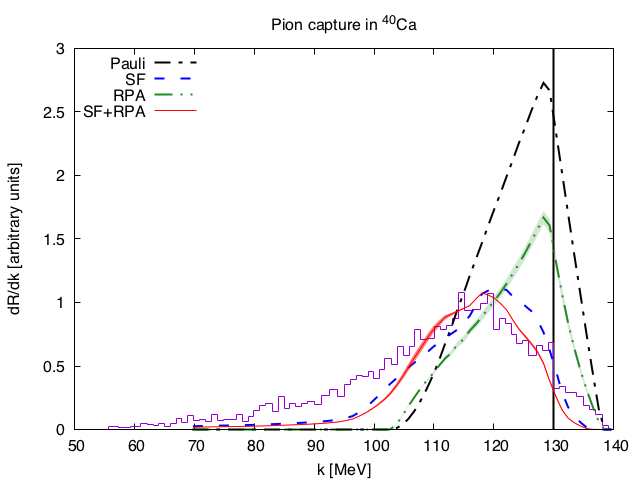}
\caption{Photon energy distributions (arbitrary units) from pion capture in $^{12}$C (left) and $^{40}$Ca (right).  Experimental spectra are taken from Ref.~\cite{Bistirlich:1972jy}. Theoretical SF+RPA curves  were adjusted to data  in the peak, other curves (Pauli, RPA, SF) were scaled by the same factor. Error bands on the RPA predictions show MC 68\% CL intervals derived from the uncertainties in the $ph$($\Delta$h)--$ph$($\Delta$h) effective interaction. The vertical lines show the maximum photon energy for the continuum contribution, $(A_Z -\pi^-)_{\rm bound} \rightarrow \gamma + n+ (A-1)_{Z-1}$, where the final nucleus is left in its ground state.}
\label{fig:11}
\end{figure}
\begin{table*}[h]
\centering
\begin{tabular}{|| l c c c  | c c c c c |c||}
  \hline
\hline
Nucleus & $nl$\ \ \ \ & $w_{nl}$\ \ & $\Gamma^{abs}_{nl}$ [keV]& Pauli [eV]\ \ & RPA [eV] \ \  & SF [eV] \ \ & SF+RPA [eV] \ \ & \\
\hline
\hline
$^{12}$C & 1s &  0.1 & $3.14\pm 0.14$ & 88.9 & $48.3\pm2.1$ &58.6 &  $50.6\pm1.3$ & \\
 & 2p &  0.9 & $0.00136\pm 0.00020$ &  18.3$\times10^{-3}$  &$(11.1\pm0.4) \times10^{-3}$ &12.2$\times10^{-3}$ & $(11.1\pm0.2) \times10^{-3}$& \\
\hline
$^{40}$Ca & 2p & 0.7 & $1.59\pm 0.02$ & 41.5 & $24.3\pm0.9$ & 23.9 & $21.5\pm0.5$&\\
 & 3d &0.3 & $0.0007\pm 0.0003$ & 20.9$\times10^{-3}$ & $(13.8\pm0.4)\times10^{-3}$ &11.7$\times10^{-3}$ & $(11.1\pm0.1)\times10^{-3}$&\\
  \hline
 \hline
\end{tabular}
\caption{Inclusive radiative pion capture widths from the $1s$ and $2p$  and the $2p$ and $3d$ levels in $^{12}$C and $^{40}$Ca, respectively. Theoretical errors in the RPA predictions show MC 68\% CL intervals derived from the uncertainties on the $ph$($\Delta$h)--$ph$($\Delta$h) effective interaction. Within the SF+RPA scheme, we obtain ratios $R^{(\gamma)}$ of $(0.9\pm 0.1)\%$ and 
$(1.4\pm 0.2)\%$ for carbon and calcium, respectively. The experimental values reported in Ref.~\cite{Bistirlich:1972jy} for these ratios are $(1.92\pm 0.20)\%$ for $^{12}$C and $(1.94\pm 0.18)\%$ for $^{40}$Ca. In this latter reference, in the case of carbon, the contributions of transitions to the $^{12}$B ground and excited states turned to be around 20-25\% of the total ratio. Thus, the continuum contribution for $^{12}$C was estimated to be    $(1.50\pm 0.15)\%$ \cite{Bistirlich:1972jy}. }
\label{table:2pion}
\end{table*}
Let us analyze how the total decay width changes when we include additional nuclear effects to Pauli blocking, implemented through the imaginary part of the Lindhard function calculated for a non-interacting LFG of nucleons. Neither SF effects, nor the correct energy balance in the reaction were considered in the previous work of Ref.~\cite{Chiang:1989ni}, where this formalism (LFG+RPA)  was used for the first time. Experimentally, it is rather difficult to distinguish between radiative pion capture processes from different
pionic atom orbits. Indeed, only the weighted ratio
\begin{equation}
\frac{d R^{(\gamma)}}{d|\vec{k}\,|} = \sum_{nl} \frac{w_{nl}}{\Gamma^{abs}_{nl}} \frac{d \Gamma_{nl}^{(\gamma)}}{d|\vec{k}\,|}
\end{equation}
can be measured. In the above equation $|\vec{k}\,|$ is the outgoing photon energy, $w_{nl}$ (are
normalized to the unity) gives the absorption probability from each $nl$ pionic level,
taking into account the electromagnetic transitions and the strong absorption. $\Gamma^{abs}_{nl}$ is the total pion absorption width from the orbit $nl$ and $\Gamma_{nl}^{(\gamma)}$
is the width due to the radiative capture of the pion from the orbit $nl$. We will present results for carbon and calcium, and we use the same values for $w_{nl}$  and $\Gamma^{abs}_{nl}$ as in Ref.~\cite{Chiang:1989ni}, which are collected in Table \ref{table:2pion}. Our predictions are also given in the same table, while the differential decay branching ratios are displayed in Fig.~\ref{fig:11}.

Let us first notice  that also here the use of interacting SFs produces a quenching of the QE peak. Actually,  the in-medium dispersion relations shift the position of the peak about 10 MeV towards lower outgoing photon energies (higher transferred energies to the nucleus), and generate  a tail which goes into the low  photon energy region. The width of the particle-nucleon (see diagram of Fig.~\ref{fig:2p2h}) also contributes to this tail. This $10$ MeV difference between the position of the peaks, which was almost unnoticed for intermediate energies, here plays an important role.

In the case of $^{40}$Ca we  see that the position of the QE peak for the SF+RPA stays in very good agreement with the data. However, and despite the improvement due to the use of realistic SFs, we observe a
clear discrepancy with experiment at photon energies below 100 MeV. In our microscopic description, the origin
of the distribution comes from the motion of the nucleons in the nucleus. Mechanisms where two nucleons are simultaneously excited with the $\gamma$ creation would give rise to
photons with less energy\footnote{As mentioned,  the particle-nucleon width included in the particle SF  contributes to the tail. Note however, there exist other 2p2h mechanisms, involving meson-exchange-currents or the excitation of the $\Delta(1232)$ (see the discussion of Sect. 8 of Ref.~\cite{Chiang:1989ni}).} (these are different mechanisms than final state interaction of the struck nucleon in one body processes because the photon has already been created and does not modify its energy). It was argued in Ref.~\cite{Chiang:1989ni} that such contributions could  explain the observed discrepancies at low photon energies. This was confirmed in \cite{Amaro:1997ed}, where two-body mechanisms were taken into account using a semi-phenomenological approach. The SF+RPA decay width distribution also underestimates the data for photon energies above 130 MeV (marked with a vertical line in Fig.~\ref{fig:11}), this is to say above the $^{39}{\rm K}+n$ threshold. This region cannot be properly described with the 
present 
formalism, because it can only be filled in by discrete transitions (delta-like peaks convoluted with the experimental photon energy resolution, which is around 2 MeV~\cite{Bistirlich:1972jy}) of the type,
\begin{equation}
 (^{40}{\rm Ca} -\pi^-)_{\rm bound} \rightarrow \gamma + ^{40}{\rm K}^*
\end{equation}
where the final $^{40}$K nucleus is left either in the ground or in an excited state. These contributions are not properly included in the present approach, and their evaluation requires certainly a good description of the nuclear states of the initial and final nuclei.  The contribution above 130 MeV is moderately small, but together with the deficiencies discussed above at low energies  explain why the current model underestimates by around a 30\% the measured ratio $R^{(\gamma)}$ (see caption of Table \ref{table:2pion}). 

For $^{12}$C the situation is somehow different and the discrete transitions play a more important role and they are clearly visible in the spectrum.  In this case, the  $^{11}{\rm B}+n$ threshold is located at $|\vec{k}\,|\sim 122$ MeV, and peaks above a continuum are observed at 117, 120,
and 125 MeV. The peak at 125 MeV can be associated with production of the $^{12}$B ground state, while the another two peaks are related to transitions to excited states of $^{12}{\rm B}^*$~\cite{Bistirlich:1972jy}. 
Except for the high photon energy region, clearly dominated by these peaks, and the low energy tail, where two-nucleon mechanisms need to be included, the SF+RPA distribution provides also in carbon 
a reasonable description of the spectrum. It is remarkably better than that obtained when these nuclear effects are not taken into account. With respect to integrated ratios, and for meaningful comparison of our predictions with data, it is necessary to subtract the discrete contributions. The integrated ratio accounting for the one neutron knock-out contribution is estimated to be $R^{(\gamma)}_{\rm exp; cont} \sim (1.50\pm 0.15)\%$ in \cite{Bistirlich:1972jy}, that is around a 40\% higher than our prediction. The difference should be partially attributed to the low energy tail, but other source of the deviations comes from the experimental absorption
widths used in the present calculation and those  in which the experimental set--up was based on.
The deviations may also be due to the uncertainty of the values of $w_{nl}$. It would be interesting to disentangle experimentally the 
capture from different atomic states to allow a direct comparison with the theory, free of the assumptions made on the values of $w_{nl}$.

\subsubsection{Inclusive muon capture} \label{sec:muoncapt}
The analysis of the inclusive muon capture results is similar to that presented in the previous subsection for the radiative pion capture. The most important difference is that obviously the outgoing neutrino distributions have  not been measured. In addition, the interaction vertex is also different, and  the transferred energy to the nucleus, and thus the maximum momentum transfer, is around 35 MeV (mass difference between the pion and the muon) smaller than in the case of pion capture. This different kinematics influences the effects produced by the non-free SFs, as shown in Fig. \ref{fig:profile}.
\begin{figure}[h!]
\centering
\includegraphics[scale=0.35]{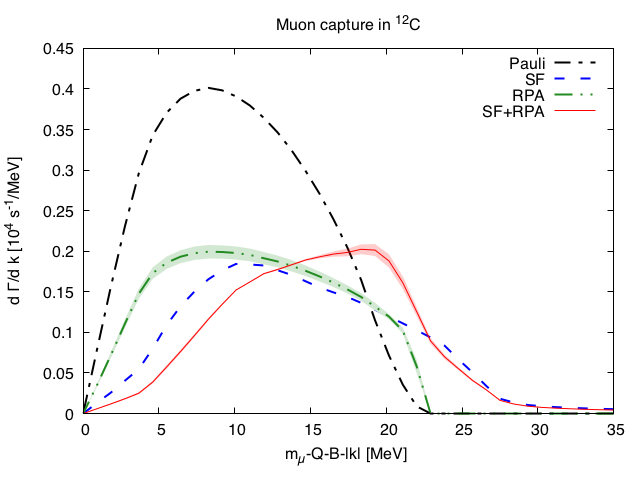}
\includegraphics[scale=0.35]{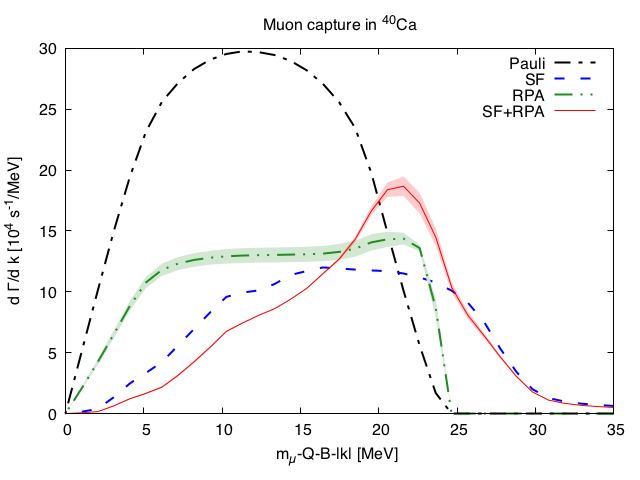}
\caption{Inclusive muon capture differential widths for
 $^{12}$C (left) and $^{40}$Ca (right), as a function of the energy transfer. Error bands on the RPA predictions show MC 68\% CL intervals derived from the uncertainties in the $ph$($\Delta$h)--$ph$($\Delta$h) effective interaction.}
\label{fig:muon-capture}
\end{figure}
The results for muon capture are shown in Table \ref{table:2}. We do not study heavy nuclei, like $^{208}$Pb, because our SFs were evaluated for symmetric nuclear matter. Our predictions stay in a very good agreement with the data, however the actual description could be likely poorer since, in principle, discrete contributions have not been properly taken into account, as we discussed for the case of pion capture. Nevertheless, the results of Table \ref{table:2} clearly show that RPA and  SF effects provide  a much better description of the data. RPA correlations induce modifications on the SF integrated decay rates significantly less important than those appreciated in the free LFG results. However, the RPA collective effects significantly modify the shape of the decay width distributions, as can be seen in Fig.~\ref{fig:muon-capture}, producing a shift of the maximum position, which is moved towards (higher) energies transferred to the nucleus of around 20  MeV. Indeed the RPA produces an enhancement of 
the distribution in this region of excitation energies, which can be related to the nuclear giant resonances (see 
for instance Refs.~\cite{Speth:1980kw, rpa_par2, Kamerdzhiev:2003rd,Botrugno:2005kn}).  A similar situation could be also seen  in Fig.~\ref{fig:11} for the case of pion capture, where we also see that the RPA correlations increase the SF results for photon energies of around 100 (110) MeV in carbon (calcium). Note however,  the individual giant resonances would show up as narrow peaks in the decay width distributions, while in the present approach, the RPA correlations provide only an enhanced signature, which likely will give a reasonable  description of  the integrated distributions. 
\begin{table*}[h]
\centering
\begin{tabular}{|| l | c c c c | c||}
  \hline
Nucleus & Pauli ($10^{4}$ s$^{-1}$) & RPA ($10^{4}$ s$^{-1}$) & SF ($10^{4}$ s$^{-1}$) & SF+RPA ($10^{4}$ s$^{-1}$) &  Exp. ($10^{4}$ s$^{-1}$)\\
\hline
\hline
$^{12}$C & 5.76 & $3.37\pm0.16$ &3.22  &   $3.19\pm0.06$ & $3.79 \pm 0.03$\\
$^{16}$O & 18.7& $10.9\pm 0.4$ & 10.6 & $ 10.3\pm 0.2$ & $10.24\pm 0.06$ \\
$^{18}$O & 13.8 & $8.2\pm 0.4$ & 7.0 & $8.7 \pm 0.1$& $8.80\pm 0.15$\\
$^{23}$Na & 64.5 & $37.0\pm 1.5$ &  30.9 & $34.3 \pm0.4$ & $37.73\pm0.14$\\
$^{40}$Ca & 498 & $272 \pm11$ &  242 & $242\pm 6$ & $252.5\pm0.6$\\
  \hline
 \hline
\end{tabular}
\caption{Experimental and theoretical total muon
  capture widths for different nuclei. Data are taken from
  Ref.~\protect\cite{Suzuki:1987jf}, and when more than one measurement is
  quoted in \protect\cite{Suzuki:1987jf}, we use a weighted average:
  $\overline{\Gamma}/\sigma^2 = \sum_i \Gamma_i/\sigma_i^2$, with
  $1/\sigma^2 = \sum_i 1/\sigma_i^2$. Theoretical errors in the RPA predictions show MC 68\% CL intervals derived from the uncertainties on the $ph$($\Delta$h)--$ph$($\Delta$h) effective interaction.}
\label{table:2}
\end{table*}

\subsubsection{The inclusive $^{12}{\rm C}(\nu_\mu,\mu^-)X$
 and $^{12}{\rm C}(\nu_e,e^-)X$ reactions near threshold} \label{sec:lowcross}
The low energy pion and muon capture decay rates  discussed in the previous subsections were measured with a good precision and  certainly provide an important test for our model. Here we will compare our results with other existing experimental neutrino low energy data. One of the characteristics of neutrino experiments is that the beams are not monochromatic  and thus the nuclear cross section should  be folded with 
the neutrino energy-flux $F(E_{\nu})$,
\begin{equation}
\sigma =\frac{1}{N} \int_{ E_{\nu}^{min}}^{E_{\nu}^{max}}dE_{\nu} \sigma(E_{\nu}) F(E_{\nu}), \quad N = \int_{ E_{\nu}^{min}}^{E_{\nu}^{max}} dE_{\nu} F(E_{\nu})
\end{equation}
The flux depends on the neutrino source and  for the  experiments (LAMPF, KARMEN, LSND) that we will consider in this subsection, electron neutrinos were produced from the  muon decay at rest ($\mu^+ \rightarrow \nu_e+\bar{\nu}_{\mu} + e^-$), and in this case the flux is approximately described by the Michel distribution,
\begin{equation}
F(E_{\nu}) \propto E_{\nu}^2(E_{\nu}^{max}-E_{\nu}), \quad E_{\nu}^{max} = \frac{m_{\mu}^2-m_e^2}{2m_{\mu}}\approx 53\, {\rm MeV}, \quad E_{\nu}^{min}=0.
\end{equation}
In the LSND experiment at Los Alamos, the inclusive
$^{12}{\rm}C(\nu_\mu,\mu^-)X$ cross section was measured using a pion
decay in flight $\nu_\mu$ beam, with energies ranging from zero\footnote{The neutrino laboratory threshold energy $E_{\nu}^{min}$  is around 123 MeV.} to 300
MeV (distribution is given in \cite{Albert:1994xs}).
\begin{figure*}[h]
\centering
\includegraphics[scale=0.3]{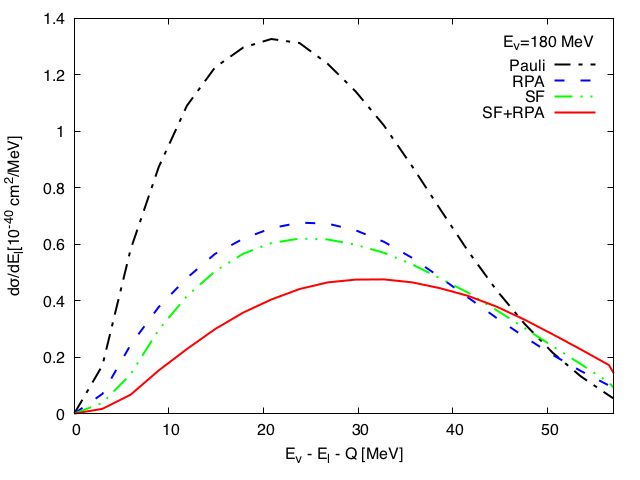}
\includegraphics[scale=0.3]{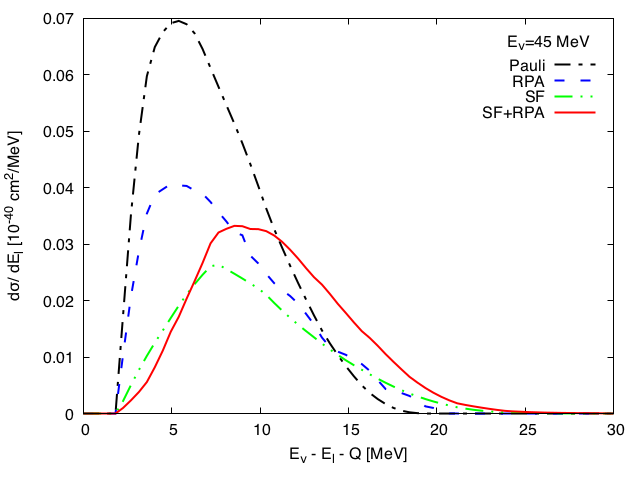}
\caption{Left (right) panel: Differential cross section for  inclusive CC muon (electron) neutrino scattering from $^{12}$C at 180 (45) MeV. The calculations have been done using  non-relativistic kinematics and with/without SF and RPA effects.}
\label{fig:22}
\end{figure*}
\begin{figure}[h]
\centering
\includegraphics[scale=0.3]{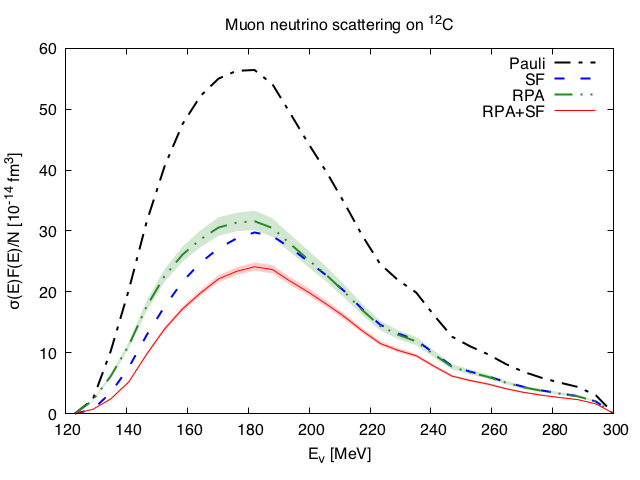}
\includegraphics[scale=0.3]{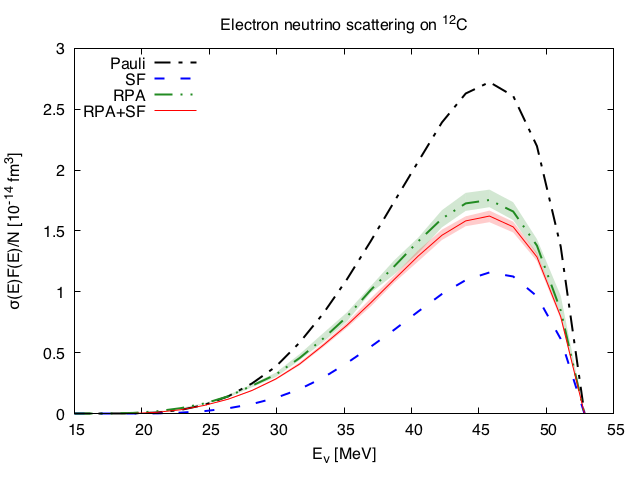}
\caption{ Predictions for the LSND measurement of the
$^{12}$C $(\nu_\mu,\mu^-)X$ reaction (left panel) and the $^{12}$C
$(\nu_e,e^-)X$ reaction near threshold (right panel). Neutrino cross sections have been convoluted with the corresponding  flux. Error bands on the RPA predictions show MC 68\% CL intervals derived from the uncertainties in the $ph$($\Delta$h)--$ph$($\Delta$h) effective interaction. }
\label{fig:21}
\end{figure}
The electron neutrino flux distribution has a maximum around $35$ MeV, while for the muon neutrino beam, over $80\%$ of the flux has an energy below $180$ MeV. Thus, these processes involve very low energy transfers,  as can be seen in Fig. \ref{fig:22}, especially in the electron neutrino case where the excitation energies are only of few MeV, and hence we are facing  the limit of applicability of the model.  Nevertheless, the results compiled in Table \ref{table:2K} (flux-weighted distributions shown in Fig. \ref{fig:21})  stay in surprisingly  good agreement with the data from LSND, KARMEN and LAMPF experiments. Nuclear effects (SF+RPA)  turn out to be essential and clearly improve the results obtained by only imposing  Pauli blocking and the correct energy balance in the reactions (results denoted as Pauli in Fig. \ref{fig:21} and  Table \ref{table:2K}).  In the table, a few selected theoretical calculations [large
basis shell model (SM) results of Refs.~\cite{Hayes:1999ew,Volpe:2000zn} and the CRPA ones from Ref.~\cite{Kolbe:2003ys}] are also compiled.
Our approach might look simplified with respect to the ones
just mentioned, but it incorporates both RPA and SF corrections and provides a description of these low energy cross sections as good, if not better, that any of them.

With all kinds of precautions, minding the low excitation energies involved, the LFG model of interacting nucleons, supplemented with a proper energy balance and  RPA collective effects, 
provides a more than reasonable combined description of the
inclusive muon capture in $^{12}$C and of the measurements of the
$^{12}$C $(\nu_\mu,\mu^-)X$ and $^{12}$C $(\nu_e,e^-)X$ reactions near threshold. 
\begin{table*}[h!]
\centering
\begin{tabular}{|r | c c| c c|ccc| c c c||}
\hline
 & Pauli & RPA & SF & SF+RPA & SM & SM & CRPA & & Experiment & \\
\hline
& & & & & \cite{Hayes:1999ew}& \cite{Volpe:2000zn} & \cite{Kolbe:2003ys} & LSND~\cite{Albert:1994xs} & LSND~\cite{Athanassopoulos:1997rn} & LSND~\cite{Auerbach:2002iy} \\[5pt]
$\bar{\sigma}(\nu_{\mu},\mu^-)$ & 23.1 &$13.2\pm0.7$ & 12.2 & $9.7\pm0.3$ &  13.2 & 15.2 & 19.2 & $8.3\pm0.7\pm1.6$ & $11.2\pm0.3\pm 1.8$ & $10.6\pm 0.3\pm 1.8$\\
  \hline
  & & & & & & & &KARMEN~\cite{Bodmann:1994py} & LSND~\cite{Athanassopoulos:1997rm} & LAMPF~\cite{Krakauer:1991rf} \\[5pt]
$\bar{\sigma}(\nu_{e},e^-)$ & 0.200 & 0.$143\pm0.006$ & 0.086 &$0.138\pm0.004$ & 0.12 & 0.16 & 0.15 &  $0.15\pm 0.01\pm 0.01$ & $0.15\pm0.01$& $0.141\pm0.023$ \\
 \hline
\end{tabular}
\caption{ Experimental and theoretical flux averaged
  $^{12}{\rm C}(\nu_\mu,\mu^-)X$ and $^{12}{\rm C}(\nu_e,e^-)X$ cross
  sections in 10$^{-40}$ cm$^2$ units. Theoretical errors in the RPA predictions show MC 68\% CL intervals derived from the uncertainties on the $ph$($\Delta$h)--$ph$($\Delta$h) effective interaction. We also quote results from other calculations (see text for details).}
\label{table:2K}
\end{table*}

\section{Conclusions}
\label{sec.concl}
We have presented a theoretical description of various QE processes within the many-body model used in \cite{Nieves:2004wx}, focusing on the effect produced by the inclusion of SFs, which account for the change of the dispersion relations of the interacting nucleons embedded in a nuclear medium. SFs are responsible for the quenching of the QE peak, produce a spreading of the strength of the response functions to higher energy transfers and shift the peak position in the same direction. The overall result is a decrease of the integrated cross sections and a considerable change of the differential shapes. RPA effects in integrated decay rates or cross sections  become significantly smaller when SF  corrections are also taken into account, in sharp contrast to the case of a free LFG where they lead to large reductions, even of around 40\%. This interesting result was mentioned already in \cite{Nieves:2004wx}, and  it is mainly due to 
the change of the nucleon dispersion relation in the medium (effects of the real parts of the particle and hole nucleon self-energies). Moreover, 
this  is also in agreement with the findings of Refs.~\cite{Pandey:2014tza, Pandey:2016jju}, from which one can conclude that RPA effects on top of the HF results are moderately small for sufficiently large values of $|-q^2|$, far from  the giant-resonance regime.

The final results for low energy processes (including both RPA and full SF effects for the very first time), although subject to some theoretical errors (originated from the RPA parameters uncertainty and the possible contribution of discrete states), describe data with a good precision, and provide a clear improvement of the poor description obtained by only imposing  Pauli blocking and the correct energy balance in the reactions. For  radiative pion capture,  we observe that the use of realistic SFs places the QE peak in a reasonable position and changes the  shape of the differential decay width, making it definitely more accurate than that obtained from the  LFG or the RPA predictions. However, the description is obscured by a discrete spectrum of resonances  not taken into account in the model. For muon capture, we only have at our disposal data of integrated widths; these rates  are well recovered by our model for various symmetric nuclei. These results, along with the LSND, KARMEN and LAMPF neutrino 
cross sections 
on carbon near threshold, which also stay in  agreement with our SF+RPA predictions, confirm the reliability of the model derived in \cite{Nieves:2004wx}. This also ensures the accuracy of the 
predictions obtained within this model for intermediate energy  neutrino scattering cross sections of interest for oscillation experiments, which are also given, and that for the first time have been obtained considering full SFs for both particle and hole nucleons, as well. We also show that errors on the $\sigma_\mu/\sigma_e$ ratio are much smaller than 5\%, and also  much smaller than the SF+RPA nuclear corrections,  which produce significant effects, not only in the individual cross sections, but also in their ratio for neutrino energies below 
400 MeV. These latter nuclear corrections, beyond Pauli blocking, turn out to be thus essential  to perform a correct analysis of  appearance  neutrino oscillation events in long-baseline experiments.

%
%
%
\begin{acknowledgments}
 We acknowledge enlightening discussions with A. Lovato and N. Rocco. This research
 has been supported by the Spanish Ministerio de Econom\'\i a y
 Competitividad and European FEDER funds under  contracts FIS2014-51948-C2-1-P and SEV-2014-0398, by
 Generalitat Valenciana under Contract PROMETEOII/2014/0068.
 \end{acknowledgments}
%
\bibliography{neutrino}
\end{document}